\documentclass{aa}  
\usepackage{txfonts}
\usepackage{booktabs} 
\usepackage{graphicx}
\usepackage{amssymb}
\usepackage{amsmath}
\usepackage{float}
\usepackage{gensymb}
\usepackage{hyperref}
\usepackage{subfig}
\usepackage{threeparttable}
\usepackage{multicol}
\usepackage{multirow}
\usepackage{longtable}
\usepackage{rotating}
\usepackage{tabularx}
\usepackage{afterpage}
\usepackage{placeins}
\usepackage{ulem} 
\usepackage{xcolor}
\usepackage{csvsimple}
\usepackage{afterpage}
\usepackage{caption}
\usepackage[switch]{lineno}
\usepackage{lipsum}
\usepackage{url}

\begin{document} 

\authorrunning{Li et al. 2025}

\titlerunning{The morphological stability of open clusters}

\title{The morphological stability of open clusters: A new 2D perspective}

\author{Yuting Li \inst{1}, Qingshun Hu \inst{1, 2}, Yufei Cai \inst{1}, Yu Dai \inst{1}, Mingfeng Qin \inst{3,4} \and Yangping Luo \inst{1}}
	
\institute{School of Physics and Astronomy, China West Normal University, No. 1 Shida Road, Nanchong 637002, People's Republic of China (\email{qingshun0801@163.com}\label{inst1}) \and Laboratoire d'Astrophysique de Bordeaux, Univ. Bordeaux, CNRS, B18N, all\'ee Geoffroy Saint-Hilaire, 33615 Pessac, France \label{inst2} \and Institute for Frontiers in Astronomy and Astrophysics, Beijing Normal University, Beijing 102206, People's Republic of China \label{inst3} \and School of Physics and Astronomy, Beijing Normal University, Beijing 100875, People's Republic of China \label{inst4}}

\date{Received 29 December 2025; Accepted 22 February 2026}

  \abstract
   {Open clusters (OCs) usually evolve gradually as the number of their members changes, which can be manifested in their morphological characteristics. Therefore, the morphological study of OCs lays the foundation for a better understanding of their formation and evolutionary processes.}
   {We aim to investigate the morphological stability of 1,490 OCs and further explore the potential change of morphological stability of the OCs at different spatial positions, using the OC catalog from the literature.}
   {We delineate the two-dimensional (2D) morphology of OCs quantitatively in the projection perpendicular to the Galactic disk plane by the rose diagram and analyze the slope changes between the morphological stabilities ($\mathrm{S}_{\mathrm{core}}$/$\mathrm{S}_{\mathrm{outer}}$ and N$_{\mathrm{core}}$/$\mathrm{N}_{\mathrm{outer}}$) and the number of members (N) within tidal radii to investigate the influence of the external environment on the OCs at different spatial positions.}
   {We define for the first time a new morphological stability parameter N$_{\mathrm{core}}$/$\mathrm{N}_{\mathrm{outer}}$, a ratio of member numbers between cluster core and outer areas within tidal radii, which has a significant positive correlation against N, with a slope of 1.140~$\pm$~0.039, significantly steeper than the 0.720~$\pm$~0.026 measured for $\mathrm{S}_{\mathrm{core}}$/$\mathrm{S}_{\mathrm{outer}}$. This demonstrates that the stellar density in the core is a more sensitive tracer for morphological stability than geometry. Spatially, the radial sample OCs have larger slopes of N$_{\mathrm{core}}$/$\mathrm{N}_{\mathrm{outer}}$ and $\mathrm{S}_{\mathrm{core}}$/$\mathrm{S}_{\mathrm{outer}}$ against N, with 1.083~$\pm$~0.116 and 0.733~$\pm$~0.080, respectively, whereas those in the tangential direction 1.013~$\pm$~0.110 and 0.529~$\pm$~0.075, respectively, which means that the impact on sample OCs from tidal forces directed toward the Galactic center is possibly stronger than that from the shear force caused by the differential rotation of the Galactic disk. Moreover, the sample OCs within 90$^{\degree}$ of the Galactic center, closer to the bar, exhibit slopes below 0.6 of $\mathrm{S}_{\mathrm{core}}$/$\mathrm{S}_{\mathrm{outer}}$ against N, indicating heightened external perturbations and diminished stability. But the opposite is true for the side greater than 90$^{\degree}$. Thus, this illustrates that the influence of the external environment on our sample OCs is asymmetrical. Besides, the sample OCs younger than 30 Myr display a shallow slope of 0.751~$\pm$~0.166, with those older than 800 Myr (1.442~$\pm$~0.128), reflecting that young OCs likely endure both internal disruptions, such as early dynamical heating weakening core binding and more severe external disturbances, compared to older OCs.}
  {The morphological stability of OCs is not only determined by their gravitational binding, but also strongly modulated by the external environment in which they are located.}

   
\maketitle

\section{Introduction} 

\hspace{2em}{Star clusters} are typically groups of stars bound together by gravitational interactions \citep{Bergond+2001, Lada+2003, Piskunov+2006}, simply classified into two main types based on their morphology: Globular clusters (GCs) \citep{Fujii+2024} and {Open clusters (OCs)} \citep{Perryman+2025}. {OCs} that form from the same giant molecular cloud (GMCs) \citep[e.g.,][]{Meynet+1993, Alves+2012, Megeath+2016} are initially in a nesting phase, characterized by high-density gas accompanied by strong gas loss. After that, the number of members of OCs gradually changes over time, usually decreasing. For instance, two-body relaxation \citep[e.g.,][]{Bergond+2001, Mamon+2019, Broggi+2024} will lead to energy redistribution, causing some low-mass member stars to migrate to the tidal tail or even escape from the OCs \citep{Portegies_Zwart+2010, Tang+2019, Akhmetali+2025}. Since the morphology of OCs is concerned with the number of their members, changes in the number of member stars will inevitably affect the OC morphology. This demonstrates that the morphology of OCs is closely related to their evolution. Therefore, studying the morphology of OCs can enhance intuitive understanding of their evolution \citep[e.g.,][]{Meingast+2019, Dalessandro+2021, Pang+2021, Hu+2021b, Hu+2021a, Tarricq+2021, Tarricq+2022, Pang+2022, Della+2023}.

The release of high-precision astrometry data from the Gaia satellite\footnote{\url{https://www.cosmos.esa.int/web/gaia/dr3}} \citep{GAIADR2+2016, GAIADR3+2023} has enabled the precise reconstruction of the two-dimensional (2D) or three-dimensional (3D) structure of OCs. Studies have revealed that some OCs exhibit morphological asymmetries, core–halo structures \citep{Kadam+2024}, extended tidal tails \citep{Sharma+2025}, and even signs of imminent disintegration \citep{Almeida+2025}. Moreover, there is also an extensive body of literature on the statistical morphology of OCs. For example, \cite{Zhong+2022} conducted a structural analysis of 256 OCs, revealing the radial density profile \citep{Camargo+2012} in their outer halo regions deviating significantly from the King model profile \citep{King+1962}. Subsequently,  \citet{Hu+2023} employed the rose diagram overlay method to study the morphological structures of 88 OCs, primarily analyzing their three 2D projections in the heliocentric Cartesian coordinate system to investigate layered structures and found that 74 of the 88 sample OCs present clear 3D layered structures. In 2024, \citet{Hu+2024} defined the ``morphological coherence'' for the first time to quantify the 3D morphological difference between the inner and outer regions of 132 OCs located within 1 kpc of the Sun, and investigated, in a Cartesian reference frame, the full 3D manifestation of this ``morphological coherence''. \citet{Hu+2025a} examined the structural parameters of 105 OCs, suggesting that OCs with more member stars within their tidal radii possess stronger gravitational binding forces. In the same year, \citet{xu+2025} performed a 3D principal component analysis (3D PCA) on eight OCs, revealing that the tidal structure orientation of certain OCs is perpendicular to their direction of motion. In addition, \citet{Lang+2025} performed N-body simulations of 279 nearby OCs, correlating member star spatial distributions with layered structures, and suggested that OCs with fewer members typically lack such structures.

Although significant progress has been made in studying the morphological structure of OCs in recent years, most studies have been concerned with 3D spatial distributions, limited by sample size, especially the research on the morphological stability of OCs. As there are symmetrical errors in the parallax measurement of Gaia \citep{GAIADR2+2016, GAIADR3+2023}, a direct propagation to distant distance through inverting parallax (1/$\omega$) could lead to a significant artificial elongation of the 3D spatial distribution of OCs along the line of sight \citep[see, e.g.,][]{Smith+1996, Bailer-Jones+2015, Luri+2018, Carrera+2019, Hu+2023}. Generally, to recover the true 3D spatial distribution of OCs, one restricted the OCs to d~$\leq$~500~pc precisely to keep their distance uncertainties below $\sim$~6~pc after distance correction \citep[e.g.,][]{Hu+2023}. However, OC samples within 500 pc are limited. Therefore, to extend the OC sample and avoid the fake elongation, we adopt, in this work, a 2D projection in the Galactic spherical coordinate system to explore the morphological stability of a large number of OCs that are almost beyond 500 pc.

While this study did not directly investigate the morphological stability of OCs in 3D space, we can still obtain and study their morphological stability in specific projection planes in 3D space through the projection effect. In this paper, we systematically analyzed the morphological stability of 1,490 OCs by the rose diagram construction, which was more than 10 times the sample number in previous studies \citep{Hu+2023, Hu+2024, Hu+2025a}.

The paper is organized as follows. In Section~\ref{sec:data}, we introduce the data used in our work. We then describe the methodology adopted in this study in Section~\ref{sec:methods}. Section~\ref{sec:results} presents all results of our entire work. We discuss the sensitivities and potential systematic biases of our approaches in Section~\ref{Discussion}. In the end, Section~\ref{sec:summary and conclusion} provides a summary and conclusion.

\section{Data}
\label{sec:data}
 
\hspace{2em}Based on Gaia DR3 data, \citet{van+2023} provides a catalog of OC member stars complete to G~>~18 mag. \citet{Hunt+2024} recently determined 5,647 reliably bound OCs, marked by `o' symbol, from a sample of  6,956 star cluster candidates. These two catalogs serve as the primary data sources for this work. We selected all OCs flagged with `o' from \citet{Hunt+2024} and cross-matched them with the corresponding member star catalog from \citet{van+2023} by cluster names in \texttt{topcat} \footnote{\url{https://en.wikipedia.org/wiki/TOPCAT_(software)}}. In order to obtain reliable member stars, we retained only those members with membership \texttt{Pmemb~=~1} from \citet{van+2023}. Previous studies and catalogs \citep [e.g.,][]{Cantat-Gaudin+2018, Cantat-Gaudin+2022, Hu+2023} have typically considered OCs with at least 50 members to be robust for analysis. Thus, we limited the sample OCs to those with the number of member stars greater than or equal to 50 members, and a total of 1,490 OCs remained. Finally, the member stars of 1,490 OCs were cross-matched with the Gaia DR3 data to obtain their fundamental parameters.

\section{Methods}
\label{sec:methods}

\hspace{2em}In order to investigate the morphological stability of OCs, we first calculated their morphological stability parameters by the rose diagram construction. The specific approaches are given below.

\subsection{Construction of the rose diagram}
\label{sec:rose-diagram}

\hspace{2em}We adopted the rose diagram method developed by \citet{Hu+2025a} to quantify the morphologies of all OCs in our sample. We first determined a plane distributed by its members for each OC. Due to the inclination of the equatorial plane of the celestial coordinate system relative to the Galactic midplane, taking an OC as an example, we used the position coordinates and parallax of its members and their uncertainties (\texttt{RA\_ICRS}, \texttt{DE\_ICRS}, \texttt{Plx}, \texttt{e\_RAdeg}, \texttt{e\_DEdeg}, \texttt{e\_Plx}) to transform them into the Galactic coordinates. We employed a Monte Carlo sampling to properly propagate the observational uncertainties in this coordinate transformation. For every member of the OC, we performed 100 samples by drawing \texttt{RA\_ICRS}, \texttt{DE\_ICRS}, and \texttt{Plx} from their respective error distributions to obtain their Galactic spherical coordinates and uncertainties (\texttt{l}, \texttt{b}, \texttt{e\_l}, \texttt{e\_b}) via the \texttt{astropy}\footnote{\url{https://docs.astropy.org/}} package \citep{Astropy+2013, Astropy+2018}. For every sampling, we took the median of the Galactic coordinates (\texttt{l}, \texttt{b}) as the centric coordinate (\texttt{$\text{l}_{0}$}, \texttt{$\text{b}_{0}$}) of the OC. At the same time, the median of sampled parallax (\texttt{Plx}) was converted into distance (\texttt{d}).

\begin{figure}[h!]
	\centering
    \includegraphics[angle=0,width=88mm]{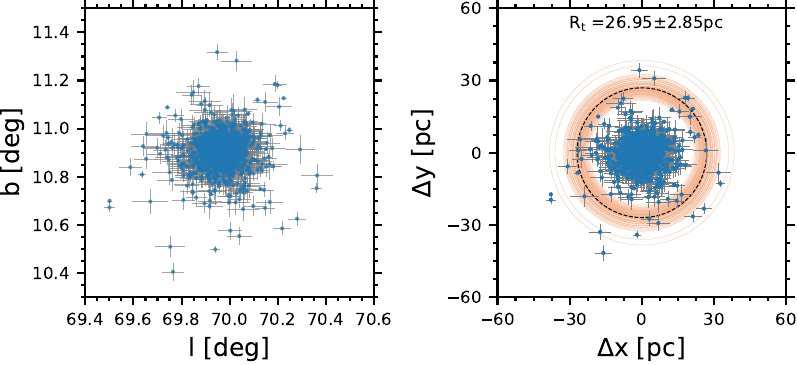}
	\caption{Distribution of \texttt{NGC\_6791} in the Galactic coordinate system (left panel) and its flattened projection distribution (right panel). The blue scattered points in each panel represent the member stars of the OC, with the gray bars being the errors of the position coordinates of its members. The orange circles correspond to the tidal radii obtained from 100 Monte Carlo sampling runs. The black dashed circle represents the mean value of the 100 results.}
	\label{fig: Galactic_Plane}
\end{figure}

Next, the converted coordinates and distances were substituted into Eq.~\ref{eq:coordinate} to project the positions onto the Galactic projection that is perpendicular to the line of sight:

\begin{equation}
\left\{
\label{eq:coordinate}
\begin{array}{l}
\Delta \mathrm{x} = \text{d} \cdot \sin(\text{l} - \text{l}_{0}) \cdot \cos \text{b} \\[12pt]
\Delta \mathrm{y} = \text{d} \cdot \bigl(\cos \text{b}_{0} \cdot \sin \text{b} - \sin \text{b}_{0} \cdot \cos \text{b} \cdot \cos(\text{l} - \text{l}_{0})\bigr)
\end{array}
\right.
\end{equation}

where $\text{l}_{0}$ and $\text{b}_{0}$ represent the coordinates of the center of the OC with $\texttt{d}$ being the distance from the OC. The mean and standard deviation of the 100 samples of ($\Delta \mathrm{x}$, $\Delta \mathrm{y}$) are taken as the final coordinates and their uncertainties, respectively. Figure~\ref{fig: Galactic_Plane} shows the Galactic distribution (left) and the flattened projection distribution of the members of \texttt{NGC\_6791} in physical space (right).

It is well known that only member stars within the tidal radius are bound by gravitation from their host OC. Therefore, we determined the tidal radius of each OC in the flattened spatial projection by the \texttt{King} model density profile fitting \citep{King+1962}:

\begin{equation}
\rho(r)=k\left[
\frac{1}{\sqrt{1+(r/r_{\mathrm{c}})^{2}}}
-\frac{1}{\sqrt{1+(r_{\mathrm{t}}/r_{\mathrm{c}})^{2}}}
\right]^{2}+c
\end{equation}

where $r_{\mathrm{c}}$ is the core radius, $r_{\mathrm{t}}$ is the tidal radius, and $k$ and $c$ are model fitting parameters. The fitting was performed by the \texttt{king\_model} function from the \texttt{diagnostics} subpackage of \texttt{gaia\_oc\_amd}\footnote{\url{https://github.com/MGJvanGroeningen/gaia_oc_amd}}. This process was repeated for each set of 100 samples, gaining a distribution of tidal radii for each OC. As shown in Fig.~\ref{fig: Galactic_Plane}, the mean value (marked by a black dashed circle in the right panel) and standard deviation of tidal radii from the 100 sampling for \texttt{NGC\_6791} give a tidal radius of $\mathrm{R}_{\mathrm{t}}$ = 26.95~$\pm$~2.85~pc.

The rose diagram construction is a quantitative approach to analyze the morphological structure of OCs \citep{Hu+2023, Hu+2025a}. By partitioning and statistically analyzing the distribution of OCs' members on a 2D projection plane according to azimuth angles within the spatial extent defined by the tidal radius ($\mathrm{R}_{\mathrm{t}}$) and combining it with polar coordinate visualization, one can effectively identify the morphological characteristics and asymmetries of OCs. The rose diagram method was first proposed by \citet{Hu+2023}. For the specific operation, readers can refer to this literature.

Prior to applying the rose diagram construction, we decomposed the flattened spatial projection planes of all sample OCs onto the planes perpendicular to the Galactic disk plane by multiplying the vertical coordinates of members on the flattened projections by the cosine of the median Galactic latitude of these members. In this paper, we obtained the rose diagram of sample OCs through the following steps. We first centered the coordinates in the plane corrected above of each OC with the Kernel Density Estimation (\texttt{KDE}) \citep{Seleznev+2016} density peak, then divided the projected plane into 12 sectors of 30$^{\degree}$ each. For each sector, we calculated two normalized parameters: the number of member stars $n$ and the median radial distance $d$. The radius of each sector was then calculated as $\mathrm{R} = \sqrt{n^2 + d^2}$, denoted by $\mathrm{R}_{\mathrm{i}}$, which is a dimensionless normalized quantity in the $\mathrm{i}th$ sector. The minimum value among these 12 radii was defined as the core radius ($\mathrm{R}_{\mathrm{core}}$), and the corresponding area $\pi \mathrm{R}_{\mathrm{core}}^2$ represents the core area ($\mathrm{S}_{\mathrm{core}}$). The total area of the irregular outer region ($\mathrm{S}_{\mathrm{outer}}$) was calculated as the sum of all areas in the sectors minus the core area. Their equations are as follows:

\begin{equation} \label{eqt1}
 \text{R}_{\mathrm{core}} = \min(\mathrm{R}_{\mathrm{i}}),  \mathrm{i}=[1,12]
\end{equation}
 
\begin{equation} \label{eqt2}
\text{S}_{\mathrm{core}} = \pi \cdot \mathrm{R}_{\mathrm{core}}^2
\end{equation}

\begin{equation} \label{eqt3}
\text{S}_{\mathrm{outer}} = \sum_{\mathrm{i}=1}^{n=12} \frac{\pi \cdot \mathrm{R}_\mathrm{i}^2}{12} - \mathrm{S}_{\mathrm{core}}
\end{equation}

Specifically, based on the mean tidal radius $\mathrm{R}_{\mathrm{t}}$ of each sample OC and its error $\sigma_{\mathrm{R}_{\mathrm{t}}}$, we constructed a normal distribution $\mathcal{N}(\mathrm{R}_{\mathrm{t}}, \sigma_{\mathrm{R}_{\mathrm{t}}}^{2})$. From this distribution, we independently drew 10 tidal radius values. For each sampling iteration $\mathrm{j}$, we used $\mathrm{R}_{\mathrm{t}}^{(\mathrm{j})}$ as the boundary to re-select OC member stars, calculated their projected coordinates relative to the cluster center, and quantified their 2D spatial distribution through the rose diagram method. Repeating this process 10 times, we obtained 10 independent estimates of the core radius R$_{\mathrm{core}}$, core area $\mathrm{S}_{\mathrm{core}}$, and outer area $\mathrm{S}_{\mathrm{outer}}$. Finally, we reported the sample mean and sample standard deviation of these parameters as the best estimates of the sample OC's morphological parameters and their uncertainties, as compiled in Table~\ref{tab:catalog}.

\begin{figure}[h!]
	\centering
    \includegraphics[angle=0,width=56mm]{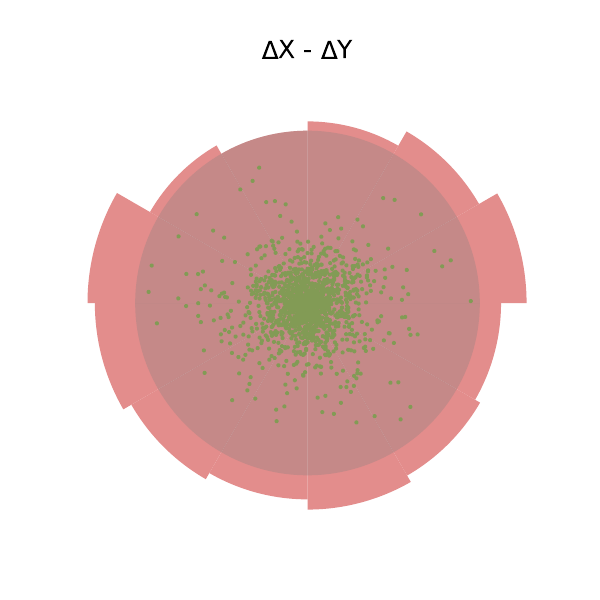}
	\caption{Rose diagram of the \texttt{NGC\_6791} on the 2D corrected projection, derived from one of ten samplings. The rose-colored sectors represent the normalized directional radial radius values obtained by the Rose diagram method. Each sector spans 30$^{\degree}$, with the first sector starting at 0$^{\degree}$ corresponding to the horizontal line on the right of this diagram. The second sector begins at a position 30$^{\degree}$ counterclockwise from the first sector, and subsequent sectors follow this pattern. The gray semi-transparent sectors are the average extension levels in each direction, with the green scattered points indicating the member stars.
}
	\label{fig: NGC_6791_rose}
\end{figure}

By the processes mentioned above, we can obtain the rose diagrams of all sample OCs. Figure~\ref{fig: NGC_6791_rose} shows the rose diagram of \texttt{NGC\_6791}, with $\mathrm{S}_{\mathrm{core}}$ marked in dark gray.

\subsection{The measurement of morphological stabilities}

\hspace{2em}\citet{Hu+2025a} first proposed the ratio of the core to the outer areas ($\mathrm{S}_{\mathrm{core}}/\mathrm{S}_{\mathrm{outer}}$, hereafter referred to as the morphological stability parameter I) to describe the morphological stability of the OCs. The essence of this parameter is consistent with the instability discussed by \citet{Hu+2023}, as both characterize the stability of an OC based on the structural ratio between the outer and the core region.

For an evolving OC, two-body relaxation \citep{Portegies+2001, Angelo+2025}, combined with the tidal stripping of weakly bound stars from the periphery, drives both the development of mass segregation and the global mass loss of the OC \citep{zquez+2017, Michael+2018}. The macroscopic outcome is an overall contraction of the cluster scale, with the central region becoming increasingly dense and gravitationally dominant relative to the outer envelope. A subvirial core ($\eta$~$<$~1) in a contracting state deepens the central gravitational potential \citep{Foster+2015}, accelerating the inward migration of massive stars and further enhancing the central density \citep{Della+2025}. Simultaneously, the outer envelope, heated by two-body relaxation and truncated by tidal forces, is progressively depleted as member stars approaching the escape velocity are preferentially stripped. Based on this theoretical expectation, we propose a direct observational diagnostic: the ratio of the number of member stars in the core to that in the outer region, $\mathrm{N}_{\mathrm{core}}/\mathrm{N}_{\mathrm{outer}}$ (hereafter referred to as the morphological stability parameter II). This parameter is designed to quantify the degree of structural tightening a cluster has undergone due to internal relaxation and external stripping. A high $\mathrm{N}_{\mathrm{core}}/\mathrm{N}_{\mathrm{outer}}$ value signifies: a core that is gravitationally bound strongly enough to retain its stars, and an outer region that the combined effects of relaxation and tides have significantly depleted.

The morphological stability parameter I, $\mathrm{S}_{\mathrm{core}}/\mathrm{S}_{\mathrm{outer}}$, relies on determining the areas of the core and outer regions. The morphological stability parameter II, $\mathrm{N}_{\mathrm{core}}/\mathrm{N}_{\mathrm{outer}}$, is calculated by directly counting the number of member stars located within these same respective regions.

\begin{table*}[htbp]

\centering
\caption{Complete catalog of parameters for 1490 OCs.}
\label{tab:catalog}
\renewcommand{\arraystretch}{1.2}
\resizebox{\textwidth}{!}{
\begin{tabular}{l c c c c c c c c c c}
\hline\hline
\rule{0pt}{3ex} Cluster & RA\_ICRS & DE\_ICRS & $\mathrm{R}_{\mathrm{t}}$ & R$_{\mathrm{core}}$  & N & $\mathrm{S}_{\mathrm{core}}$ & $\mathrm{S}_{\mathrm{outer}}$ & N$_{\mathrm{core}}$ &$\mathrm{N}_{\mathrm{outer}}$ \\ 

   &  (deg)  &  (deg) &  (pc) &  &  &  &  &  &  \\ 
\midrule
  (1) &  (2)  &  (3) &  (4)&  (5)&  (6)&  (7)&  (8)&  (9)&  (10)\\ 
\midrule

Alessi\_10 & $301.23$ & $-10.51$ & $143.43\pm69.86$ & $0.40\pm0.03$ & $100.00\pm0.00$ & $0.50\pm0.00$ & $1.17\pm0.00$ & $89.00\pm0.00$ & $11.00\pm0.00$ \\
Alessi\_12 & $310.86$ & $23.85$ & $33.93\pm7.64$ & $0.63\pm0.01$ & $302.00\pm1.10$ & $1.23\pm0.05$ & $1.86\pm0.01$ & $279.40\pm0.60$ & $21.60\pm0.60$ \\
Alessi\_19 & $274.79$ & $12.21$ & $22.48\pm23.53$ & $0.24\pm0.03$ & $91.70\pm7.18$ & $0.18\pm0.03$ & $1.51\pm0.26$ & $42.60\pm1.70$ & $49.40\pm1.70$ \\
Alessi\_2 & $71.64$ & $55.18$ & $36.83\pm14.61$ & $0.55\pm0.02$ & $192.00\pm0.00$ & $0.96\pm0.00$ & $1.65\pm0.00$ & $175.00\pm0.00$ & $17.00\pm0.00$ \\
Alessi\_20 & $2.60$ & $58.70$ & $79.40\pm36.80$ & $0.35\pm0.04$ & $297.00\pm0.00$ & $0.38\pm0.00$ & $1.78\pm0.00$ & $248.00\pm0.00$ & $49.00\pm0.00$ \\
Alessi\_21 & $107.71$ & $-9.45$ & $44.01\pm22.09$ & $0.48\pm0.02$ & $225.00\pm0.00$ & $0.72\pm0.00$ & $1.19\pm0.00$ & $191.00\pm0.00$ & $34.00\pm0.00$ \\
Alessi\_24 & $260.76$ & $-62.69$ & $9.85\pm1.11$ & $0.58\pm0.03$ & $147.40\pm0.92$ & $1.05\pm0.03$ & $1.64\pm0.03$ & $125.90\pm0.38$ & $21.10\pm0.38$ \\
Alessi\_3 & $109.17$ & $-46.19$ & $38.89\pm11.90$ & $0.53\pm0.02$ & $194.40\pm1.28$ & $0.88\pm0.02$ & $2.06\pm0.04$ & $179.80\pm0.20$ & $15.20\pm0.20$ \\
ASCC\_101 & $288.39$ & $36.34$ & $191.92\pm29.41$ & $0.26\pm0.02$ & $134.00\pm0.00$ & $0.22\pm0.02$ & $2.05\pm0.00$ & $95.00\pm0.00$ & $39.00\pm0.00$ \\
ASCC\_105 & $295.60$ & $27.28$ & $53.54\pm43.47$ & $0.37\pm0.02$ & $218.00\pm0.00$ & $0.42\pm0.02$ & $1.90\pm0.00$ & $131.00\pm0.00$ & $87.00\pm0.00$ \\
... & ... & ... & ... & ... & ...& ... & ... & ...& ... \\
\bottomrule
\end{tabular}}
\tablefoot{Column~(1) represents the cluster name; Columns~(2) and~(3) represent position coordinates of each OC. Column~(4) represents the mean and error of the tidal radius obtained from 100 samplings. Column~(5) represents the core radius and error obtained by rose diagram method, with Column~(6) denoting the number of member stars within the tidal radius of each OC and its error. Columns~(7) and~(8) respectively indicate the area and its error within the core region and outer region of each OC. Columns~(9) and~(10) represent the number of member stars in the core and the outer regions of each OC, respectively. This is a machine-readable table, which can be accessed through CDS.}
\end{table*}

\section{Results} 
\label{sec:results}

\subsection{Distribution of sample clusters}

\begin{figure}[h!]
	\centering
    \includegraphics[angle=0,width=88mm]{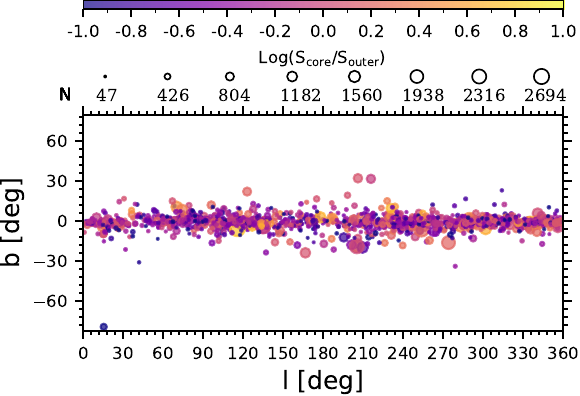}
	\caption{Map of the sample OCs in the Galactic coordinate system. The size of each circle color-filled is proportional to the number of members of each OC, with its color being coded by the morphological stability parameter I.}
	\label{fig: Galactic_latitude}
\end{figure}

\hspace{2em}Figure~\ref{fig: Galactic_latitude} shows the distribution of our sample OCs in Galactic coordinates. The majority of the sample OCs are concentrated at $|b|~\leq~20^{\degree}$. Moreover, owing to the complex and dynamically active environment of the thin disk in which these OCs reside, most among them have only a few members and weak morphological stability. On the contrary, the sample OCs that are far from the disk usually have a relatively larger number of member stars and exhibit stronger morphological stability. This means that in addition to the number of members, the morphological stability of OCs seems to be related to the spatial positions.

\begin{figure}[h!]
	\centering
    \includegraphics[angle=0,width=88mm]{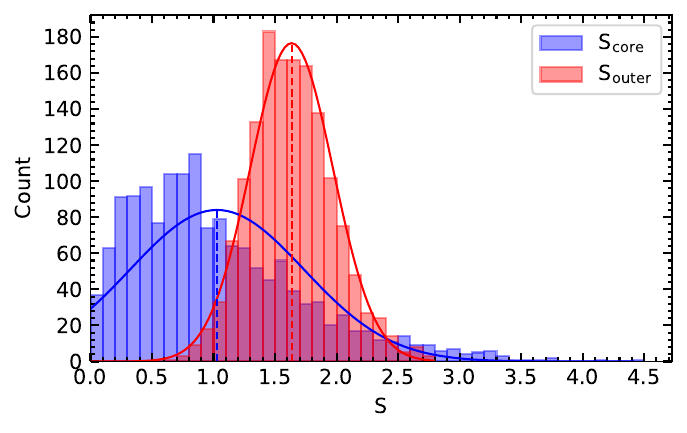}
	\caption{Histogram of the areas of the core ($\mathrm{S}_{\mathrm{core}}$, blue) and the outer ($\mathrm{S}_{\mathrm{outer}}$, red) of our sample OCs. Each set of data was fitted with a normal distribution (colored solid lines), with the mean positions being marked with vertical dashed lines.
}
	\label{fig: Distribution_Plot}
\end{figure}

Figure~\ref{fig: Distribution_Plot} shows the histograms of the core and outer areas of the sample OCs. Here, we have only considered the histogram of areas because the sample OCs differ greatly in the number of members in the core regions. We found from this figure that the core areas ($\mathrm{S}_{\mathrm{core}}$) and outer areas ($\mathrm{S}_{\mathrm{outer}}$) of our samples span a range from 0 to 4.5. The outer areas of most OCs lay between 1.0 and 2.5, while the majority of core areas were below 1.5. This suggested that for most clusters in our sample, the ratio $\mathrm{S}_{\mathrm{core}}/\mathrm{S}_{\mathrm{outer}}$ is less than 1, which was consistent with \citet{Hu+2025a}, who reported that more than 70 \% of their sample OCs had a 2D morphological stability parameter below 1. This also indicated that most of the OCs have lower morphological stability. It can be supported by the sample distribution in Fig.~\ref{fig: Galactic_latitude}, especially those with Log($\mathrm{S}_{\mathrm{core}}/\mathrm{S}_{\mathrm{outer}}$)~$<$~0. Moreover, it was apparent that the mean area of OC cores was smaller than that of OC outer regions, which suggested that a typical OC has a compact core and an extensive external structure.

We note that in this study, the sample OCs with $\mathrm{S}_{\mathrm{core}}$~=~0 or $\mathrm{N}_{\mathrm{outer}}$~=~0 should be excluded from our samples, because the morphological stability parameters I and II derived from them are devoid of physical meaning. \citep{Hu+2025a} established that when the core area of an OC is zero, it signifies that the OC exhibits no discernible core structure in the projected plane. We found that our sample does not contain cases where $\mathrm{S}_{\mathrm{core}}$~=~0, but it does contain cases where $\mathrm{N}_{\mathrm{outer}}$~=~0. Finally, a total of 16 OCs with $\mathrm{N}_{\mathrm{outer}}$~=~0 were excluded. Their parameters can be found in Table~\ref{tab:catalog}.

\subsection{Relationships between the morphological stabilities and the number of members}

\hspace{2em}\citet{Hu+2025a} has established a linear relationship between the morphological stability parameter I ($\mathrm{S}_{\mathrm{core}}/\mathrm{S}_{\mathrm{outer}}$) and the number of member stars (N) of OCs. In this section, we verified this relationship and extended it further by exploring the correlation between the morphological stability parameter II ($\mathrm{N}_{\mathrm{core}}/\mathrm{N}_{\mathrm{outer}}$) and N.

\begin{figure*}[htbp]
    \centering
    \includegraphics[angle=0,width=45mm]{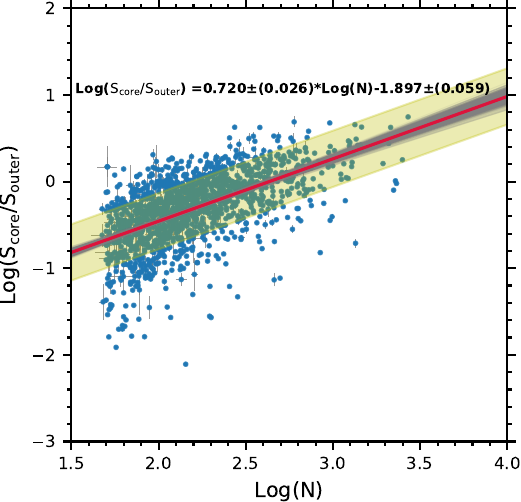}
    \includegraphics[angle=0,width=45mm]{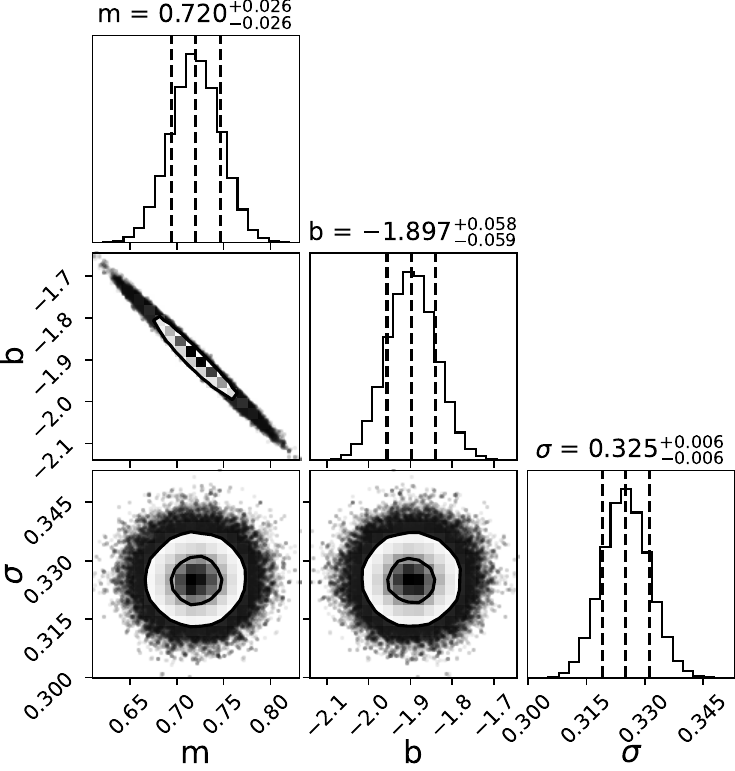}
    \includegraphics[angle=0,width=45mm]{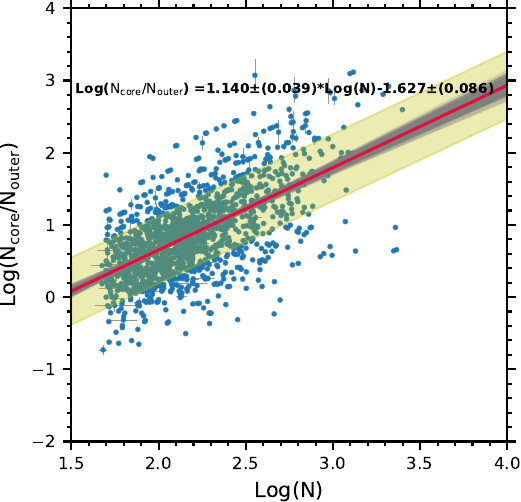}
    \includegraphics[angle=0,width=45mm]{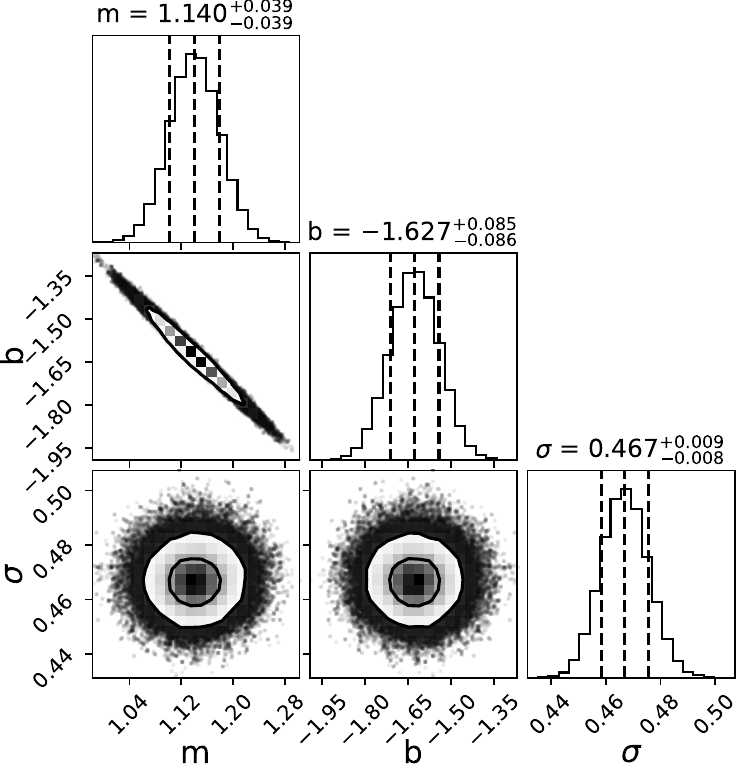}
    \caption{Relationships between the morphological stability parameters ($\mathrm{S}_{\mathrm{core}}$/$\mathrm{S}_{\mathrm{outer}}$: the first two panels from left to right; N$_{\mathrm{core}}$/$\mathrm{N}_{\mathrm{outer}}$: the last two panels) and the number of members (N). For ease of visualization, a logarithmic scale was adopted. Blue scatters in linear fitting panels represent our sample OCs, with the gray bars denoting their parameter errors. The faint gray lines show 1000 MCMC samples, with the thick red lines (the median values) and the yellow shaded bands ($1\sigma$) corresponding to our best parameter estimates. The two corner plots display the marginal posterior distributions (diagonal) and the 2-D joint posterior contours (below diagonal) of the fitted parameters ($m$, $b$, $\sigma$). Contours enclose the $1\sigma$ and $2\sigma$ confidence regions, and the median together with the 16th and 84th percentiles are quoted in each panel.}
    \label{fig:all_N_s}
\end{figure*}

To this end, we performed the linear fitting for them by Bayesian linear fitting model \citep[e.g.,][]{Anders+2017, HU+2025c} and the Markov Chain Monte Carlo (MCMC) method \citep[e.g.,][]{Foreman-Mackey+2013, Alfonso+2024, Lang+2025, Qiu+2025}. The Bayesian linear model is detailed in the Appendix~\ref{sec:mcmmc}. The fitting results are displayed in Fig.~\ref{fig:all_N_s}. We can see from this figure that there is indeed a strong linear correlation between $\mathrm{S}_{\mathrm{core}}$/$\mathrm{S}_{\mathrm{outer}}$ and N for our sample OCs. Moreover, the correlation of N$_{\mathrm{core}}$/$\mathrm{N}_{\mathrm{outer}}$ with N suggests the number of core member stars (N$_{\mathrm{core}}$) is a new metric for the gravitational binding in the morphological stability of OCs. As the number of member stars of OCs increases, particularly the number of member stars in the core, the gravitational binding capacity of the OCs strengthens, which conforms with the previous finding by \citet{Hu+2025a}. Furthermore, the slope derived from the morphological stability parameter II is steeper than that from stability parameter I, with nearly twice the slope within the margin of error of their slopes. And the \texttt{Pearson} correlation coefficients are 0.5721~$\pm$~0.0020 and 0.6052~$\pm$~0.0006, which were derived from the median and standard deviation of Pearson regression of the fitted line 1000 times for parameters I and II, respectively. Meanwhile, their P\_value are separately $2.84~\times~10^{-130}~\pm~1.72~\times~10^{-126}$ and $5.35~\times~10^{-148}~\pm~2.61~\times~10^{-147}$, indicating both two correlations being strong. Thus, the ratio of member stars in the core to those in the outer seems to be a somewhat more sensitive probe of the morphological stability of OCs than the ratio of core area to outer area, although the difference between them within uncertainties is modest.

\begin{figure}[h!]
    \centering
    \includegraphics[angle=0,width=65mm]{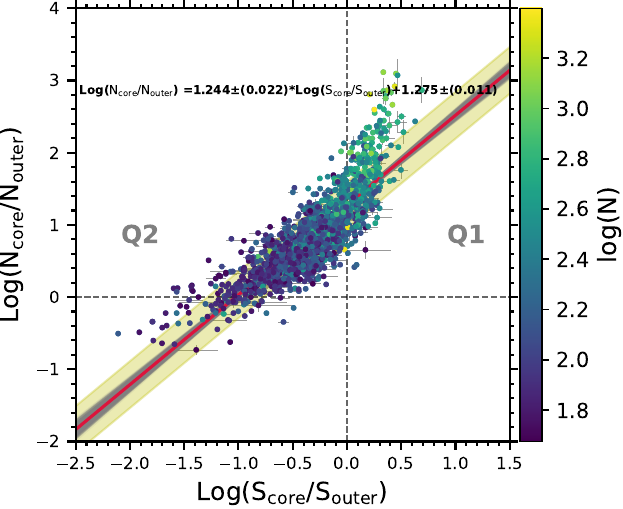}
    \includegraphics[angle=0,width=65mm]{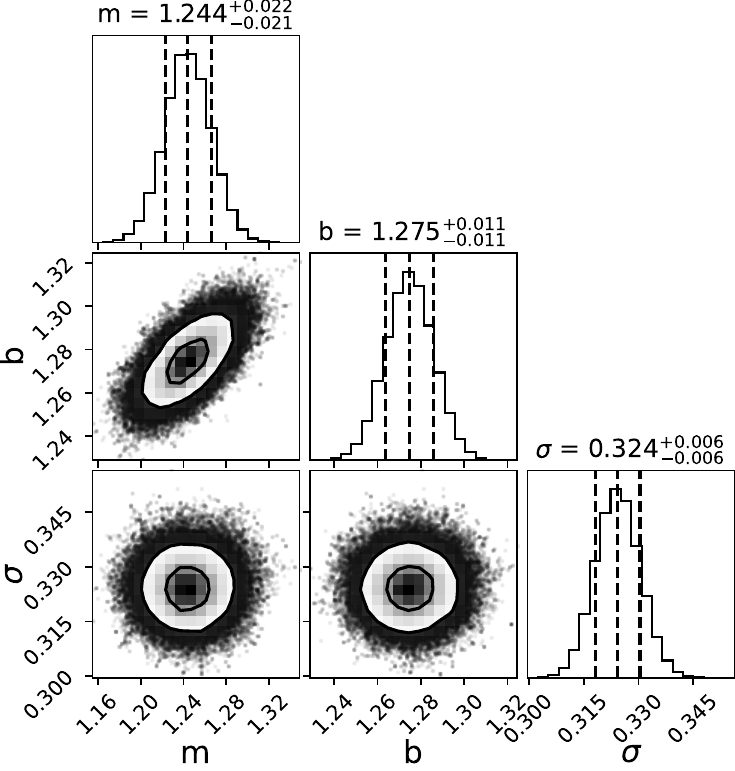}
    \caption{Relationship between $\mathrm{S}_{\mathrm{core}}$/$\mathrm{S}_{\mathrm{outer}}$ and N$_{\mathrm{core}}$/$\mathrm{N}_{\mathrm{outer}}$ for the whole sample through a logarithmic scale. The black dotted lines in the top panel correspond respectively to the zero positions on the horizontal and vertical axes, representing the scenario where both morphological stability parameters I and II are equal to 1. The color bar is coded by the number of member stars of each sample OC within its tidal radius. Corner legends in bottom panel follow Fig.~\ref{fig:all_N_s}.}
    \label{fig:6}
\end{figure}

Next, we further explored the relationship between the morphological stability parameters I and II. We fitted linear parameter II against parameter I for the entire sample, as shown in Fig.~\ref{fig:6}. The relationship, Log(N$_{\mathrm{core}}$/$\mathrm{N}_{\mathrm{outer}}$) = (1.244~$\pm$~0.022)~$\times$~Log($\mathrm{S}_{\mathrm{core}}$/$\mathrm{S}_{\mathrm{outer}}$) + (1.275~$\pm$~0.011), displays that the stellar density in the core increases more rapidly than the core area itself. This can support the hypothesis of the clusters with a more concentrated structure possessing stronger gravitational binding in their central regions.

\begin{figure*}[htbp]
    \centering
    \includegraphics[angle=0,width=60mm]{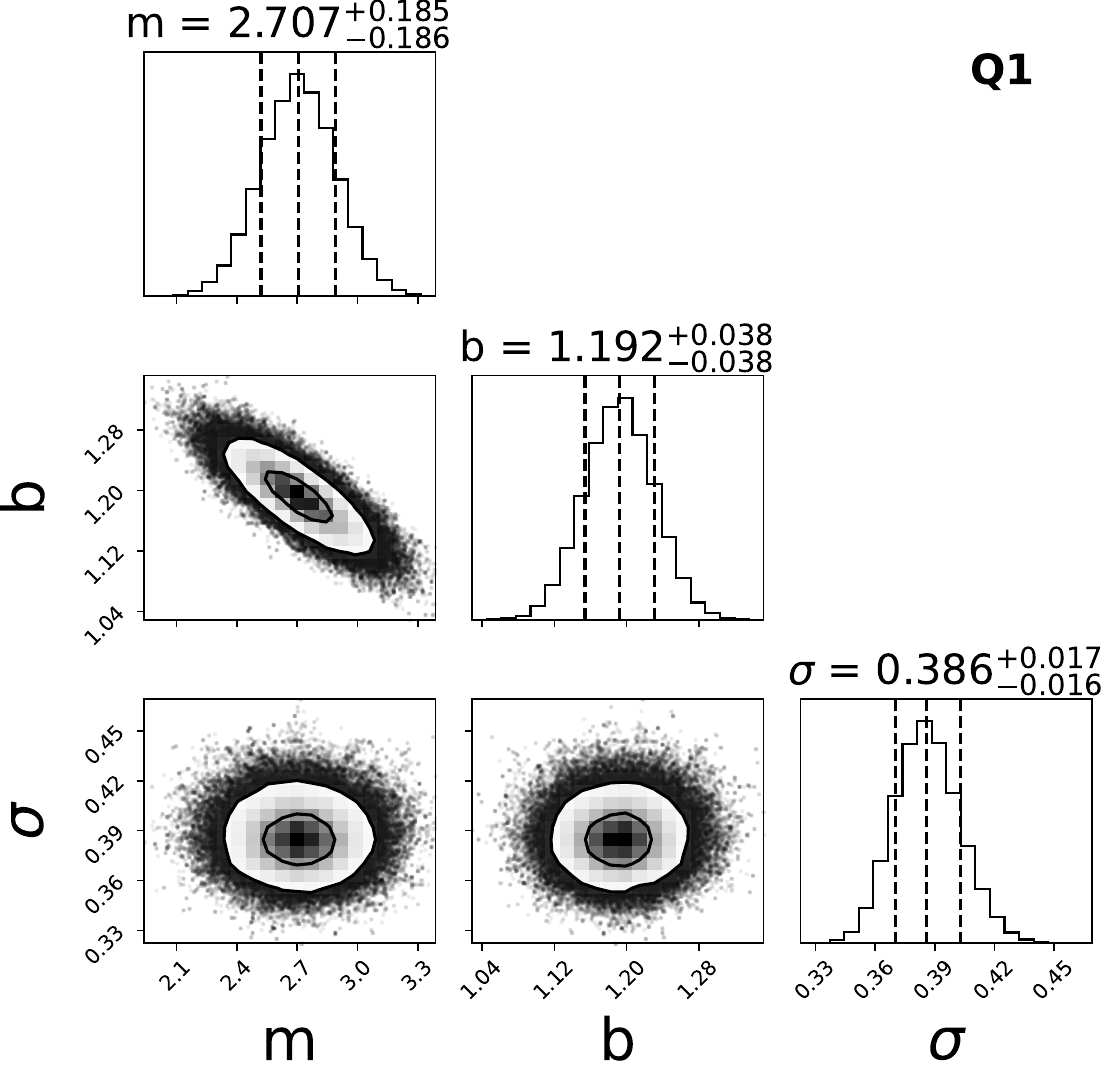}
    \includegraphics[angle=0,width=60mm]{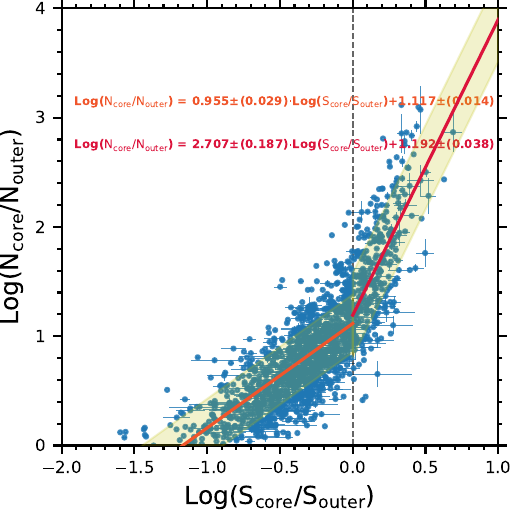}
    \includegraphics[angle=0,width=60mm]{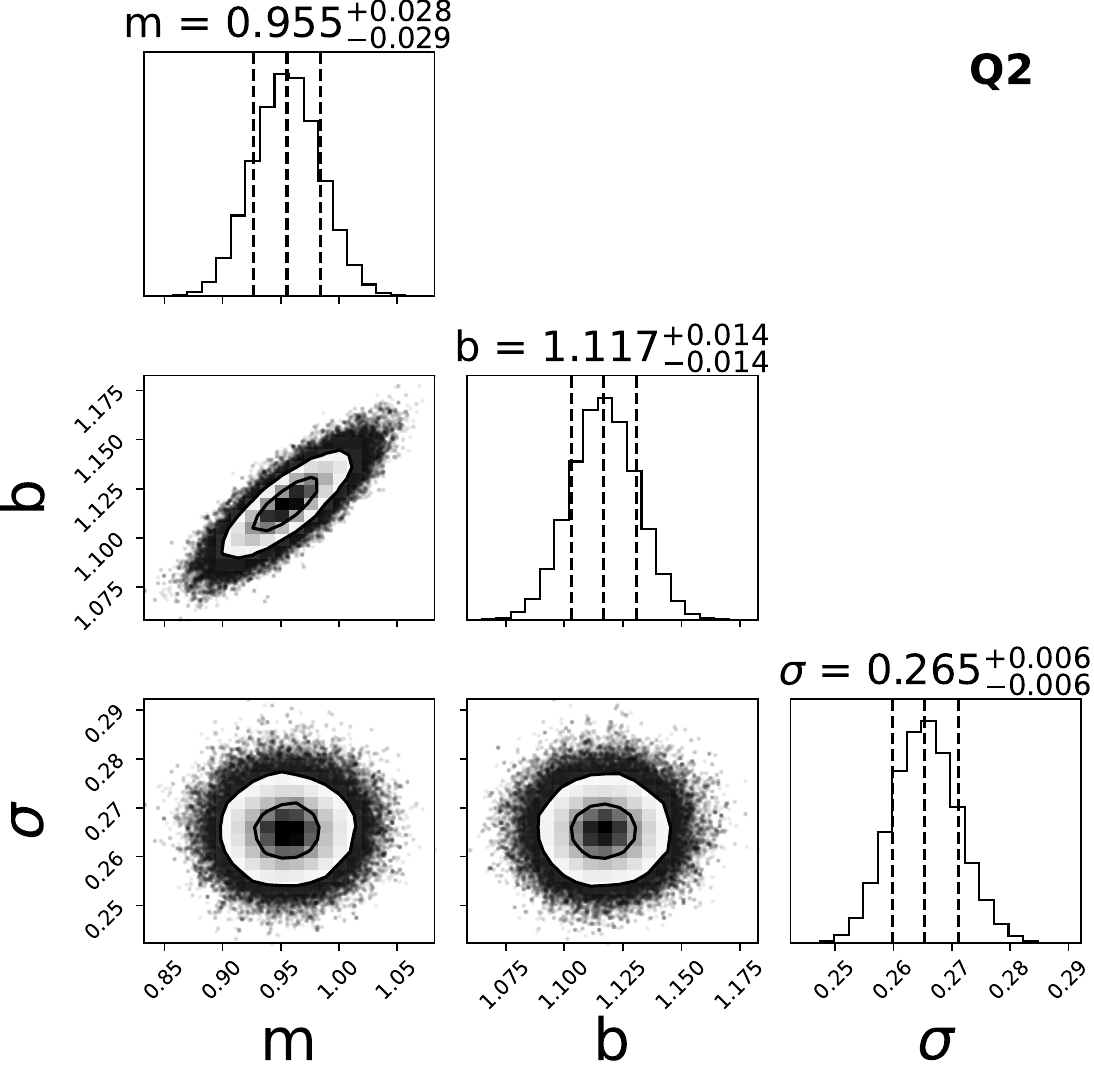}
    \caption{
        Relationships between Log(N$_{\mathrm{core}}$/$\mathrm{N}_{\mathrm{outer}}$) and Log($\mathrm{S}_{\mathrm{core}}$/$\mathrm{S}_{\mathrm{outer}}$) (Quadrant-wise Fitting). The panels are arranged from left to right as: \texttt{Q1} corner, the linear fit plot, and \texttt{Q2} corner. Corner panels' legends follow Fig.~\ref{fig:all_N_s}.}
    \label{fig:q1_q2}
\end{figure*}

Interestingly, from the top panel of Fig.~\ref{fig:6}, we can observe that the samples with the logarithm of the parameter I greater than 0 exhibit a noticeable deviation from the linear fitting line. This tells that the correlation in this region may actually be stronger than that in other regions where the logarithm of the parameter I is less than 0. In addition, Most of our samples have the logarithm of the parameter II greater than 0. Based on this, we divided the sample into four quadrants defined by the signs of Log($\mathrm{S}_{\mathrm{core}}$/$\mathrm{S}_{\mathrm{outer}}$) and Log(N$_{\mathrm{core}}$/$\mathrm{N}_{\mathrm{outer}}$), marked by two dashed lines in the top panel of Fig.~\ref{fig:6}. Due to the lack of samples in the quadrants with the logarithm of the parameter II less than 0, this work mainly focused on two quadrants with the logarithm of the parameter II greater than 0: Quadrants I (\texttt{Q1}, right upper) and II (\texttt{Q2}, left upper). It can be seen that the OC samples in \texttt{Q1} have a large core area and numerous core members, while those in \texttt{Q2} have a small core area and numerous core members. Compared to the samples in \texttt{Q1}, the OCs from \texttt{Q2} probably possess a dense core surrounded by a loose envelope and may be in an advanced stage of disruption.

\begin{figure*}[htbp]
    \centering
    \includegraphics[angle=0,width=45mm]{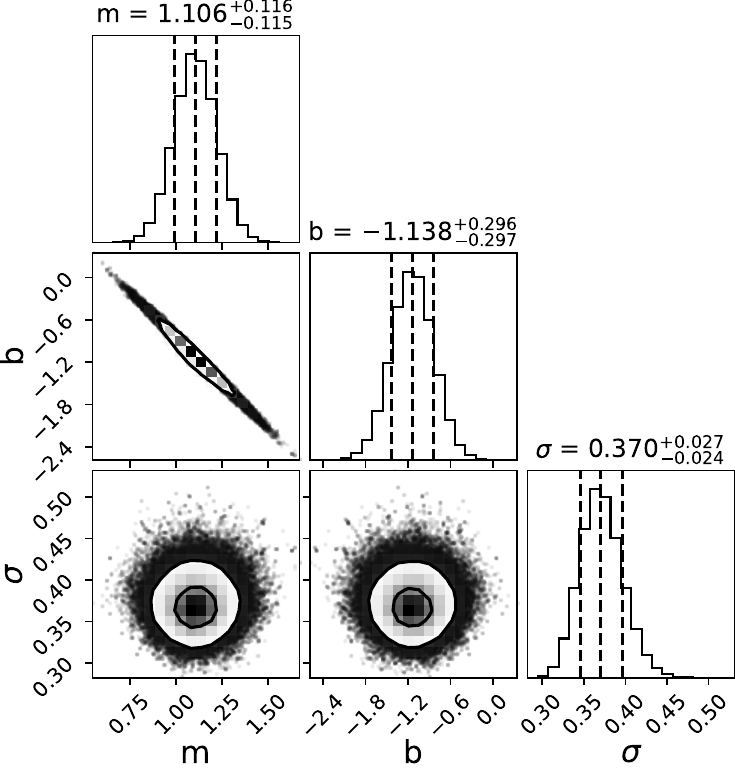}
    \includegraphics[angle=0,width=45mm]{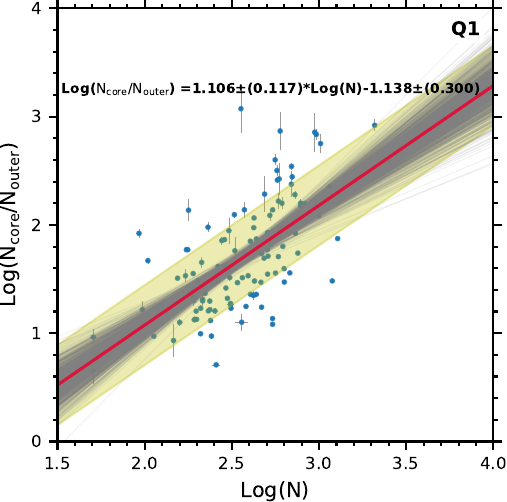}
    \includegraphics[angle=0,width=45mm]{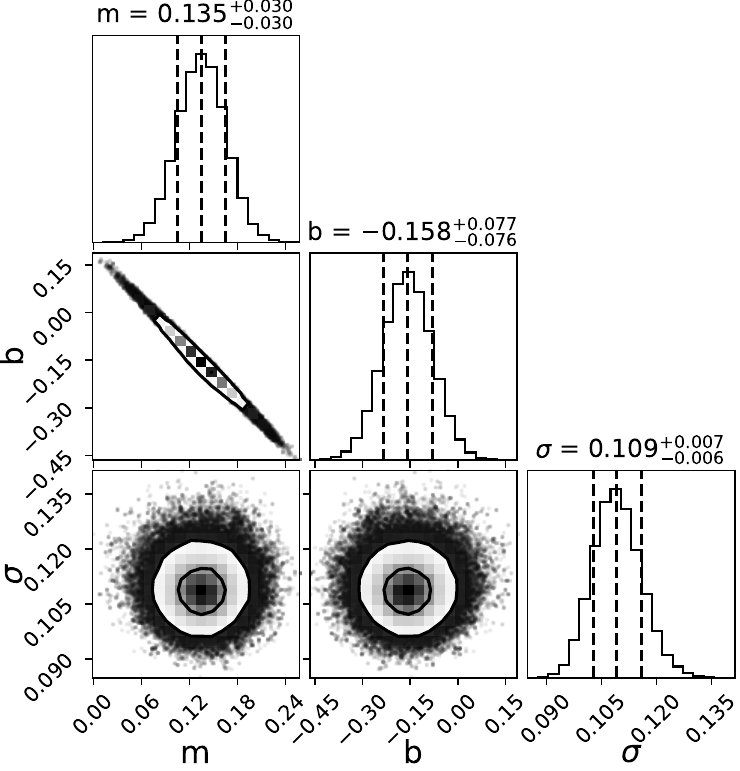}
    \includegraphics[angle=0,width=45mm]{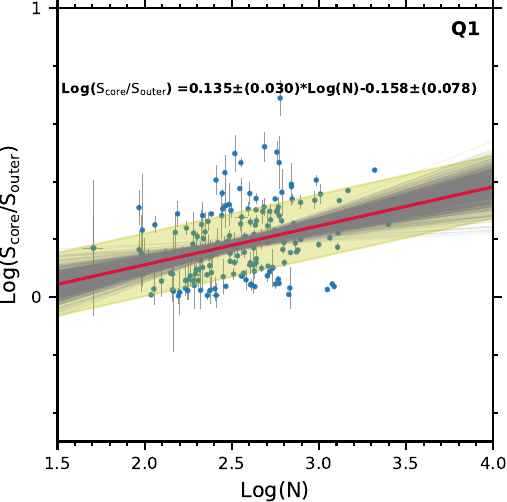}
    \caption{Relationships between the morphological stability parameters and the number of members (N) in the \texttt{Q1} region: $\mathrm{S}_{\mathrm{core}}$/$\mathrm{S}_{\mathrm{outer}}$ (the first two panels from left to right) and N$_{\mathrm{core}}$/$\mathrm{N}_{\mathrm{outer}}$ (the last two panels). Other panels' legends follow Fig.~\ref{fig:all_N_s}.}
    \label{fig:Q1}
\end{figure*}

Furthermore, in order to figure out how different they are, we performed a quantitative analysis for the OCs from \texttt{Q1} and \texttt{Q2}. It comes out that the slope of \texttt{Q1} is approximately 2.8 times steeper than that of \texttt{Q2}, as shown in Fig.~\ref{fig:q1_q2}. This indicates that the structural concentration presented in \texttt{Q1} samples is much more responsive to the density change in core members because of some processes like dynamical heating \citep{Camacho+2025} or tidal stripping \citep{Escala+2025} that are altering their structure.

\begin{figure*}[htbp]
    \centering
    \includegraphics[angle=0,width=45mm]{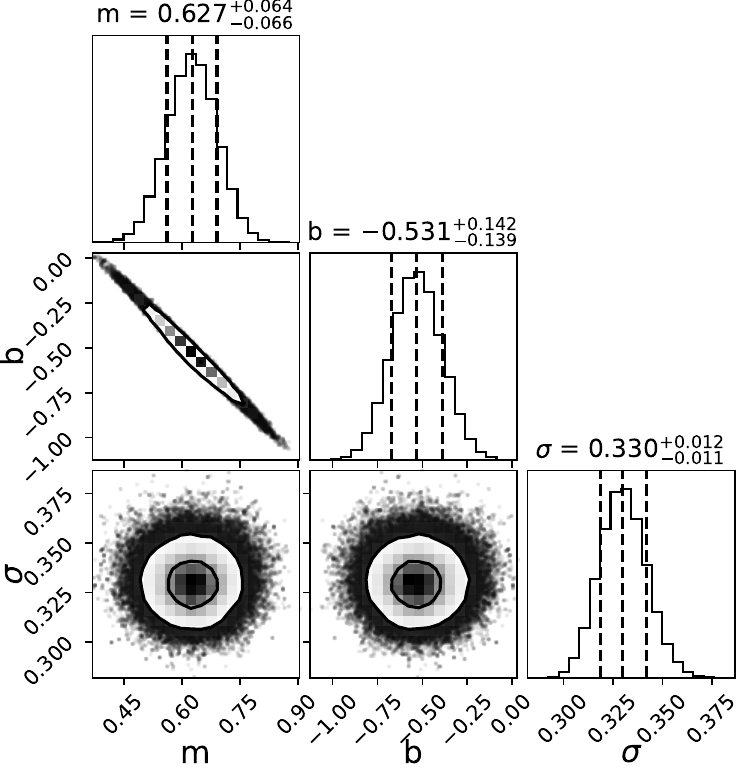}
    \includegraphics[angle=0,width=45mm]{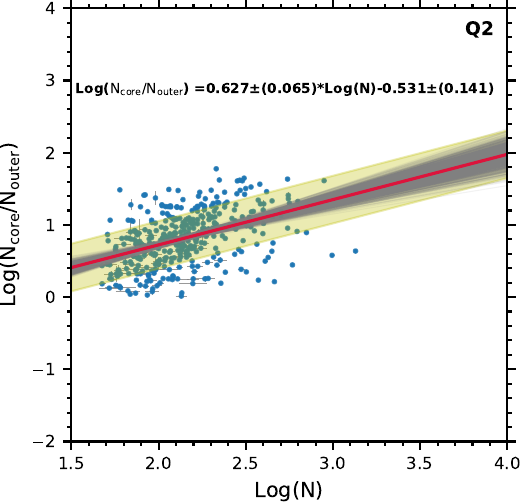}
    \includegraphics[angle=0,width=45mm]{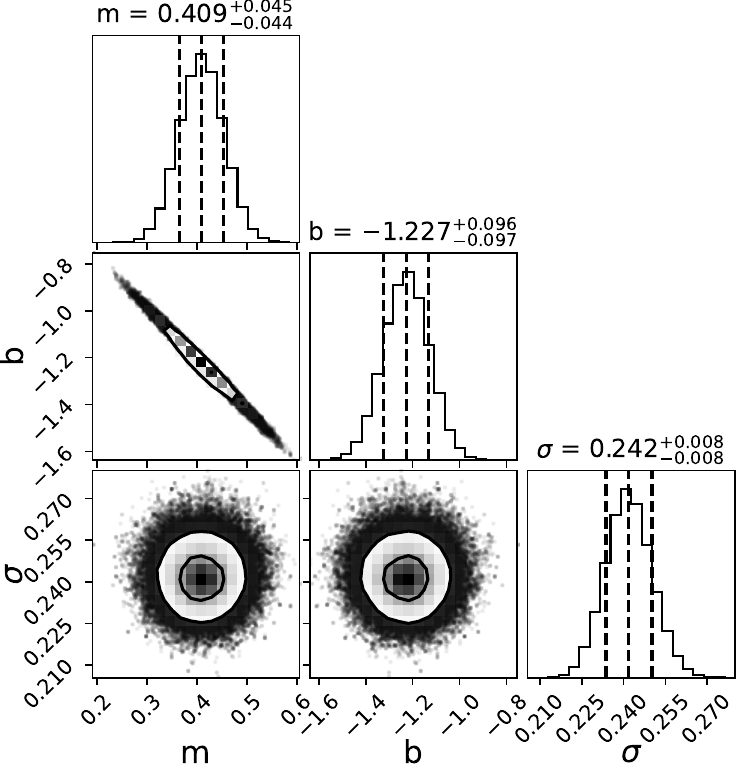}
    \includegraphics[angle=0,width=45mm]{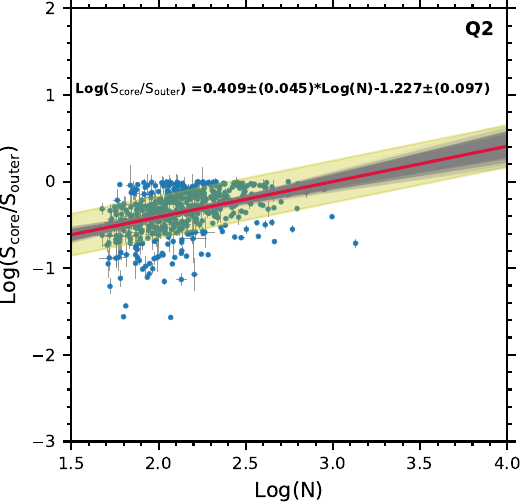}
    \caption{Relationships between the morphological stability parameters and the number of members (N) in the \texttt{Q2} region, in the same style as Fig.~\ref{fig:Q1}.}
    \label{fig:Q2}
\end{figure*}

Moreover, the turn-off position along the morphological stability parameter I in Fig.~\ref{fig:q1_q2} that we took to divide samples, Log($\mathrm{S}_{\mathrm{core}}$/$\mathrm{S}_{\mathrm{outer}}$)~=~0, illustrates that when the ratio of the core area to the outer area is greater than 1, the trend of the members' number in the core greater than that in the outer would be obvious. Meanwhile, when the morphological stability parameter I is greater than 1, the OC samples are typically with strong morphological stability \citep{Hu+2025a}; the opposite holds true otherwise. However, regardless of whether parameter I is greater than 1 or less than 1, the morphological stability parameter II almost remains greater than 1. This indicates that the morphological stability parameter I possibly relies on the premise of the number of core members exceeding that of external members, since a larger number of core members within an OC enhances its potential capacity for self-constraint.

Theoretically, a typical OC with fewer members in its core than in its outer region is almost impossible to possess sufficient binding potential to maintain its structure, leading to eventual dissolution. This is why there is a lack of such sample OCs, as illustrated by the field below the horizontal dashed line in Fig.~\ref{fig:6}. Even if such OCs exist, the estimated errors for their morphological stability parameters are large, most likely arising from the uncertainties of observational data and the extremely sparse distributions of their members.

After understanding the difference between \texttt{Q1} and \texttt{Q2}, we further performed a linear fitting on the parameters I and II for the samples from these two regions against the number of members, as shown in Figs.~\ref{fig:Q1} and \ref{fig:Q2}. All fitting parameters are listed in Table \ref{tab:Tab.2}. Figure~\ref{fig:Q1} displays the linear fitting of the parameters I and II with respect to the number of members for the \texttt{Q1} sample. It can be found that the slope of Log(N$_{\mathrm{core}}$/$\mathrm{N}_{\mathrm{outer}}$) vs. Log(N) is 1.106~$\pm$~0.117 with the one of Log($\mathrm{S}_{\mathrm{core}}$/$\mathrm{S}_{\mathrm{outer}}$) vs. Log(N) being 0.135~$\pm$~0.030. This illustrates that as the number of member stars increases, the proportion of core member stars increases sharply, supporting strong gravitational binding. However, unlike it, the core region may have approached its maximum physical density, limiting further contraction and resulting in slow area expansion.

The same way was applied to the \texttt{Q2} sample, as shown in Fig.~\ref{fig:Q2}. Due to the \texttt{Q2} sample having Log($\mathrm{S}_{\mathrm{core}}$/$\mathrm{S}_{\mathrm{outer}}$) less than 0, the slope of Log(N$_{\mathrm{core}}$/$\mathrm{N}_{\mathrm{outer}}$) vs. Log(N) is smaller than that of \texttt{Q1} sample, with 0.627~$\pm$~0.065, indicating that the proportion of member stars of the core continues to increase, but at a slower rate than in Group \texttt{Q1}. This can be explained by the fact that although the number of core members is still increasing, the spatial extent of the core grows faster, resulting in a decreasing stellar density. This behavior is commonly associated with internal dynamical interactions or external tidal perturbations, as \citet{Della+2024} demonstrated that a large fraction of young OCs experience pronounced global expansion during the first 30 Myr of their evolution, driven by violent relaxation and expulsion of residual gas, and accompanied by a substantial loss of mass from their outer regions. In addition, the slope of Log($\mathrm{S}_{\mathrm{core}}$/$\mathrm{S}_{\mathrm{outer}}$) vs. Log(N) is 0.409~$\pm$~0.045. The combination of a moderate increase in core stars coupled with a faster relative expansion of the core area is consistent with the scenario potentially driven by dynamical interactions or tidal forces \citep{Krumholz+2019}.

In conclusion, our analysis confirms that the morphological stability parameter II (N$_{\mathrm{core}}$/$\mathrm{N}_{\mathrm{outer}}$) exhibits a stronger dependence on the number of members (N) than parameter I. The disparity between \texttt{the Q1} and \texttt{Q2} samples validates these parameters to diagnose the stability and evolutionary state of the OCs. The clear linear relationship between Log(N$_{\mathrm{core}}$/$\mathrm{N}_{\mathrm{outer}}$) and Log(N) across the sample further underscores the critical role of the core stellar population in maintaining the structural integrity of OCs.

\begin{table}[htbp]
\centering
\caption{Quantitative relationships between parameters in whole sample, \texttt{Q1} sample, and \texttt{Q2} sample.}
\resizebox{\linewidth}{!}{
\label{tab:Tab.2}
\renewcommand{\arraystretch}{1.2}
\begin{tabular}{lcccc}
\hline\hline
\rule{0pt}{3ex} Groups &  Variables  &  Slopes &   N \\ 
\midrule
(1) &  (2)  &  (3) &   (4) \\ 
\midrule
All sample & Log(N$_{\mathrm{core}}$/$\mathrm{N}_{\mathrm{outer}}$) vs. Log($\mathrm{S}_{\mathrm{core}}$/$\mathrm{S}_{\mathrm{outer}}$) &  $1.244 \pm 0.022$ & 1474\\
 & Log(N$_{\mathrm{core}}$/$\mathrm{N}_{\mathrm{outer}}$) vs. Log(N) &  $1.140 \pm 0.039$ & 1474\\
 & Log($\mathrm{S}_{\mathrm{core}}$/$\mathrm{S}_{\mathrm{outer}}$) vs. Log(N) &  $0.720 \pm 0.026$ & 1490\\

\texttt{Q1} & Log(N$_{\mathrm{core}}$/$\mathrm{N}_{\mathrm{outer}}$) vs. Log($\mathrm{S}_{\mathrm{core}}$/$\mathrm{S}_{\mathrm{outer}}$) &  $2.707 \pm 0.187$ & 295 \\
 & Log(N$_{\mathrm{core}}$/$\mathrm{N}_{\mathrm{outer}}$) vs. Log(N) &  $1.106\pm 0.117$ & 295\\
 & Log($\mathrm{S}_{\mathrm{core}}$/$\mathrm{S}_{\mathrm{outer}}$) vs. Log(N) &  $0.135 \pm 0.030$ & 295 \\

\texttt{Q2} & Log(N$_{\mathrm{core}}$/$\mathrm{N}_{\mathrm{outer}}$) vs. Log($\mathrm{S}_{\mathrm{core}}$/$\mathrm{S}_{\mathrm{outer}}$) &  $0.955 \pm 0.029$ &1113\\
 & Log(N$_{\mathrm{core}}$/$\mathrm{N}_{\mathrm{outer}}$) vs. Log(N) &  $0.627\pm 0.065$ &1113\\
 & Log($\mathrm{S}_{\mathrm{core}}$/$\mathrm{S}_{\mathrm{outer}}$) vs. Log(N) &  $0.409 \pm 0.045$ &1113 \\

\bottomrule
\end{tabular}}
\tablefoot{Column (1) represents data groups, with Column (2) indicating the names of variables. Columns (3) and (4) respectively display the slope and the number of fitted clusters of each group.}
\end{table}

\subsection{Potential variations in the morphological stabilities across different spatial locations}

\hspace{2em}Our sample OCs are almost embedded within the Galactic disk, as shown in Fig.~\ref{fig: Galactic_latitude}, and undoubtedly, they are influenced by the gravitational potential at their positions within the disk, with the most significant effect being Galactic tidal forces \citep{Balaguer+2020}. Therefore, in this section, we explored whether the morphological stability of the sample OCs changes in different spatial positions.

\citet{Hu+2025a} reported that the morphological stability parameter I of OCs exhibits a correlation with the Y-axis in the Y-Z projection of the heliocentric Cartesian coordinate system. This suggests possible changes in the morphological stability along the Y-axis, which aligns with the radial direction under the Galactic Cartesian coordinate system. {In this work, the morphology of sample OCs with a low height ($|Z|$) in the radial direction is, in fact, similar to that in the Y-Z projection, while the one in the tangential direction is almost the same as that in the X-Z projection. Thus, we first investigated potential variations in the morphological stability of sample OCs in the radial and tangential directions.

To this end, we grouped the sample OCs according to their locations in the Galactic Cartesian coordinate system that is centered on the Galactic center, where the negative X-axis points towards the Sun, the positive Y-axis points in the direction of Galactic rotation, and the positive Z-axis points northward from the Galactic plane. The Sun’s position was taken from \citet{xu+2025} as [X, Y, Z] = (-8200.0, 0.0, 14.0)~pc. The positional parameters of the sample OCs were taken directly from \citet{van+2023}. To ensure that the radial and tangential sample projections correspond as closely as possible to the morphologies observed in the Y-Z and X-Z projection planes, respectively, we restricted all samples to the range $|Z|~\leq 200~\mathrm{pc}$. We initially divided the solar neighborhood into two key regions: the radial region, including a stripe region with $-250~\mathrm{pc}~\le~\mathrm{Y}~\le 250~\mathrm{pc}$ and the tangential region representing a range of $-8450~\mathrm{pc}~\le~\mathrm{X}~\le~7950~\mathrm{pc}$.

To verify whether the radial and tangential sample projections are almost equal to the morphologies observed in the Y-Z and X-Z projection planes, respectively, we compared the morphological stability parameter I in this study with those calculated by \citet{Hu+2025a}. By cross-matching, we obtained 36 and 45 common OCs between their samples and our samples in the radial and tangential directions, respectively, as shown in Fig.~\ref{fig: radial_tangential_comparison}. We can see from this figure a relatively good agreement between our samples and their samples. But there is still some dispersion around the red dashed lines of Fig.~\ref{fig: radial_tangential_comparison}. This may be because our sample projection corrected still exhibits a certain tilt angle relative to the X-Z and Y-Z projection planes. Nevertheless, our approach remains overall effective.

Figure~\ref{fig: parallax} shows the distributions of the two subsamples in the X-Y (the Galactic disk plane) and the X-Z planes. The size of the solid circles is proportional to the number of members, with their color being coded by the logarithm of the morphological stability parameter I. This distribution does not show any discernible trend. However, it is certain that the morphological stability of OCs is inevitably influenced by external forces at their spatial locations. How can we measure this influence? We already know that an OC's morphological stability is directly proportional to the number of its member stars, as demonstrated by the linear relationship presented earlier. If the morphological stability of OCs at certain spatial locations is affected, this will inevitably weaken this linear proportional relationship. Based on this speculation, we performed the linear fitting between the morphological stability parameters I and II and the number of members for the radial and tangential samples separately. The parameters of their linear relationships are compiled in Table~\ref{tab:slope_para}.

As we expected, the linear relationships mentioned above change. For morphological stability parameters I and II, the radial samples show stronger correlations than the tangential samples. This demonstrates that the morphological stabilities of the sample OCs vary at different spatial positions, in which the influence of external forces on their morphology is variable. The radial samples in this study primarily show us their morphology in the Y-Z projection plane, which is predominantly influenced by shear forces from the disk’s differential rotation. In contrast, what we can see for the tangential samples is their morphology in the X-Z projection plane, which is mainly shaped by tidal forces directed roughly toward the Galactic center. Based on this, it can be concluded that the shear forces generated by the differential rotation possibly exert a weaker influence on the morphological stability of the sample OCs than the tidal forces directed toward the Galactic center.

\begin{figure}
	\centering
    \includegraphics[angle=0,width=88mm]{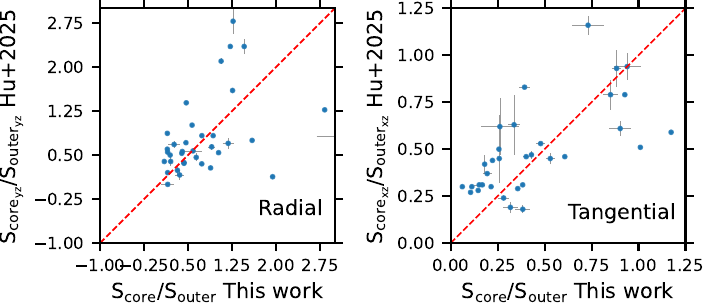}
	\caption{Left panel: radial $\mathrm{S}_{\mathrm{core}}$/$\mathrm{S}_{\mathrm{outer}}$ comparison; Right panel: tangential $\mathrm{S}_{\mathrm{core}}$/$\mathrm{S}_{\mathrm{outer}}$ comparison. The comparative samples are drawn from \citep{Hu+2025a}, comprising 36 in the radial direction (Y-Z plane) and 45 in the tangential direction (X-Z plane). The red dotted line in each panel represents the y = x reference line, with the error bars indicating the dispersions of $\mathrm{S}_{\mathrm{core}}$/$\mathrm{S}_{\mathrm{outer}}$.}
	\label{fig: radial_tangential_comparison}
\end{figure}

In addition to the radial and tangential samples primarily influenced by shear forces and tidal forces, respectively, samples at other locations must be subjected to these two external forces simultaneously. To more precisely reveal the influence of other spatial positions on the morphological stability of the sample OCs, we thus divided the samples based on azimuthal angles (e.g., in 15$^{\degree}$ intervals) centered on the Sun and extracted the samples located at angular intervals of 0$^{\degree}$ (radial), 15$^{\degree}$, 30$^{\degree}$, 45$^{\degree}$, 60$^{\degree}$, 75$^{\degree}$,  90$^{\degree}$ (tangential), 105$^{\degree}$, 120$^{\degree}$, 135$^{\degree}$, 150$^{\degree}$, and 165$^{\degree}$. And the number of sample OCs for most groups is around 100, as shown in Table~\ref{tab:slope_para}.

\begin{figure}[h!]
	\centering
    \includegraphics[angle=0,width=88mm]{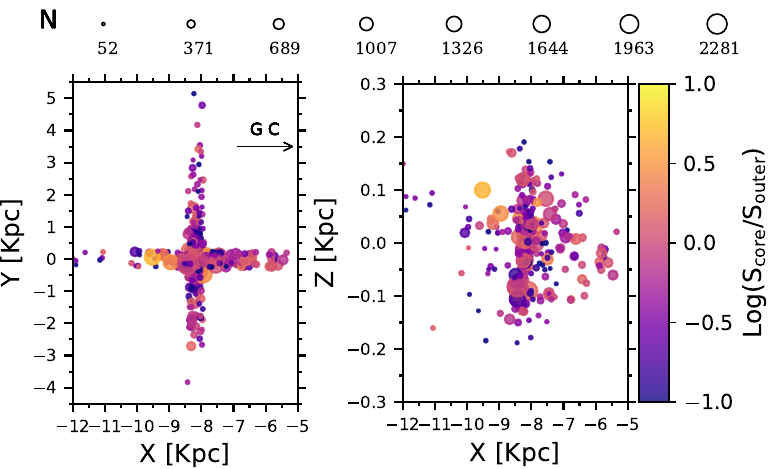}
	\caption{Distributions of the radial and tangential sample OCs in the Milky Way: X-Y plane distribution (left panel); X-Z plane distribution (right panel). The size of the solid circles indicates the number of member stars (N), with their colors being coded by the Log($\mathrm{S}_{\mathrm{core}}$/$\mathrm{S}_{\mathrm{outer}}$). The arrow marks the direction of the Galactic center (\text{GC}).}
	\label{fig: parallax}
\end{figure}

We then first studied the samples along the two directions (45$^{\degree}$ named by ``tilt-right'' and 135$^{\degree}$ named by ``tilt-left''), as shown in Fig.~\ref{fig: tilt_parallax}, because they theoretically are subject to roughly equal tidal and shear forces. If the Galactic disk is symmetrical, the morphological stability of the sample OCs in both directions is equally affected by external forces. To test this hypothesis, we performed linear fits of two morphological stability parameters against the number of members for both samples. The fitted parameters are listed in Table~\ref{tab:slope_para}. We found that the slopes of the morphological stability parameters with the number of members for the ``tilt-right'' sample are different from those for the ``tilt-left'' sample. Therefore, in general, the morphological stability of the samples in these two directions may be affected in varying degrees, indicating that the external force environment influencing the morphology of sample OCs is asymmetric in these two directions.

\begin{figure}
	\centering
    \includegraphics[angle=0,width=88mm]{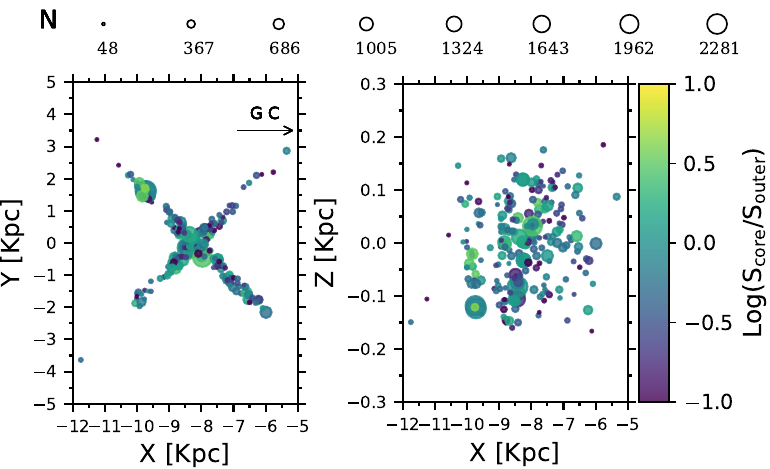}
	\caption{Distributions of the right-tilt and left-tilt sample OCs in the Milky Way: X-Y (left panel), X-Z (right panel), in the same style as figure~\ref{fig: parallax}, except for color.}
	\label{fig: tilt_parallax}
\end{figure}

We next systematically fitted all the other subgroups. All results are also summarized in Table \ref{tab:slope_para}, and their trends are clearly displayed in Fig.~\ref{fig: Group_comparison}. We found that the slopes of the radial sample in both the morphological stability parameters I and II are relatively large within the error range. On the basis of previous speculation, the morphological stability of these sample OCs is relatively stable. However, the slope of the tangential samples is relatively small among all of the samples, suggesting that the morphological stability of the tangential sample is comparatively weaker than that of samples oriented in other directions. Moreover, from the overall trend in Fig.~\ref{fig: Group_comparison}, the slopes of samples at different positions vary, indicating that the external force environment on the Galactic disk region around the Sun is actually asymmetric.

\begin{table}[htbp]
\centering
\caption{The relationship between the morphological stability parameters of OCs at different locations and the number of members.}
\resizebox{\linewidth}{!}{
\label{tab:slope_para}
\renewcommand{\arraystretch}{1.2}
\begin{tabular}{lcccc}
\hline\hline
\rule{0pt}{3ex} Groups &  Variables  &  Slopes &   N \\ 
\midrule
(1) &  (2)  &  (3) &   (4) \\ 
\midrule
Radial & Log($\mathrm{S}_{\mathrm{core}}$/$\mathrm{S}_{\mathrm{outer}}$) vs. Log(N) &  $0.733 \pm 0.080$ & 140\\
 & Log(N$_{\mathrm{core}}$/$\mathrm{N}_{\mathrm{outer}}$) vs. Log(N) &  $1.083 \pm 0.116$ & 138 \\

 Tilt\_15$^{\degree}$ & Log($\mathrm{S}_{\mathrm{core}}$/$\mathrm{S}_{\mathrm{outer}}$) vs. Log(N) &  $0.540 \pm 0.080$ & 125 \\
 & Log(N$_{\mathrm{core}}$/$\mathrm{N}_{\mathrm{outer}}$) vs. Log(N) &  $0.989 \pm 0.122$ & 124  \\

 Tilt\_30$^{\degree}$ & Log($\mathrm{S}_{\mathrm{core}}$/$\mathrm{S}_{\mathrm{outer}}$) vs. Log(N) &  $0.505 \pm 0.093$ & 127 \\
 & Log(N$_{\mathrm{core}}$/$\mathrm{N}_{\mathrm{outer}}$) vs. Log(N) &  $0.886 \pm 0.140$ & 127  \\
 
Right tilt & Log($\mathrm{S}_{\mathrm{core}}$/$\mathrm{S}_{\mathrm{outer}}$) vs. Log(N) &  $0.551 \pm 0.099$ & 106 \\
 & Log(N$_{\mathrm{core}}$/$\mathrm{N}_{\mathrm{outer}}$) vs. Log(N) &  $0.979 \pm 0.147$ & 106  \\

Tilt\_60$^{\degree}$ & Log($\mathrm{S}_{\mathrm{core}}$/$\mathrm{S}_{\mathrm{outer}}$) vs. Log(N) &  $0.537 \pm 0.101$ & 90 \\
 & Log(N$_{\mathrm{core}}$/$\mathrm{N}_{\mathrm{outer}}$) vs. Log(N) &  $1.134 \pm 0.160$ & 90  \\

Tilt\_75$^{\degree}$ & Log($\mathrm{S}_{\mathrm{core}}$/$\mathrm{S}_{\mathrm{outer}}$) vs. Log(N) &  $0.604 \pm 0.117$ & 47  \\
 & Log(N$_{\mathrm{core}}$/$\mathrm{N}_{\mathrm{outer}}$) vs. Log(N) &  $1.026 \pm 0.198$ & 47  \\
 
Tangential & Log($\mathrm{S}_{\mathrm{core}}$/$\mathrm{S}_{\mathrm{outer}}$) vs. Log(N) &  $0.529 \pm 0.075$ & 174  \\
 & Log(N$_{\mathrm{core}}$/$\mathrm{N}_{\mathrm{outer}}$) vs. Log(N) &  $1.013 \pm 0.110$ & 174 \\

 Tilt\_105$^{\degree}$ & Log($\mathrm{S}_{\mathrm{core}}$/$\mathrm{S}_{\mathrm{outer}}$) vs. Log(N) &  $0.650 \pm 0.143$ & 47 \\
 & Log(N$_{\mathrm{core}}$/$\mathrm{N}_{\mathrm{outer}}$) vs. Log(N) &  $1.350\pm 0.185$ & 47 \\

Tilt\_120$^{\degree}$ & Log($\mathrm{S}_{\mathrm{core}}$/$\mathrm{S}_{\mathrm{outer}}$) vs. Log(N) &  $0.692 \pm 0.095$ & 102 \\
 & Log(N$_{\mathrm{core}}$/$\mathrm{N}_{\mathrm{outer}}$) vs. Log(N) &  $1.314\pm 0.135$ & 102 \\

Left tilt & Log($\mathrm{S}_{\mathrm{core}}$/$\mathrm{S}_{\mathrm{outer}}$) vs. Log(N) &  $0.744 \pm 0.084$ & 125 \\
 & Log(N$_{\mathrm{core}}$/$\mathrm{N}_{\mathrm{outer}}$) vs. Log(N) &  $1.035 \pm 0.117$ & 122 \\

Tilt\_150$^{\degree}$ & Log($\mathrm{S}_{\mathrm{core}}$/$\mathrm{S}_{\mathrm{outer}}$) vs. Log(N) &  $0.691 \pm 0.071$ & 144\\
 & Log(N$_{\mathrm{core}}$/$\mathrm{N}_{\mathrm{outer}}$) vs. Log(N) &  $1.230\pm 0.097$ & 143 \\

 Tilt\_165$^{\degree}$ & Log($\mathrm{S}_{\mathrm{core}}$/$\mathrm{S}_{\mathrm{outer}}$) vs. Log(N) &  $0.709 \pm 0.073$ & 160\\
 & Log(N$_{\mathrm{core}}$/$\mathrm{N}_{\mathrm{outer}}$) vs. Log(N) &  $1.227\pm 0.105$ & 159 \\
 
\bottomrule
\end{tabular}}
\tablefoot{Such as table~\ref{tab:Tab.2}.}
\end{table}

\begin{figure}
	\centering
    \includegraphics[angle=0,width=88mm]{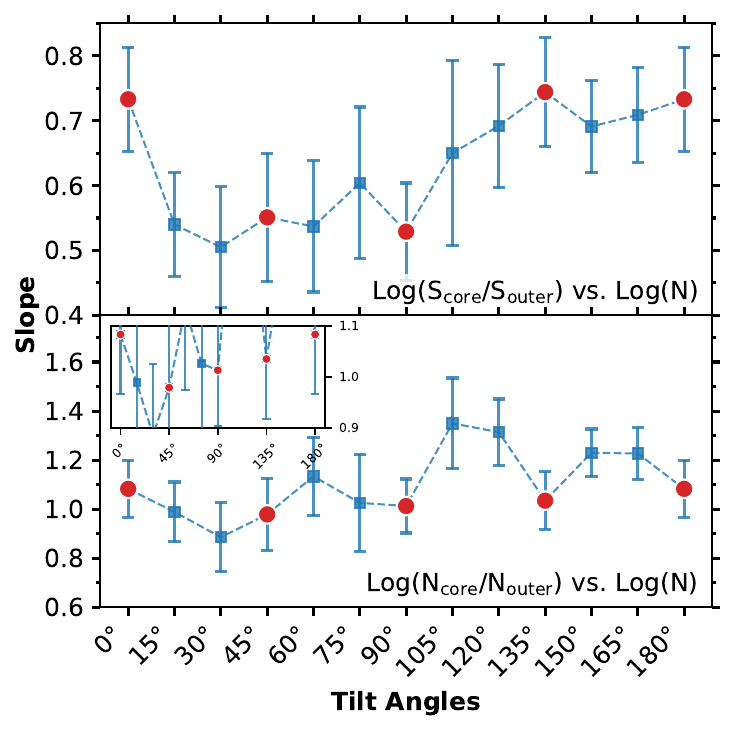}
	\caption{Comparative graph of slopes (top panel: Log($\mathrm{S}_{\mathrm{core}}$/$\mathrm{S}_{\mathrm{outer}}$) vs. Log(N), bottom panel: Log(N$_{\mathrm{core}}$/$\mathrm{N}_{\mathrm{outer}}$) vs. Log(N)) across different groups in Table~\ref{tab:slope_para}. The inset is for a better understanding of the changing trend of the slopes.}
    \label{fig: Group_comparison}
\end{figure}

Furthermore, we can also see from Fig.~\ref{fig: Group_comparison} that the slopes of the sample OCs within tilt angles less than 90$^{\degree}$ are almost smaller than those of other samples. Why is this happening?  To explore whether environmental factors are associated with this asymmetry, we plotted the map of our sample on the Galactic disk, as shown in Fig.~\ref{fig:1490_OCs_distribute}, colored by the logarithm of their morphological stability II. Based on this figure, it can be easily explained that the OC samples within tilt angles less than 90$^{\degree}$ are closer to the Galactic bar of the Milky Way than those with tilt angles greater than 90$^\circ$, and suffer from more disruption of the external force environment. The OCs with tilt angles less than 90$^\circ$ are concentrated in a sector region with a Galactocentric distance of 4-6 kpc and an azimuthal angle of approximately 0$^\circ$-90$^\circ$. This region highly overlaps with the orientation of the Galactic bar \citep{Joharle+2026, Kalita+2026} and the peak density area of the Gaia-Sausage-Enceladus merger remnant \citep{Helmi+2018}. Recent \texttt{lustrisTNG50} simulations further demonstrated that major mergers can leave azimuthal offsets in stellar warps that persist even after differential rotation correction, providing a dynamical memory of external perturbations lasting several Gyr \citep{Thulasidharan+2025}. Therefore, the combined perturbations from bar torque, merger potential, and local arm shear in the region where low tilt angle OCs are located may continuously strip member stars from their outer regions, leading to a systematically smaller ratio of N$_{\mathrm{core}}$/$\mathrm{N}_{\mathrm{outer}}$. Consequently, these OCs are subjected to more disruptive external environmental forces. We also found that most samples with small parameter II are distributed along the local arm, suggesting that the external force environment around the local arm is likely complicated.

It is well-known that young OCs are usually not far away from their birthplaces, which were likely close to the spiral arms. Therefore, we specifically studied the slope between the morphological stability parameter II and the number of members in OC samples of different ages. \citet{Della+2024} pointed out that 80\% of OCs under the age of ~30~Myr are in an explosive phase, leading us to speculate that young OCs have poor morphological stability. \citet[][see their Fig.~5]{hao+2021} mentioned that 72\% of the young OCs are located in the spiral
arms. \citet[][see their Fig.~6]{Qiu+2025} concluded that the further an OC is from the galactic disk, the more compact its center. Therefore, we divided these sample OCs into six age groups (age~$\leq$~30~Myr, 30~Myr~$<$~age~$\leq$~100~Myr, 100~Myr~$<$~age~$\leq$~200~Myr, 200~Myr~$<$~age~$\leq$~400~Myr, 400~Myr~$<$~age~$\leq$~800~Myr, and age~$>$~800~Myr), as displayed in Fig.~\ref{fig:age}, ensuring a balanced number of OCs in each group. And these linear fitting parameters of six groups are listed in Table~\ref{tab:Cavallo_age}, in which the weakest correlation is observed for young OCs with less than 30~Myr among them, with a fitting slope of 0.751~$\pm$~0.166 for Log(N$_{\mathrm{core}}$/$\mathrm{N}_{\mathrm{outer}}$) vs. Log(N). The overall trend is illustrated in Fig.~\ref{fig: age_comparison}. This suggests that the gravitational disturbances from their birthplaces have a significant impact on the morphology of the youngest sample OCs. As the OCs age, they are gradually moving from their birthplaces, and the influence of the gravitational disturbances from the birthplaces on their morphology is also getting weaker. As expected, the slopes of the other five groups with ages greater than 30 Myr are, on the whole and within the range of error margins, larger than that of the youngest sample OCs group. In addition, the same holds true for the slopes of the morphological stability parameter I against the number of members, as displayed in Table~\ref{tab:Cavallo_age}.

The observed slopes of the morphological stability parameters increasing with the number of members as a function of age can be consistently explained from a statistical perspective by \citet{Coenda+2025} and \citet{liu+2025}. The majority of young OCs with age~$\leq$~30~Myr, mostly exist in `grouped' configurations, with intra-group separations of only a few dozen parsecs \citep[][see the left panel of their Fig.~3]{Coenda+2025}. Their overlapping potential wells generate a rapidly time-varying internal tidal field. At the same time, these young clusters are generally located in the local spiral arm region, which is still part of a growing arm where macroscopic turbulence and density waves continuously inject kinetic energy \citep{liu+2025}. Therefore, the combined internal and external disturbances constantly break the stellar field, likely resulting in the poorest morphological stability of our young sample OCs \citep{Elmegreen+2008, Elmegreen+2018}. It can also be inferred from the work of \citep{Hu+2025b} that a young pair (\texttt{ASCC\_19} and \texttt{ASCC\_21}) is undergoing mutual tidal interaction processes.

OCs in the 30~Myr~$<$~age~$\leq$~200~Myr range exist mostly as binary clusters, because \citet{Coenda+2025} note that the median age for paired OCs is $\log(\text{Age}/\text{yr})~=~8.01~\pm~0.03$, which corresponds to approximately 100~Myr. Although tidal perturbations between the two clusters remain, their strength is significantly weaker than the multi-directional disturbances within young cluster groups. Meanwhile, as these OCs begin to move away from the most turbulent gas layers in the spiral arm  \citep[e.g.,][] {Binney+2008, Reid+2019}, and the external environmental noise also diminishes. Hence, their morphological stability is higher than that of young OCs. For OCs in the 200~Myr~$<$~age~$\leq$~400~Myr range, they have largely transitioned to more isolated states. The external tidal sources are progressively simplified, while internal dynamical processes begin to dominate the cluster evolution. The morphological stability parameters show further increase, reflecting the ongoing stabilization process. Old sample OCs exhibit the strongest morphological stability, as they mostly exist as single clusters and are located farther from the galactic disk \citep{Coenda+2025}. The external tidal sources are simplified to smooth the Galactic disk potential, while internal two-body relaxation rapidly erases any remaining substructure, ultimately forming a stabilized radial density profile. For the oldest OCs (age~$>$~800~Myr), they maintain high morphological stability as single clusters located at significant distances from the galactic plane. However, the slopes show a slight decrease compared to the 400~Myr~$<$~age~$\leq$~800~Myr group, possibly due to the long-term effects of dynamical evolution and mass segregation processes that have been operating over extended timescales \citep{Moreira+2025}. These OCs typically reside at high galactic latitudes far from the disk plane \citep{Wu+2024}, where the non-axisymmetric gravitational potential generated by the Galactic warp introduces additional kinematic perturbations to their orbital motions \citep{Viktor+2024, Thulasidharan+2025, Reshetnikov+2025}. Therefore, as the age group increases, the morphological stability of the sample OCs generally exhibits a stronger trend.

\begin{figure}
    \centering
    \includegraphics[angle=0,width=90mm]{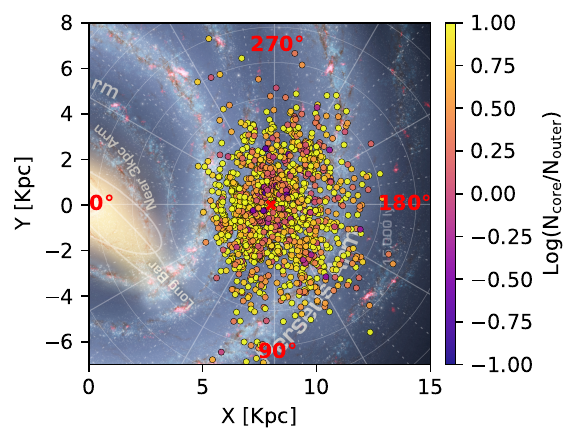}
    \caption{Morphological stability  (N$_{\mathrm{core}}$/N$_{\mathrm{outer}}$) map of sample OCs in our galaxy. The red cross marks the position of the Sun. We used the library available at mw\_plot: \url{https://milkyway-plot.readthedocs.io/en/stable/matplotlib_faceon.html\#milkyway-bird-s-eye-view}.}
    \label{fig:1490_OCs_distribute}
\end{figure}

\begin{figure*}[htbp]
    \centering
    \includegraphics[angle=0,width=60mm]{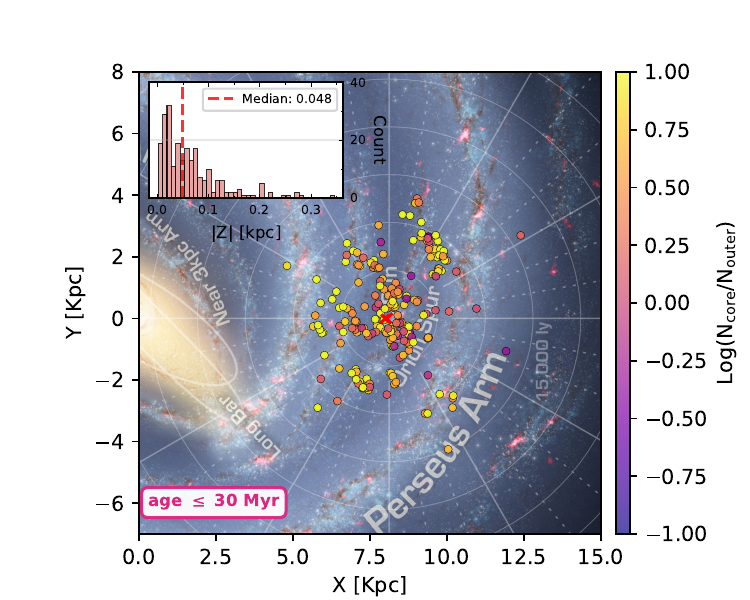}
    \includegraphics[angle=0,width=60mm]{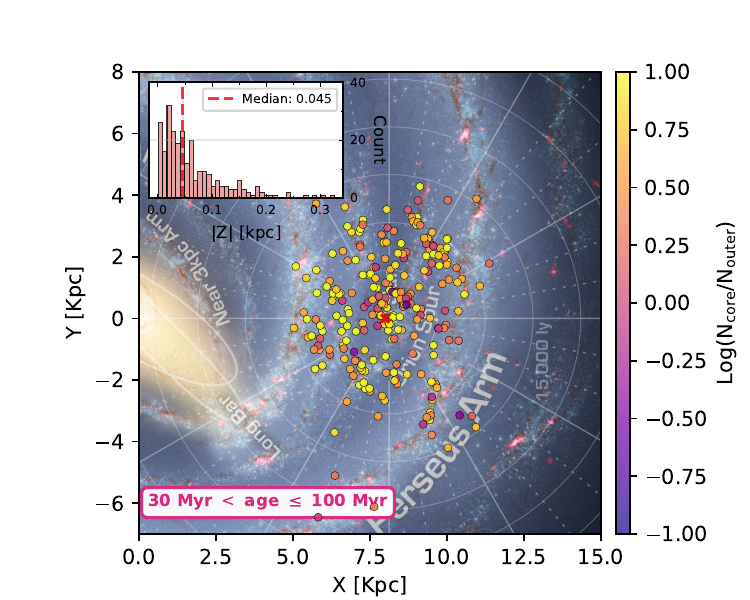}
    \includegraphics[angle=0,width=60mm]{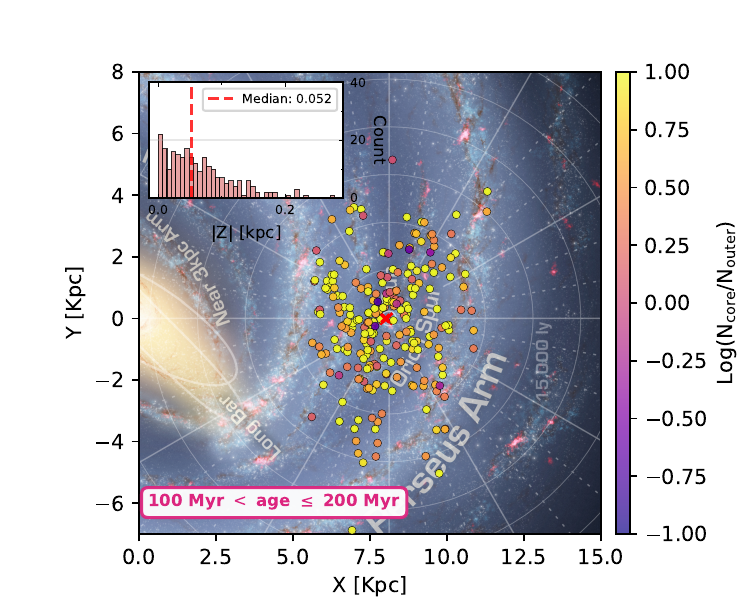}
    \includegraphics[angle=0,width=60mm]{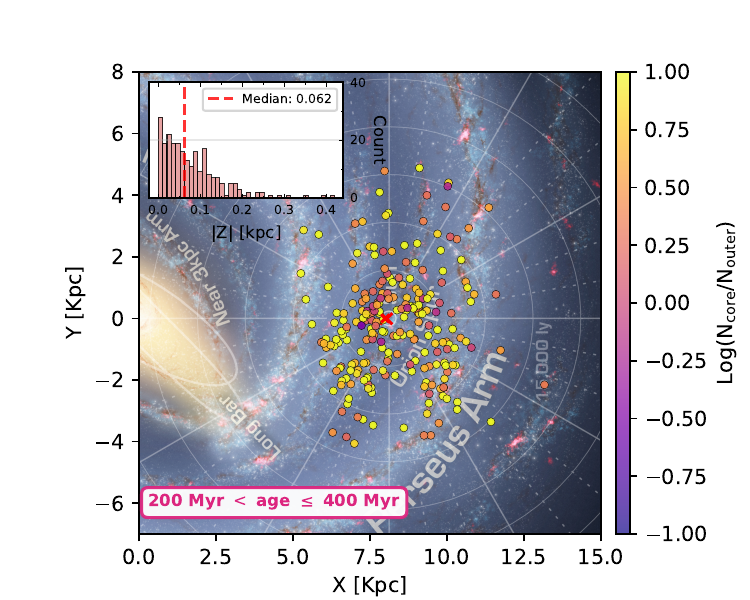}
    \includegraphics[angle=0,width=60mm]{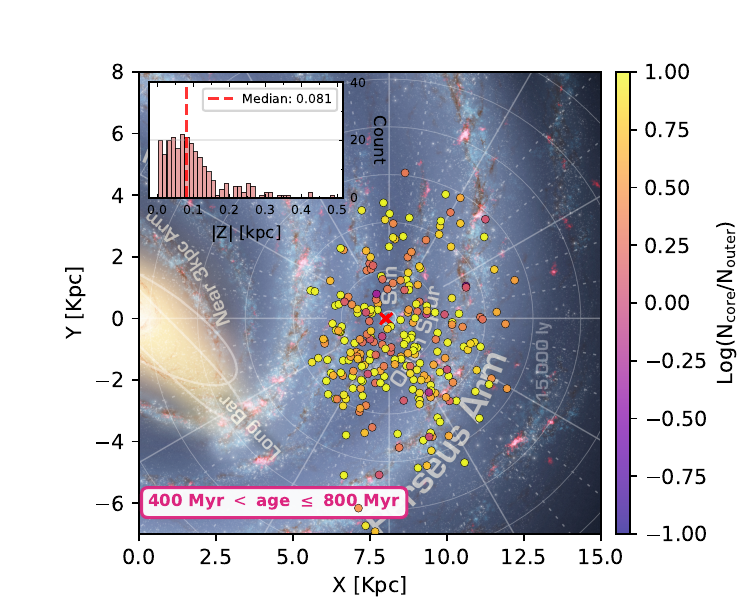}
    \includegraphics[angle=0,width=60mm]{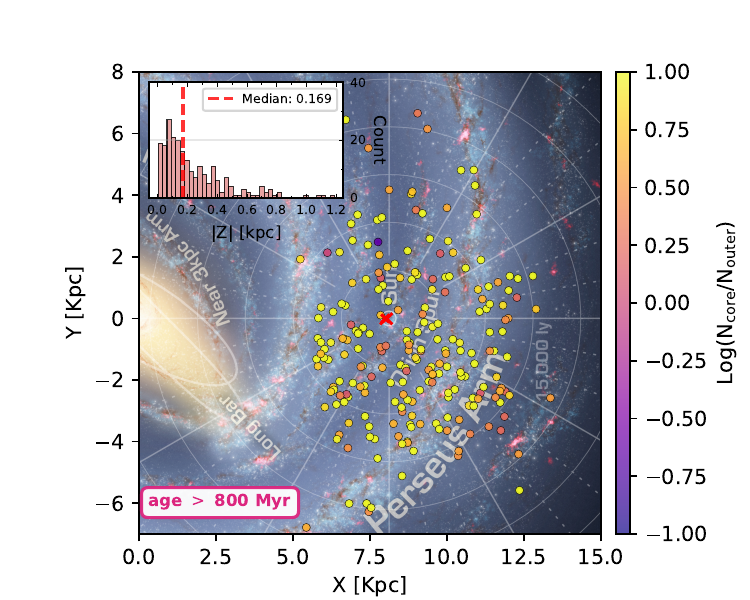}

    \caption{Distributions of sample OCs with varying ages in the Milky Way (from left to right and then from top to bottom: age $\leq 30\,\text{Myr}$, 30\, \text{Myr} $<$ age $\leq 100\,\text{Myr}$, 100\, \text{Myr} $<$ age $\leq 200\,\text{Myr}$, 200\, \text{Myr} $<$ age $\leq 400\,\text{Myr}$, 400\, \text{Myr} $<$ age $\leq 800\,\text{Myr}$, age $>$ 800\, \text{Myr}). The inset in each panel indicates the absolute distance from the Galactic plane. The age parameters used in this work were taken from \citet{Cavallo+2024}.} 
    \label{fig:age}
\end{figure*}

\begin{table*}[htbp]
\centering
\caption{The relationship between the morphological stability parameters in OCs of different ages and the number of member stars. \label{tab:Cavallo_age}}
\renewcommand{\arraystretch}{1.2}
\begin{tabular}{lccc} 
\hline\hline
\rule{0pt}{4ex} Groups & Variables & Slopes & N \\ 
\midrule
(1) & (2) & (3) & (4) \\ 
\midrule
age $\leq 30\,\text{Myr}$ 
 & Log(N$_{\mathrm{core}}$/$\mathrm{N}_{\mathrm{outer}}$) vs. Log(N) & $0.751 \pm 0.166$ &   232\\
  & Log($\mathrm{S}_{\mathrm{core}}$/$\mathrm{S}_{\mathrm{outer}}$) vs. Log(N) & $0.435 \pm 0.075$ &   232\\
30\, \text{Myr} $<$ age $\leq 100\,\text{Myr}$  
 & Log(N$_{\mathrm{core}}$/$\mathrm{N}_{\mathrm{outer}}$) vs. Log(N) & $1.153 \pm 0.166$ &   250\\
  & Log($\mathrm{S}_{\mathrm{core}}$/$\mathrm{S}_{\mathrm{outer}}$) vs. Log(N) & $0.737 \pm 0.067$ &   254\\

100\, \text{Myr} $<$ age $\leq 200\,\text{Myr}$  
 & Log(N$_{\mathrm{core}}$/$\mathrm{N}_{\mathrm{outer}}$) vs. Log(N) & $1.297 \pm 0.146$ &   240\\
  & Log($\mathrm{S}_{\mathrm{core}}$/$\mathrm{S}_{\mathrm{outer}}$) vs. Log(N) & $0.707 \pm 0.070$ &   241\\

200\, \text{Myr} $<$ age $\leq 400\,\text{Myr}$  
 & Log(N$_{\mathrm{core}}$/$\mathrm{N}_{\mathrm{outer}}$) vs. Log(N) & $1.125 \pm 0.101$ &   242\\
  & Log($\mathrm{S}_{\mathrm{core}}$/$\mathrm{S}_{\mathrm{outer}}$) vs. Log(N) & $0.857 \pm 0.062$ &   243\\

400\, \text{Myr} $<$ age $\leq 800\,\text{Myr}$  
 & Log(N$_{\mathrm{core}}$/$\mathrm{N}_{\mathrm{outer}}$) vs. Log(N) & $1.485 \pm 0.106$ &   245\\
  & Log($\mathrm{S}_{\mathrm{core}}$/$\mathrm{S}_{\mathrm{outer}}$) vs. Log(N) & $0.916 \pm 0.051$ &   250\\
    
age $>$ 800\,\text{Myr}
 & Log(N$_{\mathrm{core}}$/$\mathrm{N}_{\mathrm{outer}}$) vs. Log(N) & $1.442 \pm 0.128$ &   224\\
  & Log($\mathrm{S}_{\mathrm{core}}$/$\mathrm{S}_{\mathrm{outer}}$) vs. Log(N) & $0.853 \pm 0.051$ &   229\\
\bottomrule
\end{tabular}
\tablefoot{Such as table~\ref{tab:Tab.2}.}
\end{table*}

\begin{figure}
	\centering
    \includegraphics[angle=0,width=88mm]{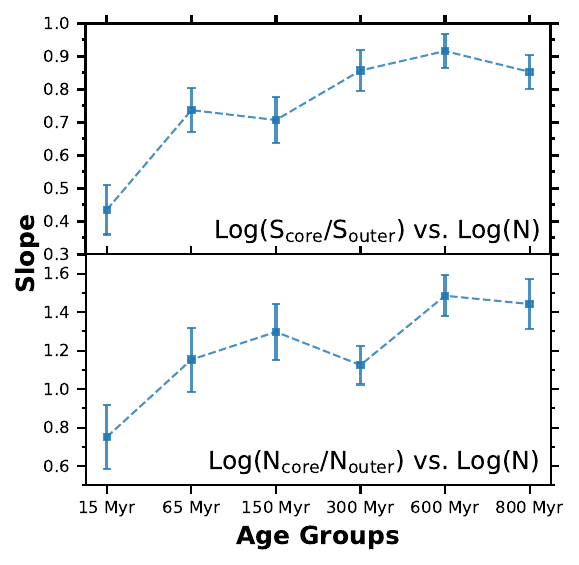}
	\caption{Same as Fig.~\ref{fig: Group_comparison}, but those from different age groups in Table~\ref{tab:Cavallo_age}. The first five age groups are labeled by their median ages, with the last group being marked by its lower age limit (800~Myr).}
        \label{fig: age_comparison}
\end{figure}

Moreover, \citet{Hu+2025a} suggested that there is no disruption for the morphological stability of OCs in the local environment around the Sun. So to investigate the influence of the local environment on our samples, we plotted the distribution of the morphological stability parameter I of ``Left tilt'' and ``Right tilt'' sample OCs against X, Y, and Z-axes, as shown in Fig.~\ref{fig:XYZ}. The results show that sample OCs closer to the Sun (marked by a vertical red dotted line) exhibit stronger morphological stability, with the strongest stability observed within approximately 200 pc. This suggests that the solar vicinity may be a relatively ``quiet" region, lacking complex additional external forces (e.g., strong gravitational perturbations from other massive structures and intense interactions with giant molecular clouds) that severely disrupt OC equilibrium \citep{Rangwal+2025}. This result is consistent with the local bubbles mentioned in \citet{Gozha+2015}. Besides, among the different directions, the morphological stability is generally higher in the Z direction (perpendicular to the Galactic plane) compared to the X and Y directions (within the plane). This is likely due to stronger gravitational restoring forces vertical to the disk, whereas in-plane directions are more susceptible to influences such as spiral arm crossings, tidal forces, and interactions with interstellar clouds.

\begin{figure}[htbp]
    \centering
    \includegraphics[angle=0,width=90mm]{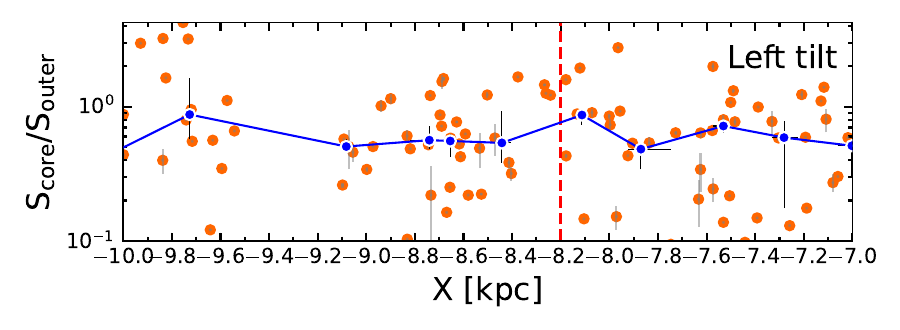}
    \includegraphics[angle=0,width=90mm]{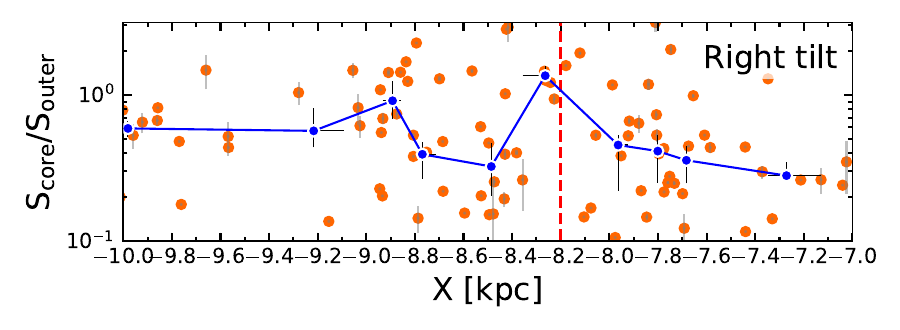}
    \includegraphics[angle=0,width=90mm]{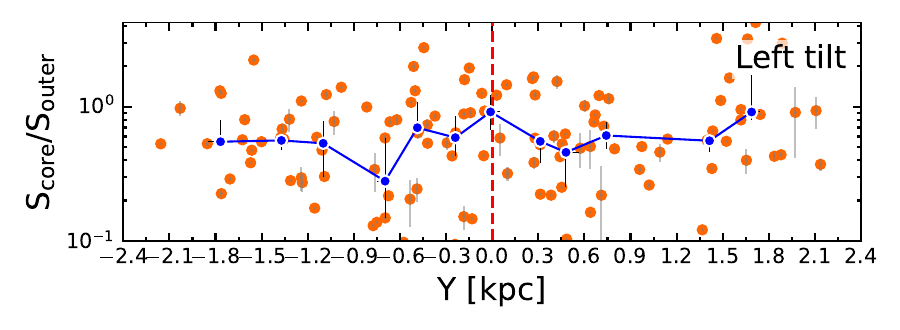}
    \includegraphics[angle=0,width=90mm]{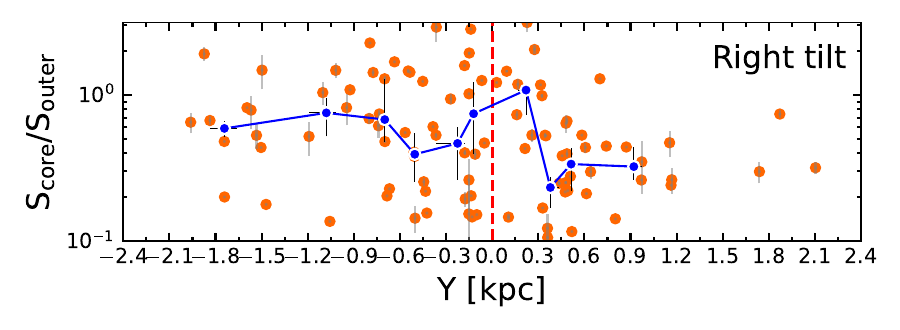}
    \includegraphics[angle=0,width=90mm]{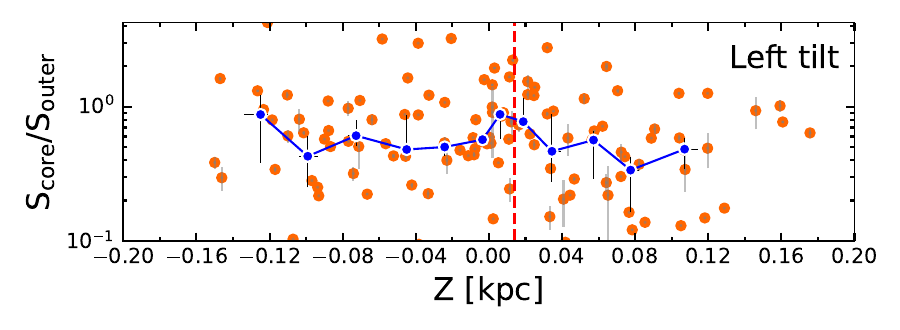}
    \includegraphics[angle=0,width=90mm]{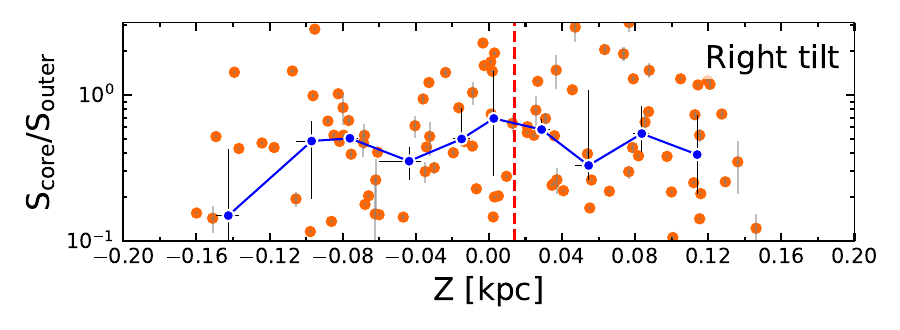}
    \caption{Distributions of the morphological stability I $\mathrm{S}_{\mathrm{core}}$/$\mathrm{S}_{\mathrm{outer}}$ in the X, Y, and Z directions for the sample OCs in the ``left-tilt" and ``right-tilt'' regions (from top to bottom: X, Y, and Z). Every ten sample OCs in each panel form a bin. Blue scattered points represent the bootstrap \citep{Kawano+1995} median estimate of each bin, with error bars indicating the 95\% confidence intervals derived from bootstrap resampling (B = 1000). The red vertical line marks the position of the Sun. Bootstrap methodology was employed to provide robust uncertainty estimates for the median values, given the non-parametric nature of the data distribution.}
    \label{fig:XYZ}
\end{figure}

\section{Discussion}\label{Discussion}

\hspace{2em}The morphological stability derived from the irregular morphology of quantitatively OCs in rose diagrams is a key parameter in this work, which depends on the calculation of the core radius in this approach. Therefore, it is necessary to discuss the systematic biases and sensitivities affecting the determination of the core radius ($\mathrm{R}_{\mathrm{core}}$).

In addition to the fluctuation of the tidal radius within the error range affecting the determination of the core radius, we also considered the influence of starting angle randomness in the rose diagram construction method on the calculation of the core radius. We conducted 10 random starting angle experiments to verify the effect of the member star distribution on the core radius calculation. The specific method is as follows: based on the mean tidal radius $\mathrm{R}_{\mathrm{t}}$ of each sample \text{OC}, we fixed $\mathrm{R}_{\mathrm{t}}$ as the boundary to re-select its members, calculated their projected coordinates relative to the cluster center, and quantified their 2D spatial distribution through the rose diagram method. During the construction of the rose diagram, a random starting angle $\theta_{\mathrm{j}} \sim \mathcal{U} (0^{\circ}, 360^{\circ})$ between $0^{\circ}$ and $360^{\circ}$ was independently generated for each iteration to rotate the reference direction of the sector division. Repeating this process 10 times, we obtained 10 independent estimates of the core radius $\mathrm{R}_{\mathrm{core}}$, core area $\mathrm{S}_{\mathrm{core}}$, and outer area $\mathrm{S}_{\mathrm{outer}}$. By this process, we can plot the sampling rose diagrams of all sample OCs, such as \texttt{NGC\_6791} presented in Fig.~\ref{fig: 10_sample_fixed_Rt_rose}. Finally, we compared the two morphological stability parameters obtained from the previous method and this way, as shown in Fig.~\ref{fig: method_comparison}.

To further comprehensively evaluate the combined impact of tidal radius uncertainty and starting angle randomness on the estimation of morphological parameters, we designed a Monte Carlo experiment integrating the two aforementioned randomization processes. The specific procedure is as follows: based on the mean tidal radius $\mathrm{R}_{\mathrm{t}}$ of each \text{OC} and its standard deviation $\sigma_{\mathrm{R}_{\mathrm{t}}}$, a normal distribution $\mathcal{N}(\mathrm{R}_{\mathrm{t}}, \sigma_{\mathrm{R}_{\mathrm{t}}}^{2})$ is constructed. In each iteration $\mathrm{k}$, a tidal radius value $\mathrm{R}_{\mathrm{t}}^{(\mathrm{k})}$ is first independently drawn from this distribution and used as the boundary to re-select cluster member stars, followed by calculating their projected coordinates. Subsequently, when quantifying their 2D spatial distribution using the rose diagram method, an independent random starting angle $\theta_{\mathrm{k}} \sim \mathcal{U} (0^{\circ}, 360^{\circ})$ is generated to determine the reference direction for sector division. This combined sampling process was repeated 10 times, yielding 10 independent sets of estimates for the core radius $\mathrm{R}_{\mathrm{core}}^{(\mathrm{k})}$, core area $\mathrm{S}_{\mathrm{core}}^{(\mathrm{k})}$, and outer area $\mathrm{S}_{\mathrm{outer}}^{(\mathrm{k})}$ under the dual randomization of both tidal radius and starting angle. This method aims to simultaneously reflect the potential influence of uncertainties in both boundary delineation and directional definition on the calculation of morphological stability parameters. All results from the combined random sampling for our samples can be obtained, such as the rose diagrams of \texttt{NGC\_6791} shown in Fig.~\ref{fig: 10_sample_rose_2}. And we also plotted the comparison of the two morphological stability parameters derived from this way and our original approach, as displayed in Fig.~\ref{fig: method_comparison_2}.

By comparing the distributions and dispersions of the morphological parameters obtained from all methods, as shown in Figs.~\ref{fig: method_comparison} and \ref{fig: method_comparison_2}, we found that the sampling analysis results of all random starting angles are consistent with the relevant conclusions based on the fixed starting angle at 0$^{\degree}$ within error ranges.

\begin{figure}
	\centering
    \includegraphics[angle=0,width=88mm]{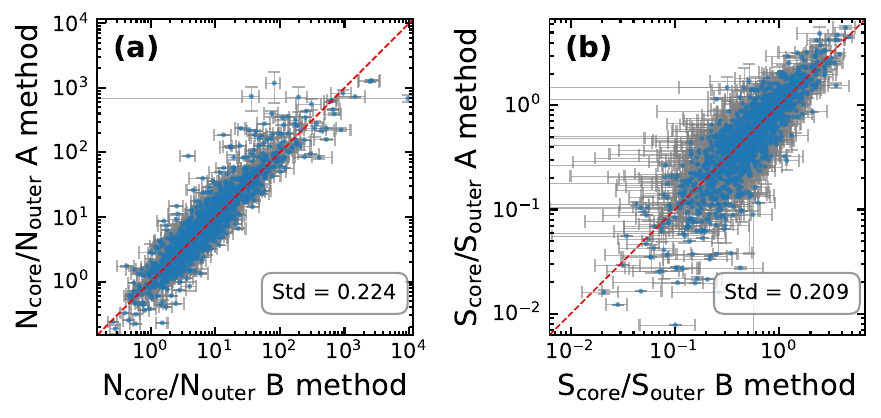}
	\caption{Comparison of the morphological stability parameters (left panel: N$_{\mathrm{core}}$/$\mathrm{N}_{\mathrm{outer}}$, right panel: $\mathrm{S}_{\mathrm{core}}$/$\mathrm{S}_{\mathrm{outer}}$) derived from two methods. Method A denotes sampling the tidal radius ten times, with Method B sampling the starting angle ten times. Red dashed lines indicate 1:1 correspondence. Error bars represent standard deviations.}
        \label{fig: method_comparison}
\end{figure}

\begin{figure}
	\centering
    \includegraphics[angle=0,width=88mm]{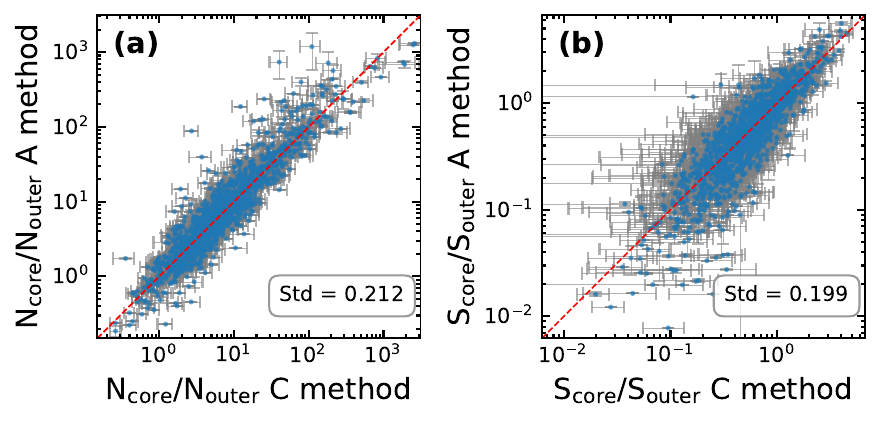}
	\caption{Such as \ref{fig: method_comparison}, but Method C samples both the tidal radius and the starting angle ten times.}
        \label{fig: method_comparison_2}
\end{figure}

\section{Summary and conclusions}

\hspace{2em} In this paper, we investigated the 2D morphological stability parameters of 1,490 OCs in the Galactic spherical coordinate frame by the rose diagram construction and explored their relationships with the number of member stars and their potential variations at different spatial locations by employing the Bayesian fitting method. The data of sample OCs used in this work are sourced from the members stars cross-matched by \citet{van+2023} from Gaia DR3. Besides, instead of stacking multiple line-of-sight projections, we projected each sample OC only once along its individual line of sight, corrected for distortions in Galactic longitude \texttt{b}, and forced the representation onto the plane that is perpendicular to the Galactic disk. Finally, we discussed the potential influence of systematic biases inherent in the rose diagram method on the results. Our main findings and conclusions are as follows:

\vspace{1em} 

1. This work defined for the first time a new morphological stability parameter of N$_{\mathrm{core}}$/$\mathrm{N}_{\mathrm{outer}}$. And within the 2D projection, both morphological stability parameters are positively correlated with the number of members. This all indicates that within a certain range, the more member stars an OC has, the stronger its binding ability will be, and the denser the structure, the stronger its stability will be.

\vspace{1em}

2. The morphological stability parameters II N$_{\mathrm{core}}$/$\mathrm{N}_{\mathrm{outer}}$ has a higher slope in \texttt{Q1} region than in \texttt{Q2} region, while the morphological stability parameters I $\mathrm{S}_{\mathrm{core}}$/$\mathrm{S}_{\mathrm{outer}}$ shows an inverse trend. This suggests that the morphological stability parameters I and II jointly characterized the structural configuration and evolutionary stage of OCs.

\vspace{1em} 

3. The two morphological stability parameters exhibit lower values in the tangential OCs, compared to the radial region, indicating that the influence of the tidal forces towards the Galactic center on the tangential sample OCs is likely stronger than that of the shear force resulting from the Galactic differential rotation on the radial sample OCs.

\vspace{1em} 

4. The slope of the morphological stability parameters against the number of members in the ``tilt-right'' region is generally lower than that in the ``tilt-left'' region. This indicates that the influence of external environments on the sample OCs is asymmetric, which may be related to their proximity to the Galactic center. The closer they are to the Galactic center, the stronger the galactic tidal force they experience.

\vspace{1em} 

5. The slopes of the morphological stability of the sample OCs against the number of members show a significant increase in the trend from young to old. This implies that compared to older sample OCs, younger sample OCs experience more severe external environmental disturbances.

\vspace{1em} 

6. The highest morphological stability parameters observed in OCs near the Sun suggest that external perturbations in this region are relatively weak, implying the existence of a locally quiescent environment.

\vspace{1em}

We established a connection between the internal evolution and external environment of the sample OCs by their morphological stability parameters. Specifically, our results reveal that a lower morphological stability parameter is indicative of a more complex and dynamically active evolutionary state for the OCs. To deepen our understanding, future studies could perform stratified analysis considering OCs' age and mass to examine whether morphological stability trends of the OCs still hold across different sub-populations in the local environment. Extending the sample to greater distances, particularly toward spiral arms or the Galactic center, would help identify critical scales at which the morphological stability changes markedly, shedding light on the global influence of Galactic structure on cluster evolution.

\label{sec:summary and conclusion}

\begin{acknowledgements}
      We thank an anonymous reviewer for valuable suggestions that have improved our methodology, results, and presentation. This work is supported by the National Natural Science Foundation of China (NSFC) under grant Nos. 12303037 and 12573035, the Natural Science Foundation of Sichuan Province (No. 2026NSFSC0742), and the Fundamental Research Funds of China West Normal University (CWNU, No.23KE024). Qingshun Hu would like to acknowledge the financial support provided by the China Scholarship Council program (Grant No. 202308510136). This study has made an indirect use of Gaia data, through large catalogues of open clusters and stellar parameters data published in the third data release. Gaia is operated by the European Space Agency (ESA) (\url{https://www.cosmos.esa.int/gaia}). The preparation of this work has made extensive use of Topcat \citep{Taylor+2005}, of the Simbad and VizieR databases at CDS, Strasbourg, France, and of NASA's Astrophysics Data System Bibliographic Services. 
\end{acknowledgements}

\bibliographystyle{aa}
\bibliography{references}

@ARTICLE{Alfonso+2024,
       author = {{Alfonso}, Jeison and {Garc{\'\i}a-Varela}, Alejandro and {Vieira}, Katherine},
        title = "{Exploring Galactic open clusters with Gaia: I. An examination in the first kiloparsec}",
      journal = {\aap},
         year = 2024,
        month = sep,
       volume = {689},
          eid = {A18},
        pages = {A18},
          doi = {10.1051/0004-6361/202450901},
archivePrefix = {arXiv},
       eprint = {2407.09407},
 primaryClass = {astro-ph.GA},
       adsurl = {https://ui.adsabs.harvard.edu/abs/2024A&A...689A..18A }
}

@ARTICLE{Almeida+2025,
       author = {{Almeida}, Duarte and {Moitinho}, Andr{\'e} and {Moreira}, Sandro},
        title = "{Open cluster dissolution rate and the initial cluster mass function in the solar neighbourhood: Modelling the age and mass distributions of clusters observed by Gaia}",
      journal = {\aap},
         year = 2025,
        month = jan,
       volume = {693},
          eid = {A305},
        pages = {A305},
          doi = {10.1051/0004-6361/202451853},
archivePrefix = {arXiv},
       eprint = {2412.19204},
 primaryClass = {astro-ph.GA},
       adsurl = {https://ui.adsabs.harvard.edu/abs/2025A&A...693A.305A }
}

@ARTICLE{Akhmetali+2025,
       author = {{Akhmetali}, Almat},
        title = "{Fractality of open clusters in singles, pairs, and groups}",
      journal = {Journal of Astrophysics and Astronomy},
         year = 2026,
        month = jan,
       volume = {47},
       number = {1},
          eid = {4},
        pages = {4},
          doi = {10.1007/s12036-025-10122-3},
archivePrefix = {arXiv},
       eprint = {2508.13980},
 primaryClass = {astro-ph.GA},
       adsurl = {https://ui.adsabs.harvard.edu/abs/2026JApA...47....4A}
}

@ARTICLE{Perryman+2025,
       author = {{Perryman}, Michael},
        title = "{Space astrometry with Gaia: Advances in understanding our Galaxy}",
      journal = {\physrep},
         year = 2026,
        month = jan,
       volume = {1150},
        pages = {1-229},
          doi = {10.1016/j.physrep.2025.09.004},
archivePrefix = {arXiv},
       eprint = {2509.10883},
 primaryClass = {astro-ph.IM},
       adsurl = {https://ui.adsabs.harvard.edu/abs/2026PhR..1150....1P}
}

@ARTICLE{Qiu+2025,
       author = {{Qiu}, Jin-Sheng and {Wan}, Zhen and {Li}, Xu-Zhi and {Zhu}, Qing-Feng and {Fan}, Lu-lu and {Xu}, Xiao-Hui and {Zhao}, Jun-Han and {Pu}, Zhi-Yong},
        title = "{The statistical analysis of the Galactic open clusters' structure}",
      journal = {\mnras},
         year = 2026,
        month = jan,
       volume = {545},
       number = {2},
          eid = {staf2134},
        pages = {staf2134},
          doi = {10.1093/mnras/staf2134},
archivePrefix = {arXiv},
       eprint = {2512.09387},
 primaryClass = {astro-ph.GA},
       adsurl = {https://ui.adsabs.harvard.edu/abs/2026MNRAS.545f2134Q}
}

@ARTICLE{Alves+2012,
       author = {{Alves}, J. and {Bouy}, H.},
        title = "{Orion revisited. I. The massive cluster in front of the Orion nebula cluster}",
      journal = {\aap},
         year = 2012,
        month = nov,
       volume = {547},
          eid = {A97},
        pages = {A97},
          doi = {10.1051/0004-6361/201220119},
archivePrefix = {arXiv},
       eprint = {1209.3787},
 primaryClass = {astro-ph.GA},
       adsurl = {https://ui.adsabs.harvard.edu/abs/2012A&A...547A..97A }
}

@ARTICLE{Anders+2017,
       author = {{Anders}, F. and {Chiappini}, C. and {Minchev}, I. and {Miglio}, A. and {Montalb{\'a}n}, J. and {Mosser}, B. and {Rodrigues}, T.~S. and {Santiago}, B.~X. and {Baudin}, F. and {Beers}, T.~C. and {da Costa}, L.~N. and {Garc{\'\i}a}, R.~A. and {Garc{\'\i}a-Hern{\'a}ndez}, D.~A. and {Holtzman}, J. and {Maia}, M.~A.~G. and {Majewski}, S. and {Mathur}, S. and {Noels-Grotsch}, A. and {Pan}, K. and {Schneider}, D.~P. and {Schultheis}, M. and {Steinmetz}, M. and {Valentini}, M. and {Zamora}, O.},
        title = "{Red giants observed by CoRoT and APOGEE: The evolution of the Milky Way's radial metallicity gradient}",
      journal = {\aap},
         year = 2017,
        month = apr,
       volume = {600},
          eid = {A70},
        pages = {A70},
          doi = {10.1051/0004-6361/201629363},
archivePrefix = {arXiv},
       eprint = {1608.04951},
 primaryClass = {astro-ph.GA},
       adsurl = {https://ui.adsabs.harvard.edu/abs/2017A&A...600A..70A}
}

@ARTICLE{Angelo+2025,
       author = {{Angelo}, M.~S. and {Santos}, J.~F.~C. and {Corradi}, W.~J.~B. and {Maia}, F.~F.~S.},
        title = "{Exploring the dynamical state of Galactic open clusters using Gaia DR3 and observational parameters}",
      journal = {\mnras},
         year = 2025,
        month = may,
       volume = {539},
       number = {3},
        pages = {2513-2536},
          doi = {10.1093/mnras/staf584},
archivePrefix = {arXiv},
       eprint = {2504.06362},
 primaryClass = {astro-ph.GA},
       adsurl = {https://ui.adsabs.harvard.edu/abs/2025MNRAS.539.2513A}
}

@ARTICLE{Astropy+2013,
       author = {{Astropy Collaboration} and {Robitaille}, Thomas P. and {Tollerud}, Erik J. and {Greenfield}, Perry and {Droettboom}, Michael and {Bray}, Erik and {Aldcroft}, Tom and {Davis}, Matt and {Ginsburg}, Adam and {Price-Whelan}, Adrian M. and {Kerzendorf}, Wolfgang E. and {Conley}, Alexander and {Crighton}, Neil and {Barbary}, Kyle and {Muna}, Demitri and {Ferguson}, Henry and {Grollier}, Fr{\'e}d{\'e}ric and {Parikh}, Madhura M. and {Nair}, Prasanth H. and {Unther}, Hans M. and {Deil}, Christoph and {Woillez}, Julien and {Conseil}, Simon and {Kramer}, Roban and {Turner}, James E.~H. and {Singer}, Leo and {Fox}, Ryan and {Weaver}, Benjamin A. and {Zabalza}, Victor and {Edwards}, Zachary I. and {Azalee Bostroem}, K. and {Burke}, D.~J. and {Casey}, Andrew R. and {Crawford}, Steven M. and {Dencheva}, Nadia and {Ely}, Justin and {Jenness}, Tim and {Labrie}, Kathleen and {Lim}, Pey Lian and {Pierfederici}, Francesco and {Pontzen}, Andrew and {Ptak}, Andy and {Refsdal}, Brian and {Servillat}, Mathieu and {Streicher}, Ole},
        title = "{Astropy: A community Python package for astronomy}",
      journal = {\aap},
         year = 2013,
        month = oct,
       volume = {558},
          eid = {A33},
        pages = {A33},
          doi = {10.1051/0004-6361/201322068},
archivePrefix = {arXiv},
       eprint = {1307.6212},
 primaryClass = {astro-ph.IM},
       adsurl = {https://ui.adsabs.harvard.edu/abs/2013A&A...558A..33A}
}

@ARTICLE{Astropy+2018,
       author = {{Astropy Collaboration} and {Price-Whelan}, A.~M. and {Sip{\H{o}}cz}, B.~M. and {G{\"u}nther}, H.~M. and {Lim}, P.~L. and {Crawford}, S.~M. and {Conseil}, S. and {Shupe}, D.~L. and {Craig}, M.~W. and {Dencheva}, N. and {Ginsburg}, A. and {VanderPlas}, J.~T. and {Bradley}, L.~D. and {P{\'e}rez-Su{\'a}rez}, D. and {de Val-Borro}, M. and {Aldcroft}, T.~L. and {Cruz}, K.~L. and {Robitaille}, T.~P. and {Tollerud}, E.~J. and {Ardelean}, C. and {Babej}, T. and {Bach}, Y.~P. and {Bachetti}, M. and {Bakanov}, A.~V. and {Bamford}, S.~P. and {Barentsen}, G. and {Barmby}, P. and {Baumbach}, A. and {Berry}, K.~L. and {Biscani}, F. and {Boquien}, M. and {Bostroem}, K.~A. and {Bouma}, L.~G. and {Brammer}, G.~B. and {Bray}, E.~M. and {Breytenbach}, H. and {Buddelmeijer}, H. and {Burke}, D.~J. and {Calderone}, G. and {Cano Rodr{\'\i}guez}, J.~L. and {Cara}, M. and {Cardoso}, J.~V.~M. and {Cheedella}, S. and {Copin}, Y. and {Corrales}, L. and {Crichton}, D. and {D'Avella}, D. and {Deil}, C. and {Depagne}, {\'E}. and {Dietrich}, J.~P. and {Donath}, A. and {Droettboom}, M. and {Earl}, N. and {Erben}, T. and {Fabbro}, S. and {Ferreira}, L.~A. and {Finethy}, T. and {Fox}, R.~T. and {Garrison}, L.~H. and {Gibbons}, S.~L.~J. and {Goldstein}, D.~A. and {Gommers}, R. and {Greco}, J.~P. and {Greenfield}, P. and {Groener}, A.~M. and {Grollier}, F. and {Hagen}, A. and {Hirst}, P. and {Homeier}, D. and {Horton}, A.~J. and {Hosseinzadeh}, G. and {Hu}, L. and {Hunkeler}, J.~S. and {Ivezi{\'c}}, {\v{Z}}. and {Jain}, A. and {Jenness}, T. and {Kanarek}, G. and {Kendrew}, S. and {Kern}, N.~S. and {Kerzendorf}, W.~E. and {Khvalko}, A. and {King}, J. and {Kirkby}, D. and {Kulkarni}, A.~M. and {Kumar}, A. and {Lee}, A. and {Lenz}, D. and {Littlefair}, S.~P. and {Ma}, Z. and {Macleod}, D.~M. and {Mastropietro}, M. and {McCully}, C. and {Montagnac}, S. and {Morris}, B.~M. and {Mueller}, M. and {Mumford}, S.~J. and {Muna}, D. and {Murphy}, N.~A. and {Nelson}, S. and {Nguyen}, G.~H. and {Ninan}, J.~P. and {N{\"o}the}, M. and {Ogaz}, S. and {Oh}, S. and {Parejko}, J.~K. and {Parley}, N. and {Pascual}, S. and {Patil}, R. and {Patil}, A.~A. and {Plunkett}, A.~L. and {Prochaska}, J.~X. and {Rastogi}, T. and {Reddy Janga}, V. and {Sabater}, J. and {Sakurikar}, P. and {Seifert}, M. and {Sherbert}, L.~E. and {Sherwood-Taylor}, H. and {Shih}, A.~Y. and {Sick}, J. and {Silbiger}, M.~T. and {Singanamalla}, S. and {Singer}, L.~P. and {Sladen}, P.~H. and {Sooley}, K.~A. and {Sornarajah}, S. and {Streicher}, O. and {Teuben}, P. and {Thomas}, S.~W. and {Tremblay}, G.~R. and {Turner}, J.~E.~H. and {Terr{\'o}n}, V. and {van Kerkwijk}, M.~H. and {de la Vega}, A. and {Watkins}, L.~L. and {Weaver}, B.~A. and {Whitmore}, J.~B. and {Woillez}, J. and {Zabalza}, V. and {Astropy Contributors}},
        title = "{The Astropy Project: Building an Open-science Project and Status of the v2.0 Core Package}",
      journal = {\aj},
         year = 2018,
        month = sep,
       volume = {156},
       number = {3},
          eid = {123},
        pages = {123},
          doi = {10.3847/1538-3881/aabc4f},
archivePrefix = {arXiv},
       eprint = {1801.02634},
 primaryClass = {astro-ph.IM},
       adsurl = {https://ui.adsabs.harvard.edu/abs/2018AJ....156..123A}
}

@ARTICLE{Balaguer+2020,
       author = {{Balaguer-N{\'u}{\~n}ez}, L. and {L{\'o}pez del Fresno}, M. and {Solano}, E. and {Galad{\'\i}-Enr{\'\i}quez}, D. and {Jordi}, C. and {Jimenez-Esteban}, F. and {Masana}, E. and {Carbajo-Hijarrubia}, J. and {Paunzen}, E.},
        title = "{Clusterix 2.0: a virtual observatory tool to estimate cluster membership probability}",
      journal = {\mnras},
         year = 2020,
        month = mar,
       volume = {492},
       number = {4},
        pages = {5811-5843},
          doi = {10.1093/mnras/stz3610},
archivePrefix = {arXiv},
       eprint = {1910.07356},
 primaryClass = {astro-ph.IM},
       adsurl = {https://ui.adsabs.harvard.edu/abs/2020MNRAS.492.5811B }
}

@ARTICLE{Bailer-Jones+2015,
       author = {{Bailer-Jones}, Coryn A.~L.},
        title = "{Estimating Distances from Parallaxes}",
      journal = {\pasp},
         year = 2015,
        month = oct,
       volume = {127},
       number = {956},
        pages = {994},
          doi = {10.1086/683116},
archivePrefix = {arXiv},
       eprint = {1507.02105},
 primaryClass = {astro-ph.IM},
       adsurl = {https://ui.adsabs.harvard.edu/abs/2015PASP..127..994B}
}

@ARTICLE{Bergond+2001,
       author = {{Bergond}, G. and {Leon}, S. and {Guibert}, J.},
        title = "{Gravitational tidal effects on galactic open clusters}",
      journal = {\aap},
         year = 2001,
        month = oct,
       volume = {377},
        pages = {462-472},
          doi = {10.1051/0004-6361:20011043},
archivePrefix = {arXiv},
       eprint = {astro-ph/0109341},
 primaryClass = {astro-ph},
       adsurl = {https://ui.adsabs.harvard.edu/abs/2001A&A...377..462B }
}

@BOOK{Binney+2008,
       author = {{Binney}, James and {Tremaine}, Scott},
        title = "{Galactic Dynamics: Second Edition}",
         year = 2008,
       adsurl = {https://ui.adsabs.harvard.edu/abs/2008gady.book.....B }
}

@INPROCEEDINGS{Broggi+2024,
       author = {{Broggi}, Luca and {Stone}, Nicholas C. and {Ryu}, Taeho and {Bortolas}, Elisa and {Dotti}, Massimo and {Bonetti}, Matteo and {Sesana}, Alberto},
        title = "{Partial tidal disruptions and two body relaxation}",
    booktitle = {EAS2024, European Astronomical Society Annual Meeting},
         year = 2024,
        month = jul,
          eid = {328},
        pages = {328},
       adsurl = {https://ui.adsabs.harvard.edu/abs/2024eas..conf..328B }
}

@ARTICLE{Camargo+2012,
  title={Towards a census of the Galactic anticentre star clusters--II. Exploring lower overdensities},
  author={Camargo, Denilso and Bonatto, Charles and Bica, Eduardo},
  journal={Monthly Notices of the Royal Astronomical Society},
  volume={423},
  number={2},
  pages={1940--1954},
  year={2012},
  publisher={Blackwell Publishing Ltd Oxford, UK}
}

@ARTICLE{Camacho+2025,
       author = {{Camacho}, Vianey and {Bonilla-Barroso}, Andrea and {Ballesteros-Paredes}, Javier and {Zamora-Avil{\'e}s}, Manuel and {Aguilar}, Luis},
        title = "{Dynamical heating of newborn stars driven by accretion-induced orbital tightening}",
      journal = {\mnras},
         year = 2025,
        month = apr,
       volume = {538},
       number = {3},
        pages = {1773-1783},
          doi = {10.1093/mnras/staf378},
archivePrefix = {arXiv},
       eprint = {2502.18016},
 primaryClass = {astro-ph.GA},
       adsurl = {https://ui.adsabs.harvard.edu/abs/2025MNRAS.538.1773C }
}

@ARTICLE{Carrera+2019,
       author = {{Carrera}, R. and {Pasquato}, M. and {Vallenari}, A. and {Balaguer-N{\'u}{\~n}ez}, L. and {Cantat-Gaudin}, T. and {Mapelli}, M. and {Bragaglia}, A. and {Bossini}, D. and {Jordi}, C. and {Galad{\'\i}-Enr{\'\i}quez}, D. and {Solano}, E.},
        title = "{Extended halo of NGC 2682 (M 67) from Gaia DR2}",
      journal = {\aap},
         year = 2019,
        month = jul,
       volume = {627},
          eid = {A119},
        pages = {A119},
          doi = {10.1051/0004-6361/201935599},
archivePrefix = {arXiv},
       eprint = {1905.02020},
 primaryClass = {astro-ph.SR},
       adsurl = {https://ui.adsabs.harvard.edu/abs/2019A&A...627A.119C}
}

@ARTICLE{Cantat-Gaudin+2022,
       author = {{Cantat-Gaudin}, Tristan},
        title = "{Milky Way Star Clusters and Gaia: A Review of the Ongoing Revolution}",
      journal = {Universe},
         year = 2022,
        month = feb,
       volume = {8},
       number = {2},
          eid = {111},
        pages = {111},
          doi = {10.3390/universe8020111},
       adsurl = {https://ui.adsabs.harvard.edu/abs/2022Univ....8..111C }
}

@dataset{Cantat-Gaudin+2018,
       author = {{Cantat-Gaudin}, T. and {Jordi}, C. and {Vallenari}, A. and {Bragaglia}, A. and {Balaguer-Nunez}, L. and {Soubiran}, C. and {Bossini}, D. and {Moitinho}, A. and {Castro-Ginard}, A. and {Krone-Martins}, A. and {Casamiquela}, L. and {Sordo}, R. and {Carrera}, R.},
        title = "{VizieR Online Data Catalog: Gaia DR2 open clusters in the Milky Way (Cantat-Gaudin+, 2018)}",
 howpublished = {VizieR On-line Data Catalog: J/A+A/618/A93. Originally published in: 2018A\&A...618A..93C},
         year = 2018,
        month = jul,
          eid = {J/A+A/618/A93},
          doi = {10.26093/cds/vizier.36180093},
       adsurl = {https://ui.adsabs.harvard.edu/abs/2018yCat..36180093C }
}

@ARTICLE{Cavallo+2024,
       author = {{Cavallo}, Lorenzo and {Spina}, Lorenzo and {Carraro}, Giovanni and {Magrini}, Laura and {Poggio}, Eloisa and {Cantat-Gaudin}, Tristan and {Pasquato}, Mario and {Lucatello}, Sara and {Ortolani}, Sergio and {Schiappacasse-Ulloa}, Jose},
        title = "{Parameter Estimation for Open Clusters using an Artificial Neural Network with a QuadTree-based Feature Extractor}",
      journal = {\aj},
         year = 2024,
        month = jan,
       volume = {167},
       number = {1},
          eid = {12},
        pages = {12},
          doi = {10.3847/1538-3881/ad07e5},
archivePrefix = {arXiv},
       eprint = {2311.03009},
 primaryClass = {astro-ph.GA},
       adsurl = {https://ui.adsabs.harvard.edu/abs/2024AJ....167...12C }
}

@ARTICLE{Coenda+2025,
       author = {{Coenda}, V. and {Baume}, G. and {Palma}, T. and {Feinstein}, C.},
        title = "{Global properties, fractality, and mass segregation in single, paired, and grouped open clusters}",
      journal = {\aap},
         year = 2025,
        month = jun,
       volume = {699},
          eid = {A15},
        pages = {A15},
          doi = {10.1051/0004-6361/202554809},
archivePrefix = {arXiv},
       eprint = {2505.14788},
 primaryClass = {astro-ph.GA},
       adsurl = {https://ui.adsabs.harvard.edu/abs/2025A&A...699A..15C }
}

@ARTICLE{Dalessandro+2021,
       author = {{Dalessandro}, Emanuele and {Varri}, A.~L. and {Tiongco}, M. and {Vesperini}, E. and {Fanelli}, C. and {Mucciarelli}, A. and {Origlia}, L. and {Bellazzini}, M. and {Saracino}, S. and {Oliva}, E. and {Sanna}, N. and {Fabrizio}, M. and {Livernois}, A.},
        title = "{First Phase Space Portrait of a Hierarchical Stellar Structure in the Milky Way}",
      journal = {\apj},
         year = 2021,
        month = mar,
       volume = {909},
       number = {1},
          eid = {90},
        pages = {90},
          doi = {10.3847/1538-4357/abda43},
archivePrefix = {arXiv},
       eprint = {2101.04133},
 primaryClass = {astro-ph.GA},
       adsurl = {https://ui.adsabs.harvard.edu/abs/2021ApJ...909...90D }
}

@ARTICLE{Della+2024,
       author = {{Della Croce}, A. and {Dalessandro}, E. and {Livernois}, A. and {Vesperini}, E.},
        title = "{Young, wild, and free: The early expansion of star clusters}",
      journal = {\aap},
         year = 2024,
        month = mar,
       volume = {683},
          eid = {A10},
        pages = {A10},
          doi = {10.1051/0004-6361/202347420},
archivePrefix = {arXiv},
       eprint = {2312.02263},
 primaryClass = {astro-ph.GA},
       adsurl = {https://ui.adsabs.harvard.edu/abs/2024A&A...683A..10D }
}

@ARTICLE{Della+2023,
       author = {{Della Croce}, A. and {Dalessandro}, E. and {Livernois}, A. and {Vesperini}, E. and {Fanelli}, C. and {Origlia}, L. and {Bellazzini}, M. and {Oliva}, E. and {Sanna}, N. and {Varri}, A.~L.},
        title = "{Ongoing hierarchical massive cluster assembly: The LISCA II structure in the Perseus complex}",
      journal = {\aap},
         year = 2023,
        month = jun,
       volume = {674},
          eid = {A93},
        pages = {A93},
          doi = {10.1051/0004-6361/202346095},
archivePrefix = {arXiv},
       eprint = {2303.15501},
 primaryClass = {astro-ph.GA},
       adsurl = {https://ui.adsabs.harvard.edu/abs/2023A&A...674A..93D }
}

@ARTICLE{Della+2025,
       author = {{Della Croce}, A. and {Dalessandro}, E. and {Vesperini}, E. and {Bellazzini}, M. and {Fanelli}, C. and {Origlia}, L. and {Sanna}, N.},
        title = "{Tracing the W3/W4/W5 and Perseus complex dynamical evolution with star clusters}",
      journal = {\aap},
         year = 2025,
        month = jun,
       volume = {698},
          eid = {A142},
        pages = {A142},
          doi = {10.1051/0004-6361/202553840},
archivePrefix = {arXiv},
       eprint = {2504.16159},
 primaryClass = {astro-ph.GA},
       adsurl = {https://ui.adsabs.harvard.edu/abs/2025A&A...698A.142D}
}

@ARTICLE{Escala+2025,
       author = {{Escala}, Ivanna and {Grion Filho}, Douglas and {Guhathakurta}, Puragra and {Gilbert}, Karoline M. and {Fardal}, Mark and {Cullinane}, L.~R. and {Tollerud}, Erik and {Quirk}, Amanda C.~N. and {Chen}, Zhuo and {Hyver}, Molly and {Williams}, Benjamin F.},
        title = "{Kinematical Modeling of the Resolved Stellar Outskirts of M32: Constraints on Tidal Stripping Scenarios}",
      journal = {\apj},
         year = 2025,
        month = sep,
       volume = {991},
       number = {1},
          eid = {31},
        pages = {31},
          doi = {10.3847/1538-4357/adf63c},
archivePrefix = {arXiv},
       eprint = {2502.20594},
 primaryClass = {astro-ph.GA},
       adsurl = {https://ui.adsabs.harvard.edu/abs/2025ApJ...991...31E }
}

@ARTICLE{Elmegreen+2008,
       author = {{Elmegreen}, Bruce G.},
        title = "{Variations in Stellar Clustering with Environment: Dispersed Star Formation and the Origin of Faint Fuzzies}",
      journal = {\apj},
         year = 2008,
        month = jan,
       volume = {672},
       number = {2},
        pages = {1006-1012},
          doi = {10.1086/523791},
archivePrefix = {arXiv},
       eprint = {0710.5788},
 primaryClass = {astro-ph},
       adsurl = {https://ui.adsabs.harvard.edu/abs/2008ApJ...672.1006E}
}

@ARTICLE{Elmegreen+2018,
       author = {{Elmegreen}, Bruce G.},
        title = "{On the Dispersal of Young Stellar Hierarchies}",
      journal = {\apj},
         year = 2018,
        month = jan,
       volume = {853},
       number = {1},
          eid = {88},
        pages = {88},
          doi = {10.3847/1538-4357/aaa252},
archivePrefix = {arXiv},
       eprint = {1712.05967},
 primaryClass = {astro-ph.GA},
       adsurl = {https://ui.adsabs.harvard.edu/abs/2018ApJ...853...88E}
}

@ARTICLE{Foreman-Mackey+2013,
       author = {{Foreman-Mackey}, Daniel and {Hogg}, David W. and {Lang}, Dustin and {Goodman}, Jonathan},
        title = "{emcee: The MCMC Hammer}",
      journal = {\pasp},
         year = 2013,
        month = mar,
       volume = {125},
       number = {925},
        pages = {306},
          doi = {10.1086/670067},
archivePrefix = {arXiv},
       eprint = {1202.3665},
 primaryClass = {astro-ph.IM},
       adsurl = {https://ui.adsabs.harvard.edu/abs/2013PASP..125..306F }
}

@article{Foreman-Mackey+2016,
  author  = {Foreman-Mackey, D.},
  title   = {corner.py: Scatterplot matrices in Python},
  journal = {JOSS},
  year    = {2016},
  volume  = {1},
  pages   = {24},
  doi     = {10.21105/joss.00024}
}

@ARTICLE{Fujii+2024,
       author = {{Fujii}, Michiko S. and {Wang}, Long and {Tanikawa}, Ataru and {Hirai}, Yutaka and {Saitoh}, Takayuki R.},
        title = "{Simulations predict intermediate-mass black hole formation in globular clusters}",
      journal = {Science},
         year = 2024,
        month = jun,
       volume = {384},
       number = {6703},
        pages = {1488-1492},
          doi = {10.1126/science.adi4211},
archivePrefix = {arXiv},
       eprint = {2406.06772},
 primaryClass = {astro-ph.GA},
       adsurl = {https://ui.adsabs.harvard.edu/abs/2024Sci...384.1488F }
}

@ARTICLE{Foster+2015,
       author = {{Foster}, Jonathan B. and {Cottaar}, Michiel and {Covey}, Kevin R. and {Arce}, H{\'e}ctor G. and {Meyer}, Michael R. and {Nidever}, David L. and {Stassun}, Keivan G. and {Tan}, Jonathan C. and {Chojnowski}, S. Drew and {da Rio}, Nicola and {Flaherty}, Kevin M. and {Rebull}, Luisa and {Frinchaboy}, Peter M. and {Majewski}, Steven R. and {Skrutskie}, Michael and {Wilson}, John C. and {Zasowski}, Gail},
        title = "{IN-SYNC. II. Virial Stars from Subvirial Cores{\textemdash}the Velocity Dispersion of Embedded Pre-main-sequence Stars in NGC 1333}",
      journal = {\apj},
         year = 2015,
        month = feb,
       volume = {799},
       number = {2},
          eid = {136},
        pages = {136},
          doi = {10.1088/0004-637X/799/2/136},
archivePrefix = {arXiv},
       eprint = {1411.6013},
 primaryClass = {astro-ph.GA},
       adsurl = {https://ui.adsabs.harvard.edu/abs/2015ApJ...799..136F}
}

@ARTICLE{GAIADR3+2023,
       author = {{Gaia Collaboration} and {Vallenari}, A. and {Brown}, A.~G.~A. and {Prusti}, T. and {de Bruijne}, J.~H.~J. and {Arenou}, F. and {Babusiaux}, C. and {Biermann}, M. and {Creevey}, O.~L. and {Ducourant}, C. and {Evans}, D.~W. and {Eyer}, L. and {Guerra}, R. and {Hutton}, A. and {Jordi}, C. and {Klioner}, S.~A. and {Lammers}, U.~L. and {Lindegren}, L. and {Luri}, X. and {Mignard}, F. and {Panem}, C. and {Pourbaix}, D. and {Randich}, S. and {Sartoretti}, P. and {Soubiran}, C. and {Tanga}, P. and {Walton}, N.~A. and {Bailer-Jones}, C.~A.~L. and {Bastian}, U. and {Drimmel}, R. and {Jansen}, F. and {Katz}, D. and {Lattanzi}, M.~G. and {van Leeuwen}, F. and {Bakker}, J. and {Cacciari}, C. and {Casta{\~n}eda}, J. and {De Angeli}, F. and {Fabricius}, C. and {Fouesneau}, M. and {Fr{\'e}mat}, Y. and {Galluccio}, L. and {Guerrier}, A. and {Heiter}, U. and {Masana}, E. and {Messineo}, R. and {Mowlavi}, N. and {Nicolas}, C. and {Nienartowicz}, K. and {Pailler}, F. and {Panuzzo}, P. and {Riclet}, F. and {Roux}, W. and {Seabroke}, G.~M. and {Sordo}, R. and {Th{\'e}venin}, F. and {Gracia-Abril}, G. and {Portell}, J. and {Teyssier}, D. and {Altmann}, M. and {Andrae}, R. and {Audard}, M. and {Bellas-Velidis}, I. and {Benson}, K. and {Berthier}, J. and {Blomme}, R. and {Burgess}, P.~W. and {Busonero}, D. and {Busso}, G. and {C{\'a}novas}, H. and {Carry}, B. and {Cellino}, A. and {Cheek}, N. and {Clementini}, G. and {Damerdji}, Y. and {Davidson}, M. and {de Teodoro}, P. and {Nu{\~n}ez Campos}, M. and {Delchambre}, L. and {Dell'Oro}, A. and {Esquej}, P. and {Fern{\'a}ndez-Hern{\'a}ndez}, J. and {Fraile}, E. and {Garabato}, D. and {Garc{\'\i}a-Lario}, P. and {Gosset}, E. and {Haigron}, R. and {Halbwachs}, J.-L. and {Hambly}, N.~C. and {Harrison}, D.~L. and {Hern{\'a}ndez}, J. and {Hestroffer}, D. and {Hodgkin}, S.~T. and {Holl}, B. and {Jan{\ss}en}, K. and {Jevardat de Fombelle}, G. and {Jordan}, S. and {Krone-Martins}, A. and {Lanzafame}, A.~C. and {L{\"o}ffler}, W. and {Marchal}, O. and {Marrese}, P.~M. and {Moitinho}, A. and {Muinonen}, K. and {Osborne}, P. and {Pancino}, E. and {Pauwels}, T. and {Recio-Blanco}, A. and {Reyl{\'e}}, C. and {Riello}, M. and {Rimoldini}, L. and {Roegiers}, T. and {Rybizki}, J. and {Sarro}, L.~M. and {Siopis}, C. and {Smith}, M. and {Sozzetti}, A. and {Utrilla}, E. and {van Leeuwen}, M. and {Abbas}, U. and {{\'A}brah{\'a}m}, P. and {Abreu Aramburu}, A. and {Aerts}, C. and {Aguado}, J.~J. and {Ajaj}, M. and {Aldea-Montero}, F. and {Altavilla}, G. and {{\'A}lvarez}, M.~A. and {Alves}, J. and {Anders}, F. and {Anderson}, R.~I. and {Anglada Varela}, E. and {Antoja}, T. and {Baines}, D. and {Baker}, S.~G. and {Balaguer-N{\'u}{\~n}ez}, L. and {Balbinot}, E. and {Balog}, Z. and {Barache}, C. and {Barbato}, D. and {Barros}, M. and {Barstow}, M.~A. and {Bartolom{\'e}}, S. and {Bassilana}, J.-L. and {Bauchet}, N. and {Becciani}, U. and {Bellazzini}, M. and {Berihuete}, A. and {Bernet}, M. and {Bertone}, S. and {Bianchi}, L. and {Binnenfeld}, A. and {Blanco-Cuaresma}, S. and {Blazere}, A. and {Boch}, T. and {Bombrun}, A. and {Bossini}, D. and {Bouquillon}, S. and {Bragaglia}, A. and {Bramante}, L. and {Breedt}, E. and {Bressan}, A. and {Brouillet}, N. and {Brugaletta}, E. and {Bucciarelli}, B. and {Burlacu}, A. and {Butkevich}, A.~G. and {Buzzi}, R. and {Caffau}, E. and {Cancelliere}, R. and {Cantat-Gaudin}, T. and {Carballo}, R. and {Carlucci}, T. and {Carnerero}, M.~I. and {Carrasco}, J.~M. and {Casamiquela}, L. and {Castellani}, M. and {Castro-Ginard}, A. and {Chaoul}, L. and {Charlot}, P. and {Chemin}, L. and {Chiaramida}, V. and {Chiavassa}, A. and {Chornay}, N. and {Comoretto}, G. and {Contursi}, G. and {Cooper}, W.~J. and {Cornez}, T. and {Cowell}, S. and {Crifo}, F. and {Cropper}, M. and {Crosta}, M. and {Crowley}, C. and {Dafonte}, C. and {Dapergolas}, A. and {David}, M. and {David}, P. and {de Laverny}, P. and {De Luise}, F. and {De March}, R.},
        title = "{Gaia Data Release 3. Summary of the content and survey properties}",
      journal = {\aap},
         year = 2023,
        month = jun,
       volume = {674},
          eid = {A1},
        pages = {A1},
          doi = {10.1051/0004-6361/202243940},
archivePrefix = {arXiv},
       eprint = {2208.00211},
 primaryClass = {astro-ph.GA},
       adsurl = {https://ui.adsabs.harvard.edu/abs/2023A&A...674A...1G }
}

@ARTICLE{GAIADR2+2016,
       author = {{Gaia Collaboration} and {Prusti}, T. and {de Bruijne}, J.~H.~J. and {Brown}, A.~G.~A. and {Vallenari}, A. and {Babusiaux}, C. and {Bailer-Jones}, C.~A.~L. and {Bastian}, U. and {Biermann}, M. and {Evans}, D.~W. and {Eyer}, L. and {Jansen}, F. and {Jordi}, C. and {Klioner}, S.~A. and {Lammers}, U. and {Lindegren}, L. and {Luri}, X. and {Mignard}, F. and {Milligan}, D.~J. and {Panem}, C. and {Poinsignon}, V. and {Pourbaix}, D. and {Randich}, S. and {Sarri}, G. and {Sartoretti}, P. and {Siddiqui}, H.~I. and {Soubiran}, C. and {Valette}, V. and {van Leeuwen}, F. and {Walton}, N.~A. and {Aerts}, C. and {Arenou}, F. and {Cropper}, M. and {Drimmel}, R. and {H{\o}g}, E. and {Katz}, D. and {Lattanzi}, M.~G. and {O'Mullane}, W. and {Grebel}, E.~K. and {Holland}, A.~D. and {Huc}, C. and {Passot}, X. and {Bramante}, L. and {Cacciari}, C. and {Casta{\~n}eda}, J. and {Chaoul}, L. and {Cheek}, N. and {De Angeli}, F. and {Fabricius}, C. and {Guerra}, R. and {Hern{\'a}ndez}, J. and {Jean-Antoine-Piccolo}, A. and {Masana}, E. and {Messineo}, R. and {Mowlavi}, N. and {Nienartowicz}, K. and {Ord{\'o}{\~n}ez-Blanco}, D. and {Panuzzo}, P. and {Portell}, J. and {Richards}, P.~J. and {Riello}, M. and {Seabroke}, G.~M. and {Tanga}, P. and {Th{\'e}venin}, F. and {Torra}, J. and {Els}, S.~G. and {Gracia-Abril}, G. and {Comoretto}, G. and {Garcia-Reinaldos}, M. and {Lock}, T. and {Mercier}, E. and {Altmann}, M. and {Andrae}, R. and {Astraatmadja}, T.~L. and {Bellas-Velidis}, I. and {Benson}, K. and {Berthier}, J. and {Blomme}, R. and {Busso}, G. and {Carry}, B. and {Cellino}, A. and {Clementini}, G. and {Cowell}, S. and {Creevey}, O. and {Cuypers}, J. and {Davidson}, M. and {De Ridder}, J. and {de Torres}, A. and {Delchambre}, L. and {Dell'Oro}, A. and {Ducourant}, C. and {Fr{\'e}mat}, Y. and {Garc{\'\i}a-Torres}, M. and {Gosset}, E. and {Halbwachs}, J.-L. and {Hambly}, N.~C. and {Harrison}, D.~L. and {Hauser}, M. and {Hestroffer}, D. and {Hodgkin}, S.~T. and {Huckle}, H.~E. and {Hutton}, A. and {Jasniewicz}, G. and {Jordan}, S. and {Kontizas}, M. and {Korn}, A.~J. and {Lanzafame}, A.~C. and {Manteiga}, M. and {Moitinho}, A. and {Muinonen}, K. and {Osinde}, J. and {Pancino}, E. and {Pauwels}, T. and {Petit}, J.-M. and {Recio-Blanco}, A. and {Robin}, A.~C. and {Sarro}, L.~M. and {Siopis}, C. and {Smith}, M. and {Smith}, K.~W. and {Sozzetti}, A. and {Thuillot}, W. and {van Reeven}, W. and {Viala}, Y. and {Abbas}, U. and {Abreu Aramburu}, A. and {Accart}, S. and {Aguado}, J.~J. and {Allan}, P.~M. and {Allasia}, W. and {Altavilla}, G. and {{\'A}lvarez}, M.~A. and {Alves}, J. and {Anderson}, R.~I. and {Andrei}, A.~H. and {Anglada Varela}, E. and {Antiche}, E. and {Antoja}, T. and {Ant{\'o}n}, S. and {Arcay}, B. and {Atzei}, A. and {Ayache}, L. and {Bach}, N. and {Baker}, S.~G. and {Balaguer-N{\'u}{\~n}ez}, L. and {Barache}, C. and {Barata}, C. and {Barbier}, A. and {Barblan}, F. and {Baroni}, M. and {Barrado y Navascu{\'e}s}, D. and {Barros}, M. and {Barstow}, M.~A. and {Becciani}, U. and {Bellazzini}, M. and {Bellei}, G. and {Bello Garc{\'\i}a}, A. and {Belokurov}, V. and {Bendjoya}, P. and {Berihuete}, A. and {Bianchi}, L. and {Bienaym{\'e}}, O. and {Billebaud}, F. and {Blagorodnova}, N. and {Blanco-Cuaresma}, S. and {Boch}, T. and {Bombrun}, A. and {Borrachero}, R. and {Bouquillon}, S. and {Bourda}, G. and {Bouy}, H. and {Bragaglia}, A. and {Breddels}, M.~A. and {Brouillet}, N. and {Br{\"u}semeister}, T. and {Bucciarelli}, B. and {Budnik}, F. and {Burgess}, P. and {Burgon}, R. and {Burlacu}, A. and {Busonero}, D. and {Buzzi}, R. and {Caffau}, E. and {Cambras}, J. and {Campbell}, H. and {Cancelliere}, R. and {Cantat-Gaudin}, T. and {Carlucci}, T. and {Carrasco}, J.~M. and {Castellani}, M. and {Charlot}, P. and {Charnas}, J. and {Charvet}, P. and {Chassat}, F. and {Chiavassa}, A. and {Clotet}, M. and {Cocozza}, G. and {Collins}, R.~S. and {Collins}, P. and {Costigan}, G.},
        title = "{The Gaia mission}",
      journal = {\aap},
         year = 2016,
        month = nov,
       volume = {595},
          eid = {A1},
        pages = {A1},
          doi = {10.1051/0004-6361/201629272},
archivePrefix = {arXiv},
       eprint = {1609.04153},
 primaryClass = {astro-ph.IM},
       adsurl = {https://ui.adsabs.harvard.edu/abs/2016A&A...595A...1G }
}

@ARTICLE{Gozha+2015,
       author = {{Gozha}, M.~L. and {Marsakov}, V.~A.},
        title = "{Properties of open clusters in the regions of the Local Bubble and the Eridanus Superbubble}",
      journal = {Baltic Astronomy},
         year = 2015,
        month = jan,
       volume = {24},
        pages = {17-23},
          doi = {10.1515/astro-2017-0198},
       adsurl = {https://ui.adsabs.harvard.edu/abs/2015BaltA..24...17G }
}

@ARTICLE{Hao+2021,
       author = {{Hao}, C.~J. and {Xu}, Y. and {Hou}, L.~G. and {Bian}, S.~B. and {Li}, J.~J. and {Wu}, Z.~Y. and {He}, Z.~H. and {Li}, Y.~J. and {Liu}, D.~J.},
        title = "{Evolution of the local spiral structure of the Milky Way revealed by open clusters}",
      journal = {\aap},
         year = 2021,
        month = aug,
       volume = {652},
          eid = {A102},
        pages = {A102},
          doi = {10.1051/0004-6361/202140608},
archivePrefix = {arXiv},
       eprint = {2107.06478},
 primaryClass = {astro-ph.GA},
       adsurl = {https://ui.adsabs.harvard.edu/abs/2021A&A...652A.102H }
}

@ARTICLE{Hu+2025a,
       author = {{Hu}, Qingshun and {Qin}, Songmei and {Luo}, Yangping and {Li}, Yuting},
        title = "{3D projection analysis: Characterizing the morphological stability of nearby open clusters}",
      journal = {\aap},
         year = 2025,
        month = jan,
       volume = {693},
          eid = {A125},
        pages = {A125},
          doi = {10.1051/0004-6361/202451243},
archivePrefix = {arXiv},
       eprint = {2412.10158},
 primaryClass = {astro-ph.GA},
       adsurl = {https://ui.adsabs.harvard.edu/abs/2025A&A...693A.125H }
}

@ARTICLE{Hu+2021a,
       author = {{Hu}, Qingshun and {Zhang}, Yu and {Esamdin}, Ali and {Liu}, Jinzhong and {Zeng}, Xiangyun},
        title = "{Deciphering Star Cluster Evolution by Shape Morphology}",
      journal = {\apj},
         year = 2021,
        month = may,
       volume = {912},
       number = {1},
          eid = {5},
        pages = {5},
          doi = {10.3847/1538-4357/abec3e},
archivePrefix = {arXiv},
       eprint = {2103.02912},
 primaryClass = {astro-ph.GA},
       adsurl = {https://ui.adsabs.harvard.edu/abs/2021ApJ...912....5H }
}

@ARTICLE{Hu+2021b,
       author = {{Hu}, Qingshun and {Zhang}, Yu and {Esamdin}, Ali},
        title = "{Decoding the morphological evolution of open clusters}",
      journal = {\aap},
         year = 2021,
        month = dec,
       volume = {656},
          eid = {A49},
        pages = {A49},
          doi = {10.1051/0004-6361/202141460},
archivePrefix = {arXiv},
       eprint = {2109.04678},
 primaryClass = {astro-ph.GA},
       adsurl = {https://ui.adsabs.harvard.edu/abs/2021A&A...656A..49H }
}

@ARTICLE{Hu+2023,
       author = {{Hu}, Qingshun and {Zhang}, Yu and {Esamdin}, Ali and {Wang}, Hong and {Qin}, Mingfeng},
        title = "{A detection of the layered structure of nearby open clusters}",
      journal = {\aap},
         year = 2023,
        month = apr,
       volume = {672},
          eid = {A12},
        pages = {A12},
          doi = {10.1051/0004-6361/202244199},
archivePrefix = {arXiv},
       eprint = {2302.02750},
 primaryClass = {astro-ph.GA},
       adsurl = {https://ui.adsabs.harvard.edu/abs/2023A&A...672A..12H }
}

@ARTICLE{Hu+2024,
       author = {{Hu}, Qingshun and {Zhang}, Yu and {Qin}, Songmei and {Zhong}, Jing and {Chen}, Li and {Luo}, Yangping},
        title = "{Exploration of morphological coherence in open clusters with a ``core-shell'' structure}",
      journal = {\aap},
         year = 2024,
        month = jul,
       volume = {687},
          eid = {A291},
        pages = {A291},
          doi = {10.1051/0004-6361/202347625},
archivePrefix = {arXiv},
       eprint = {2405.05771},
 primaryClass = {astro-ph.GA},
       adsurl = {https://ui.adsabs.harvard.edu/abs/2024A&A...687A.291H }
}

@ARTICLE{Hu+2025b,
       author = {{Hu}, Qingshun and {Li}, Yuting and {Qin}, Mingfeng and {Lv}, Chenglong and {Pan}, Yang and {Luo}, Yangping and {Ma}, Shuo},
        title = "{An In-depth Investigation of the Primordial Cluster Pair ASCC 19 and ASCC 21}",
      journal = {\aj},
         year = 2025,
        month = feb,
       volume = {169},
       number = {2},
          eid = {98},
        pages = {98},
          doi = {10.3847/1538-3881/ada443},
archivePrefix = {arXiv},
       eprint = {2408.03552},
 primaryClass = {astro-ph.SR},
       adsurl = {https://ui.adsabs.harvard.edu/abs/2025AJ....169...98H }
}

@ARTICLE{HU+2025c,
       author = {{Hu}, Qingshun and {Soubiran}, Caroline},
        title = "{Metallicities of old open clusters: A new Galactic map}",
      journal = {\aap},
         year = 2025,
        month = jul,
       volume = {699},
          eid = {A246},
        pages = {A246},
          doi = {10.1051/0004-6361/202554752},
archivePrefix = {arXiv},
       eprint = {2505.18378},
 primaryClass = {astro-ph.GA},
       adsurl = {https://ui.adsabs.harvard.edu/abs/2025A&A...699A.246H }
}

@ARTICLE{Hunt+2024,
       author = {{Hunt}, Emily L. and {Reffert}, Sabine},
        title = "{Improving the open cluster census. III. Using cluster masses, radii, and dynamics to create a cleaned open cluster catalogue}",
      journal = {\aap},
         year = 2024,
        month = jun,
       volume = {686},
          eid = {A42},
        pages = {A42},
          doi = {10.1051/0004-6361/202348662},
archivePrefix = {arXiv},
       eprint = {2403.05143},
 primaryClass = {astro-ph.GA},
       adsurl = {https://ui.adsabs.harvard.edu/abs/2024A&A...686A..42H }
}

@article{hogg+2018,
  author  = {Hogg, D.~W. and Foreman-Mackey, D.},
  title   = {Data Analysis Recipes: Using Markov Chain Monte Carlo},
  journal = {ApJS},
  year    = {2018},
  volume  = {236},
  pages   = {11},
  doi     = {10.3847/1538-4365/aaa6c3}
}

@ARTICLE{Helmi+2018,
       author = {{Helmi}, Amina and {Babusiaux}, Carine and {Koppelman}, Helmer H. and {Massari}, Davide and {Veljanoski}, Jovan and {Brown}, Anthony G.~A.},
        title = "{The merger that led to the formation of the Milky Way's inner stellar halo and thick disk}",
      journal = {\nat},
         year = 2018,
        month = oct,
       volume = {563},
       number = {7729},
        pages = {85-88},
          doi = {10.1038/s41586-018-0625-x},
archivePrefix = {arXiv},
       eprint = {1806.06038},
 primaryClass = {astro-ph.GA},
       adsurl = {https://ui.adsabs.harvard.edu/abs/2018Natur.563...85H}
}

@ARTICLE{Joharle+2026,
       author = {{Joharle}, Simran and {Nogueras-Lara}, Francisco and {Fiteni}, Karl},
        title = "{Kinematic and extinction analysis of a potential spiral arm beyond the Galactic bar}",
      journal = {\aap},
         year = 2026,
        month = jan,
       volume = {705},
          eid = {A131},
        pages = {A131},
          doi = {10.1051/0004-6361/202554294},
archivePrefix = {arXiv},
       eprint = {2511.04778},
 primaryClass = {astro-ph.GA},
       adsurl = {https://ui.adsabs.harvard.edu/abs/2026A&A...705A.131J}
}

@ARTICLE{Kalita+2026,
       author = {{Kalita}, Boris S. and {Ho}, Luis C. and {Silverman}, John D. and {Bournaud}, Fr{\'e}d{\'e}ric and {Dessauges-Zavadsky}, Miroslava and {Daddi}, Emanuele and {Puglisi}, Annagrazia and {Ding}, Xuheng and {Yu}, Si-Yue},
        title = "{Galactic Bars Are Already Mature at Cosmic Noon: Bar Strength and Flatness at z {\ensuremath{\sim}} 1.5}",
      journal = {\apj},
         year = 2026,
        month = feb,
       volume = {997},
       number = {2},
          eid = {247},
        pages = {247},
          doi = {10.3847/1538-4357/ae2755},
archivePrefix = {arXiv},
       eprint = {2512.04163},
 primaryClass = {astro-ph.GA},
       adsurl = {https://ui.adsabs.harvard.edu/abs/2026ApJ...997..247K}
}

@ARTICLE{Kadam+2024,
       author = {{Kadam}, S.~K. and {Salunkhe}, Sameer and {Vagshette}, N.~D. and {Paul}, Surajit and {Sonkamble}, S.~S. and {Pawar}, P.~K. and {Patil}, M.~K.},
        title = "{Sloshing and spiral structures breeding a putative radio mini-halo in the environment of a cool-core cluster, Abell 795}",
      journal = {\mnras},
         year = 2024,
        month = jul,
       volume = {531},
       number = {4},
        pages = {4060-4069},
          doi = {10.1093/mnras/stae1401},
archivePrefix = {arXiv},
       eprint = {2405.19750},
 primaryClass = {astro-ph.GA},
       adsurl = {https://ui.adsabs.harvard.edu/abs/2024MNRAS.531.4060K }
}

@ARTICLE{Kawano+1995,
       author = {{Kawano}, H. and {Higuchi}, T.},
        title = "{The bootstrap method in space physics: Error estimation for the minimum variance analysis}",
      journal = {\grl},
         year = 1995,
        month = feb,
       volume = {22},
       number = {3},
        pages = {307-310},
          doi = {10.1029/94GL02969},
       adsurl = {https://ui.adsabs.harvard.edu/abs/1995GeoRL..22..307K }
}

@ARTICLE{King+1962,
       author = {{King}, Ivan},
        title = "{The structure of star clusters. I. an empirical density law}",
      journal = {\aj},
         year = 1962,
        month = oct,
       volume = {67},
        pages = {471},
          doi = {10.1086/108756},
       adsurl = {https://ui.adsabs.harvard.edu/abs/1962AJ.....67..471K }
}

@ARTICLE{Krumholz+2019,
       author = {{Krumholz}, Mark R. and {McKee}, Christopher F. and {Bland-Hawthorn}, Joss},
        title = "{Star Clusters Across Cosmic Time}",
      journal = {\araa},
         year = 2019,
        month = aug,
       volume = {57},
        pages = {227-303},
          doi = {10.1146/annurev-astro-091918-104430},
archivePrefix = {arXiv},
       eprint = {1812.01615},
 primaryClass = {astro-ph.GA},
       adsurl = {https://ui.adsabs.harvard.edu/abs/2019ARA&A..57..227K}
}

@ARTICLE{Lada+2003,
       author = {{Lada}, Charles J. and {Lada}, Elizabeth A.},
        title = "{Embedded Clusters in Molecular Clouds}",
      journal = {\araa},
         year = 2003,
        month = jan,
       volume = {41},
        pages = {57-115},
          doi = {10.1146/annurev.astro.41.011802.094844},
archivePrefix = {arXiv},
       eprint = {astro-ph/0301540},
 primaryClass = {astro-ph},
       adsurl = {https://ui.adsabs.harvard.edu/abs/2003ARA&A..41...57L }
}

@ARTICLE{Lang+2025,
       author = {{Lang}, Kaixiang and {Zhang}, Yu and {Niu}, Hubiao and {Maurya}, Jayanand and {Liu}, Jinzhong and {Liu}, Guimei},
        title = "{Insights into the 3D layered structure of nearby open clusters through N-body simulations}",
      journal = {\aap},
         year = 2025,
        month = may,
       volume = {697},
          eid = {A122},
        pages = {A122},
          doi = {10.1051/0004-6361/202554066},
archivePrefix = {arXiv},
       eprint = {2505.08184},
 primaryClass = {astro-ph.GA},
       adsurl = {https://ui.adsabs.harvard.edu/abs/2025A&A...697A.122L }
}

@ARTICLE{Liu+2025,
       author = {{Liu}, Xiaochen and {He}, Zhihong and {Luo}, Yangping and {Wang}, Kun},
        title = "{The disrupting and growing open cluster spiral arm patterns of the Milky Way}",
      journal = {\mnras},
         year = 2025,
        month = mar,
       volume = {537},
       number = {3},
        pages = {2403-2411},
          doi = {10.1093/mnras/staf153},
archivePrefix = {arXiv},
       eprint = {2501.14215},
 primaryClass = {astro-ph.GA},
       adsurl = {https://ui.adsabs.harvard.edu/abs/2025MNRAS.537.2403L }
}

@ARTICLE{Mamon+2019,
       author = {{Mamon}, G.~A. and {Cava}, A. and {Biviano}, A. and {Moretti}, A. and {Poggianti}, B. and {Bettoni}, D.},
        title = "{Structural and dynamical modeling of WINGS clusters. II. The orbital anisotropies of elliptical, spiral, and lenticular galaxies}",
      journal = {\aap},
         year = 2019,
        month = nov,
       volume = {631},
          eid = {A131},
        pages = {A131},
          doi = {10.1051/0004-6361/201935081},
archivePrefix = {arXiv},
       eprint = {1901.06393},
 primaryClass = {astro-ph.GA},
       adsurl = {https://ui.adsabs.harvard.edu/abs/2019A&A...631A.131M }
}

@ARTICLE{Luri+2018,
       author = {{Luri}, X. and {Brown}, A.~G.~A. and {Sarro}, L.~M. and {Arenou}, F. and {Bailer-Jones}, C.~A.~L. and {Castro-Ginard}, A. and {de Bruijne}, J. and {Prusti}, T. and {Babusiaux}, C. and {Delgado}, H.~E.},
        title = "{Gaia Data Release 2. Using Gaia parallaxes}",
      journal = {\aap},
         year = 2018,
        month = aug,
       volume = {616},
          eid = {A9},
        pages = {A9},
          doi = {10.1051/0004-6361/201832964},
archivePrefix = {arXiv},
       eprint = {1804.09376},
 primaryClass = {astro-ph.IM},
       adsurl = {https://ui.adsabs.harvard.edu/abs/2018A&A...616A...9L}
}

@ARTICLE{Moreira+2025,
       author = {{Moreira}, Sandro and {Moitinho}, Andr{\'e} and {Silva}, Andr{\'e} and {Almeida}, Duarte},
        title = "{Modelling the evolution of the Galactic disc scale height traced by open clusters}",
      journal = {\aap},
         year = 2025,
        month = feb,
       volume = {694},
          eid = {A70},
        pages = {A70},
          doi = {10.1051/0004-6361/202450369},
archivePrefix = {arXiv},
       eprint = {2406.14661},
 primaryClass = {astro-ph.GA},
       adsurl = {https://ui.adsabs.harvard.edu/abs/2025A&A...694A..70M}
}

@ARTICLE{Megeath+2016,
       author = {{Megeath}, S.~T. and {Gutermuth}, R. and {Muzerolle}, J. and {Kryukova}, E. and {Hora}, J.~L. and {Allen}, L.~E. and {Flaherty}, K. and {Hartmann}, L. and {Myers}, P.~C. and {Pipher}, J.~L. and {Stauffer}, J. and {Young}, E.~T. and {Fazio}, G.~G.},
        title = "{The Spitzer Space Telescope Survey of the Orion A and B Molecular Clouds. II. The Spatial Distribution and Demographics of Dusty Young Stellar Objects}",
      journal = {\aj},
         year = 2016,
        month = jan,
       volume = {151},
       number = {1},
          eid = {5},
        pages = {5},
          doi = {10.3847/0004-6256/151/1/5},
archivePrefix = {arXiv},
       eprint = {1511.01202},
 primaryClass = {astro-ph.GA},
       adsurl = {https://ui.adsabs.harvard.edu/abs/2016AJ....151....5M }
}

@ARTICLE{Meingast+2019,
       author = {{Meingast}, Stefan and {Alves}, Jo{\~a}o},
        title = "{Extended stellar systems in the solar neighborhood. I. The tidal tails of the Hyades}",
      journal = {\aap},
         year = 2019,
        month = jan,
       volume = {621},
          eid = {L3},
        pages = {L3},
          doi = {10.1051/0004-6361/201834622},
archivePrefix = {arXiv},
       eprint = {1811.04931},
 primaryClass = {astro-ph.GA},
       adsurl = {https://ui.adsabs.harvard.edu/abs/2019A&A...621L...3M }
}

@ARTICLE{Meynet+1993,
       author = {{Meynet}, G. and {Mermilliod}, J. -C. and {Maeder}, A.},
        title = "{New dating of galactic open clusters.}",
      journal = {\aaps},
         year = 1993,
        month = may,
       volume = {98},
        pages = {477-504},
       adsurl = {https://ui.adsabs.harvard.edu/abs/1993A&AS...98..477M }
}

@ARTICLE{Pang+2021,
       author = {{Pang}, Xiaoying and {Li}, Yuqian and {Yu}, Zeqiu and {Tang}, Shih-Yun and {Dinnbier}, Franti{\v{s}}ek and {Kroupa}, Pavel and {Pasquato}, Mario and {Kouwenhoven}, M.~B.~N.},
        title = "{3D Morphology of Open Clusters in the Solar Neighborhood with Gaia EDR 3: Its Relation to Cluster Dynamics}",
      journal = {\apj},
         year = 2021,
        month = may,
       volume = {912},
       number = {2},
          eid = {162},
        pages = {162},
          doi = {10.3847/1538-4357/abeaac},
archivePrefix = {arXiv},
       eprint = {2102.10508},
 primaryClass = {astro-ph.GA},
       adsurl = {https://ui.adsabs.harvard.edu/abs/2021ApJ...912..162P}
}

@ARTICLE{Pang+2022,
       author = {{Pang}, Xiaoying and {Tang}, Shih-Yun and {Li}, Yuqian and {Yu}, Zeqiu and {Wang}, Long and {Li}, Jiayu and {Li}, Yezhang and {Wang}, Yifan and {Wang}, Yanshu and {Zhang}, Teng and {Pasquato}, Mario and {Kouwenhoven}, M.~B.~N.},
        title = "{3D Morphology of Open Clusters in the Solar Neighborhood with Gaia EDR 3. II. Hierarchical Star Formation Revealed by Spatial and Kinematic Substructures}",
      journal = {\apj},
         year = 2022,
        month = jun,
       volume = {931},
       number = {2},
          eid = {156},
        pages = {156},
          doi = {10.3847/1538-4357/ac674e},
archivePrefix = {arXiv},
       eprint = {2204.06000},
 primaryClass = {astro-ph.GA},
       adsurl = {https://ui.adsabs.harvard.edu/abs/2022ApJ...931..156P}
}

@ARTICLE{Michael+2018,
       author = {{Grudi{\'c}}, Michael Y. and {Guszejnov}, D{\'a}vid and {Hopkins}, Philip F. and {Lamberts}, Astrid and {Boylan-Kolchin}, Michael and {Murray}, Norman and {Schmitz}, Denise},
        title = "{From the top down and back up again: star cluster structure from hierarchical star formation}",
      journal = {\mnras},
         year = 2018,
        month = nov,
       volume = {481},
       number = {1},
        pages = {688-702},
          doi = {10.1093/mnras/sty2303},
archivePrefix = {arXiv},
       eprint = {1708.09065},
 primaryClass = {astro-ph.GA},
       adsurl = {https://ui.adsabs.harvard.edu/abs/2018MNRAS.481..688G}
}

@ARTICLE{Portegies_Zwart+2010,
       author = {{Portegies Zwart}, Simon F. and {McMillan}, Stephen L.~W. and {Gieles}, Mark},
        title = "{Young Massive Star Clusters}",
      journal = {\araa},
         year = 2010,
        month = sep,
       volume = {48},
        pages = {431-493},
          doi = {10.1146/annurev-astro-081309-130834},
archivePrefix = {arXiv},
       eprint = {1002.1961},
 primaryClass = {astro-ph.GA},
       adsurl = {https://ui.adsabs.harvard.edu/abs/2010ARA&A..48..431P}
}

@ARTICLE{Portegies+2001,
       author = {{Portegies Zwart}, Simon F. and {McMillan}, Stephen L.~W. and {Hut}, Piet and {Makino}, Junichiro},
        title = "{Star cluster ecology - IV. Dissection of an open star cluster: photometry}",
      journal = {\mnras},
         year = 2001,
        month = feb,
       volume = {321},
       number = {2},
        pages = {199-226},
          doi = {10.1046/j.1365-8711.2001.03976.x},
archivePrefix = {arXiv},
       eprint = {astro-ph/0005248},
 primaryClass = {astro-ph},
       adsurl = {https://ui.adsabs.harvard.edu/abs/2001MNRAS.321..199P}
}

@ARTICLE{Piskunov+2006,
       author = {{Piskunov}, A.~E. and {Kharchenko}, N.~V. and {R{\"o}ser}, S. and {Schilbach}, E. and {Scholz}, R. -D.},
        title = "{Revisiting the population of Galactic open clusters}",
      journal = {\aap},
         year = 2006,
        month = jan,
       volume = {445},
       number = {2},
        pages = {545-565},
          doi = {10.1051/0004-6361:20053764},
archivePrefix = {arXiv},
       eprint = {astro-ph/0508575},
 primaryClass = {astro-ph},
       adsurl = {https://ui.adsabs.harvard.edu/abs/2006A&A...445..545P }
}

@ARTICLE{Reid+2019,
       author = {{Reid}, M.~J. and {Menten}, K.~M. and {Brunthaler}, A. and {Zheng}, X.~W. and {Dame}, T.~M. and {Xu}, Y. and {Li}, J. and {Sakai}, N. and {Wu}, Y. and {Immer}, K. and {Zhang}, B. and {Sanna}, A. and {Moscadelli}, L. and {Rygl}, K.~L.~J. and {Bartkiewicz}, A. and {Hu}, B. and {Quiroga-Nu{\~n}ez}, L.~H. and {van Langevelde}, H.~J.},
        title = "{Trigonometric Parallaxes of High-mass Star-forming Regions: Our View of the Milky Way}",
      journal = {\apj},
         year = 2019,
        month = nov,
       volume = {885},
       number = {2},
          eid = {131},
        pages = {131},
          doi = {10.3847/1538-4357/ab4a11},
archivePrefix = {arXiv},
       eprint = {1910.03357},
 primaryClass = {astro-ph.GA},
       adsurl = {https://ui.adsabs.harvard.edu/abs/2019ApJ...885..131R }
}

@ARTICLE{Rangwal+2025,
       author = {{Rangwal}, Geeta and {Arya}, Aman and {Subramaniam}, Annapurni and {Singh}, Kulinder PAL and {Liu}, Xiaowei},
        title = "{Orbits and vertical height distribution of 4006 open clusters in the Galactic disk using Gaia DR3}",
      journal = {Journal of Astrophysics and Astronomy},
         year = 2025,
        month = aug,
       volume = {46},
       number = {2},
          eid = {52},
        pages = {52},
          doi = {10.1007/s12036-025-10061-z},
archivePrefix = {arXiv},
       eprint = {2410.15305},
 primaryClass = {astro-ph.GA},
       adsurl = {https://ui.adsabs.harvard.edu/abs/2025JApA...46...52R}
}

@ARTICLE{Reshetnikov+2025,
       author = {{Reshetnikov}, Vladimir P. and {Chugunov}, Ilia V. and {Marchuk}, Alexander A. and {Mosenkov}, Aleksandr V. and {Kozlov}, Matvey D. and {Savchenko}, Sergey S. and {Makarov}, Dmitry I. and {Antipova}, Aleksandra V. and {Sypkova}, Anastasia M.},
        title = "{Galactic warps: From cosmic noon to the current epoch}",
      journal = {\aap},
         year = 2025,
        month = may,
       volume = {697},
          eid = {L1},
        pages = {L1},
          doi = {10.1051/0004-6361/202554941},
archivePrefix = {arXiv},
       eprint = {2504.12403},
 primaryClass = {astro-ph.GA},
       adsurl = {https://ui.adsabs.harvard.edu/abs/2025A&A...697L...1R}
}

@ARTICLE{Sharma+2025,
       author = {{Sharma}, Ira and {Jadhav}, Vikrant V. and {Subramaniam}, Annapurni},
        title = "{Tidal Tails and Their Dynamics in Open Clusters Using Gaia DR3}",
      journal = {arXiv e-prints},
         year = 2025,
        month = sep,
          eid = {arXiv:2509.09279},
        pages = {arXiv:2509.09279},
          doi = {10.48550/arXiv.2509.09279},
archivePrefix = {arXiv},
       eprint = {2509.09279},
 primaryClass = {astro-ph.GA},
       adsurl = {https://ui.adsabs.harvard.edu/abs/2025arXiv250909279S }
}

@ARTICLE{Seleznev+2016,
       author = {{Seleznev}, Anton F.},
        title = "{Open-cluster density profiles derived using a kernel estimator}",
      journal = {\mnras},
         year = 2016,
        month = mar,
       volume = {456},
       number = {4},
        pages = {3757-3773},
          doi = {10.1093/mnras/stv2874},
archivePrefix = {arXiv},
       eprint = {1601.03898},
 primaryClass = {astro-ph.GA},
       adsurl = {https://ui.adsabs.harvard.edu/abs/2016MNRAS.456.3757S}
}

@ARTICLE{Smith+1996,
       author = {{Smith}, Jr., Haywood and {Eichhorn}, Heinrich},
        title = "{On the estimation of distances from trigonometric parallaxes}",
      journal = {\mnras},
         year = 1996,
        month = jul,
       volume = {281},
        pages = {211-218},
          doi = {10.1093/mnras/281.1.211},
       adsurl = {https://ui.adsabs.harvard.edu/abs/1996MNRAS.281..211S}
}

@INPROCEEDINGS{Taylor+2005,
       author = {{Taylor}, M.~B.},
        title = "{TOPCAT \& STIL: Starlink Table/VOTable Processing Software}",
    booktitle = {Astronomical Data Analysis Software and Systems XIV},
         year = 2005,
       editor = {{Shopbell}, P. and {Britton}, M. and {Ebert}, R.},
       series = {Astronomical Society of the Pacific Conference Series},
       volume = {347},
        month = dec,
        pages = {29},
       adsurl = {https://ui.adsabs.harvard.edu/abs/2005ASPC..347...29T }
}

@ARTICLE{Tang+2019,
       author = {{Tang}, Shih-Yun and {Pang}, Xiaoying and {Yuan}, Zhen and {Chen}, W.~P. and {Hong}, Jongsuk and {Goldman}, Bertrand and {Just}, Andreas and {Shukirgaliyev}, Bekdaulet and {Lin}, Chien-Cheng},
        title = "{Discovery of Tidal Tails in Disrupting Open Clusters: Coma Berenices and a Neighbor Stellar Group}",
      journal = {\apj},
         year = 2019,
        month = may,
       volume = {877},
       number = {1},
          eid = {12},
        pages = {12},
          doi = {10.3847/1538-4357/ab13b0},
archivePrefix = {arXiv},
       eprint = {1902.01404},
 primaryClass = {astro-ph.GA},
       adsurl = {https://ui.adsabs.harvard.edu/abs/2019ApJ...877...12T }
}

@ARTICLE{Tarricq+2022,
       author = {{Tarricq}, Y. and {Soubiran}, C. and {Casamiquela}, L. and {Castro-Ginard}, A. and {Olivares}, J. and {Miret-Roig}, N. and {Galli}, P.~A.~B.},
        title = "{Structural parameters of 389 local open clusters}",
      journal = {\aap},
         year = 2022,
        month = mar,
       volume = {659},
          eid = {A59},
        pages = {A59},
          doi = {10.1051/0004-6361/202142186},
archivePrefix = {arXiv},
       eprint = {2111.05291},
 primaryClass = {astro-ph.GA},
       adsurl = {https://ui.adsabs.harvard.edu/abs/2022A&A...659A..59T}
}

@ARTICLE{Tarricq+2021,
       author = {{Tarricq}, Y. and {Soubiran}, C. and {Casamiquela}, L. and {Cantat-Gaudin}, T. and {Chemin}, L. and {Anders}, F. and {Antoja}, T. and {Romero-G{\'o}mez}, M. and {Figueras}, F. and {Jordi}, C. and {Bragaglia}, A. and {Balaguer-N{\'u}{\~n}ez}, L. and {Carrera}, R. and {Castro-Ginard}, A. and {Moitinho}, A. and {Ramos}, P. and {Bossini}, D.},
        title = "{3D kinematics and age distribution of the open cluster population}",
      journal = {\aap},
         year = 2021,
        month = mar,
       volume = {647},
          eid = {A19},
        pages = {A19},
          doi = {10.1051/0004-6361/202039388},
archivePrefix = {arXiv},
       eprint = {2012.04017},
 primaryClass = {astro-ph.GA},
       adsurl = {https://ui.adsabs.harvard.edu/abs/2021A&A...647A..19T}
}

@ARTICLE{Thulasidharan+2025,
       author = {{Thulasidharan}, Lekshmi and {D'Onghia}, Elena and {Benjamin}, Robert},
        title = "{Azimuthal Misalignments in Stellar Warp Structure as Dynamical Tracers of Mergers in Milky Way─like Galaxies}",
      journal = {\apjl},
         year = 2025,
        month = nov,
       volume = {993},
       number = {1},
          eid = {L28},
        pages = {L28},
          doi = {10.3847/2041-8213/ae11b4},
archivePrefix = {arXiv},
       eprint = {2510.04194},
 primaryClass = {astro-ph.GA},
       adsurl = {https://ui.adsabs.harvard.edu/abs/2025ApJ...993L..28T}
}

@ARTICLE{Viktor+2024,
       author = {{J{\'o}nsson}, Viktor Hrannar and {McMillan}, Paul J.},
        title = "{The tangled warp of the Milky Way}",
      journal = {\aap},
         year = 2024,
        month = aug,
       volume = {688},
          eid = {A38},
        pages = {A38},
          doi = {10.1051/0004-6361/202449744},
archivePrefix = {arXiv},
       eprint = {2405.09624},
 primaryClass = {astro-ph.GA},
       adsurl = {https://ui.adsabs.harvard.edu/abs/2024A&A...688A..38J}
}

@ARTICLE{van+2023,
       author = {{van Groeningen}, M.~G.~J. and {Castro-Ginard}, A. and {Brown}, A.~G.~A. and {Casamiquela}, L. and {Jordi}, C.},
        title = "{A machine-learning-based tool for open cluster membership determination in Gaia DR3}",
      journal = {\aap},
         year = 2023,
        month = jul,
       volume = {675},
          eid = {A68},
        pages = {A68},
          doi = {10.1051/0004-6361/202345952},
archivePrefix = {arXiv},
       eprint = {2303.08474},
 primaryClass = {astro-ph.GA},
       adsurl = {https://ui.adsabs.harvard.edu/abs/2023A&A...675A..68V }
}

@ARTICLE{Wu+2024,
       author = {{Wu}, You and {Chen}, Jing and {Zhang}, Su and {Wei}, Xingyin and {He}, Feilong and {Zhao}, Yunbo and {He}, Xuran},
        title = "{Investigating 56 High Galactic Latitude Open Cluster Candidates in Gaia DR3}",
      journal = {\apj},
         year = 2024,
        month = apr,
       volume = {965},
       number = {2},
          eid = {131},
        pages = {131},
          doi = {10.3847/1538-4357/ad2fbf},
       adsurl = {https://ui.adsabs.harvard.edu/abs/2024ApJ...965..131W}
}

@ARTICLE{Xu+2025,
       author = {{Xu}, Ming and {Fu}, Xiaoting and {Chen}, Yang and {Li}, Lu and {Fang}, Min and {Zhao}, He and {Liu}, Penghui and {Zuo}, Yichang},
        title = "{Nearby open clusters with tidal features: Golden sample selection and 3D structure}",
      journal = {\aap},
         year = 2025,
        month = jun,
       volume = {698},
          eid = {A156},
        pages = {A156},
          doi = {10.1051/0004-6361/202554212},
archivePrefix = {arXiv},
       eprint = {2504.17744},
 primaryClass = {astro-ph.GA},
       adsurl = {https://ui.adsabs.harvard.edu/abs/2025A&A...698A.156X }
}

@ARTICLE{Zhong+2022,
       author = {{Zhong}, Jing and {Chen}, Li and {Jiang}, Yueyue and {Qin}, Songmei and {Hou}, Jinliang},
        title = "{New Insights into the Structure of Open Clusters in the Gaia Era}",
      journal = {\aj},
         year = 2022,
        month = aug,
       volume = {164},
       number = {2},
          eid = {54},
        pages = {54},
          doi = {10.3847/1538-3881/ac77fa},
archivePrefix = {arXiv},
       eprint = {2206.04904},
 primaryClass = {astro-ph.GA},
       adsurl = {https://ui.adsabs.harvard.edu/abs/2022AJ....164...54Z }
}

@ARTICLE{zquez+2017,
       author = {{V{\'a}zquez-Semadeni}, Enrique and {Gonz{\'a}lez-Samaniego}, Alejandro and {Col{\'\i}n}, Pedro},
        title = "{Hierarchical star cluster assembly in globally collapsing molecular clouds}",
      journal = {\mnras},
         year = 2017,
        month = may,
       volume = {467},
       number = {2},
        pages = {1313-1328},
          doi = {10.1093/mnras/stw3229},
archivePrefix = {arXiv},
       eprint = {1611.00088},
 primaryClass = {astro-ph.GA},
       adsurl = {https://ui.adsabs.harvard.edu/abs/2017MNRAS.467.1313V}
}
\newpage
\appendix

\newpage
\section{Bayesian linear regression with MCMC}
\label{sec:mcmmc}

We implemented a uniform Bayesian linear-regression framework to quantify empirical relations between any two observables $(\mathrm{X},\,\mathrm{Y})$ that are both subject to measurement errors and to an unknown intrinsic scatter. The method is written in \texttt{Python} and based on the affine-invariant ensemble sampler \texttt{emcee} \citep{Foreman-Mackey+2013}, also employed by \citet{Anders+2017} and \citet{HU+2025c}.

\subsection{Model parameterization}
\label{subsec:model}
The linear model is

\begin{equation}
\mathrm{Y} = m\,\mathrm{X} + b + \varepsilon,
\qquad
\varepsilon\sim\mathcal{N}(0,\,\sigma^{2}),
\end{equation} where $m$ is the slope, $b$ the intercept, and $\sigma$ the intrinsic scatter around the linear relation.

\subsection{Likelihood with heteroscedastic errors}
\label{subsec:like}

Each datum carries Gaussian uncertainties $(\sigma_{\mathrm{X},i},\,\sigma_{\mathrm{Y},i})$.
Following \citet{hogg+2018}, the total variance for point $i$ is

\begin{equation}
\sigma^{2}_{\mathrm{tot},i} =
\sigma_{\mathrm{Y},i}^{2} + (m\,\sigma_{\mathrm{X},i})^{2} + \sigma^{2}.
\end{equation} The log-likelihood is

\begin{equation}
\ln\mathcal{L}(m,b,\sigma) =
-\frac{1}{2}\sum_{i=1}^{N}
\left[
\frac{(\mathrm{Y}_{i}-m\mathrm{X}_{i}-b)^{2}}{\sigma^{2}_{\mathrm{tot},i}}
+
\ln\left(2\pi\sigma^{2}_{\mathrm{tot},i}\right)
\right].
\end{equation}

\vspace{0em} 
\subsection{Prior distributions}
\label{subsec:prior}

Uninformative uniform priors were adopted:
\begin{align}
m   &\sim \mathcal{U}(m_{\min},\,m_{\max}),\\
b   &\sim \mathcal{U}(b_{\min},\,b_{\max}),\\
\sigma &\sim \mathcal{U}(0,\,\sigma_{\max}).
\end{align}
These ranges are chosen physically wide enough to contain any realistic solution, and the values used in this work are listed in Table~\ref{tab:priors}.

\begin{table}[h]
\centering
\caption{Uniform prior boundaries for the regression parameters.}
\label{tab:priors}
\begin{tabular}{lcc}
\toprule
Parameter & Lower & Upper \\
\midrule
Slope $m$ & $-1$ & $2$ \\
Intercept $b$ & $-10$ & $10$ \\
Intrinsic scatter $\sigma$ & $0$ & $5$ \\
\bottomrule
\end{tabular}
\end{table}

\subsection{Posterior distribution}
According to Bayes' theorem, our goal is to maximize the posterior distribution. By combining the likelihood function and the prior distribution, we obtained the posterior distribution of the parameters \( m \) and \( b \). The logarithmic posterior distribution is

\begin{equation}
L_p = -\frac{1}{2} \sum_{i=1}^{N} \left[ \ln \sigma_i^2 + \frac{(y_i - \hat{y}_i)^2}{\sigma_i^2} - 2 \ln w_i \right],
\end{equation}
   \[
\begin{array}{lp{0.8\linewidth}}
- \sigma_i       & \text{the error of the } i\text{-th data point,} \\
- w_i            & \text{the weight of each data point,} \\
- \hat{y}_i      & \text{the predicted value based on the regression model.}
\end{array}
\]
   
\subsection{MCMC implementation}
\label{subsec:mcmc}

We sampled the posterior with \texttt{emcee} using 24 walkers evolved for 10\,000 steps, with the first 500 discarded as burn-in. Convergence was checked by eye and via the Gelman--Rubin statistic ($\hat{R}<1.05$). The posterior chains yield the marginalized 16th, 50th, and 84th percentiles, which define the best-fit value and its $1\sigma$ uncertainty. Visualization is performed with the \texttt{corner} package \citep{Foreman-Mackey+2016}.

\subsection{General applicability}
\label{subsec:general}
The same pipeline is applied to any $(\mathrm{X},\,\mathrm{Y})$ pair of interest. Here, we fitted three linear relationships ($\log_{10}(\mathrm{N}_{\mathrm{core}}/\mathrm{N}_{\mathrm{outer}})$ vs. $\log_{10}(\mathrm{N})$, $\log_{10}(\mathrm{S}_{\mathrm{core}}/\mathrm{S}_{\mathrm{outer}})$ vs. $\log_{10}(\mathrm{N})$, and $\log_{10}(\mathrm{N}_{\mathrm{core}}/\mathrm{N}_{\mathrm{outer}})$ vs. $\log_{10}(\mathrm{S}_{\mathrm{core}}/\mathrm{S}_{\mathrm{outer}})$) with this method without the modification of the likelihood or priors.

\begin{figure*}[htbp]

\section{Corner panels and Bayesian fit panels for each group.}
\label{appendix:corner}

    \centering
    \includegraphics[angle=0,width=45mm]{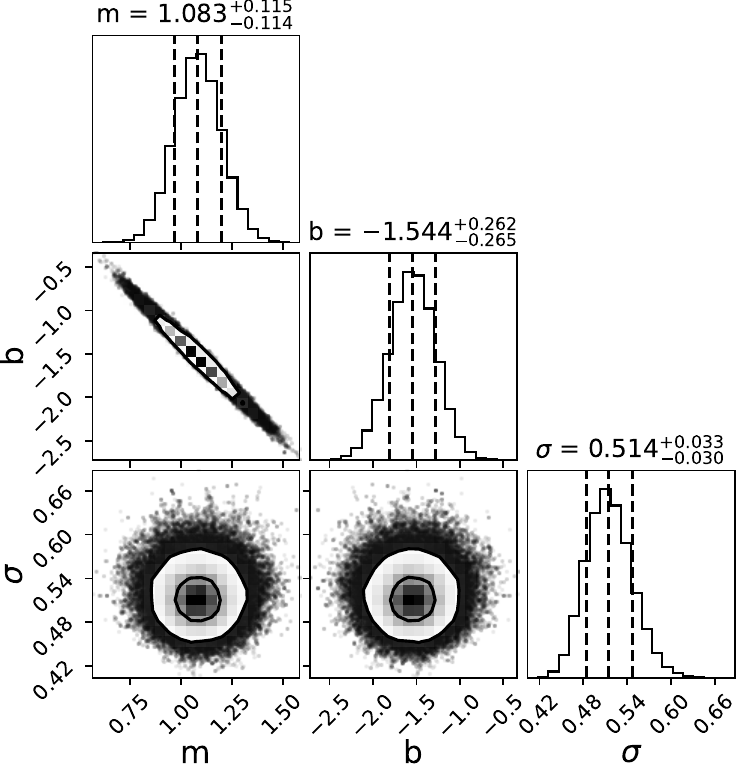}
    \includegraphics[angle=0,width=45mm]{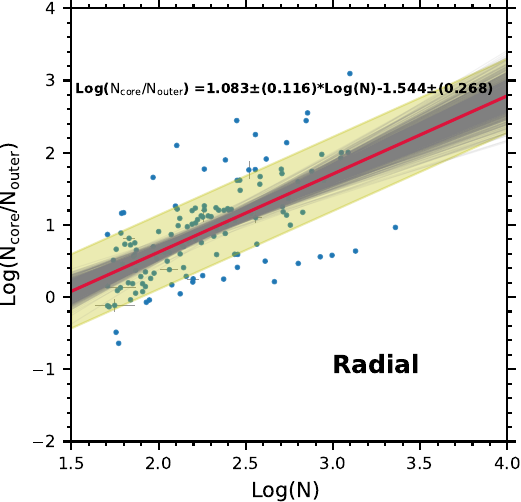}
    \includegraphics[angle=0,width=45mm]{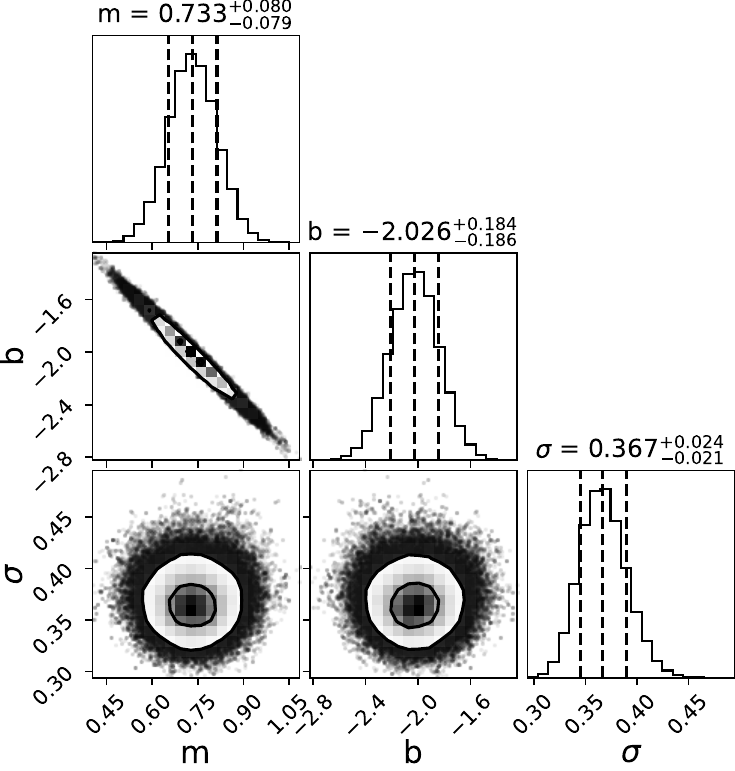}
    \includegraphics[angle=0,width=45mm]{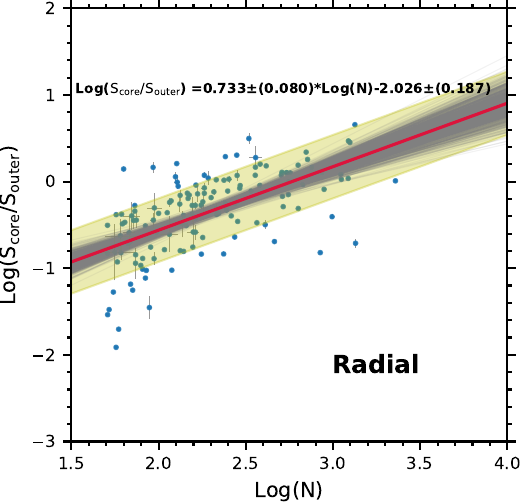}

    \includegraphics[angle=0,width=45mm]{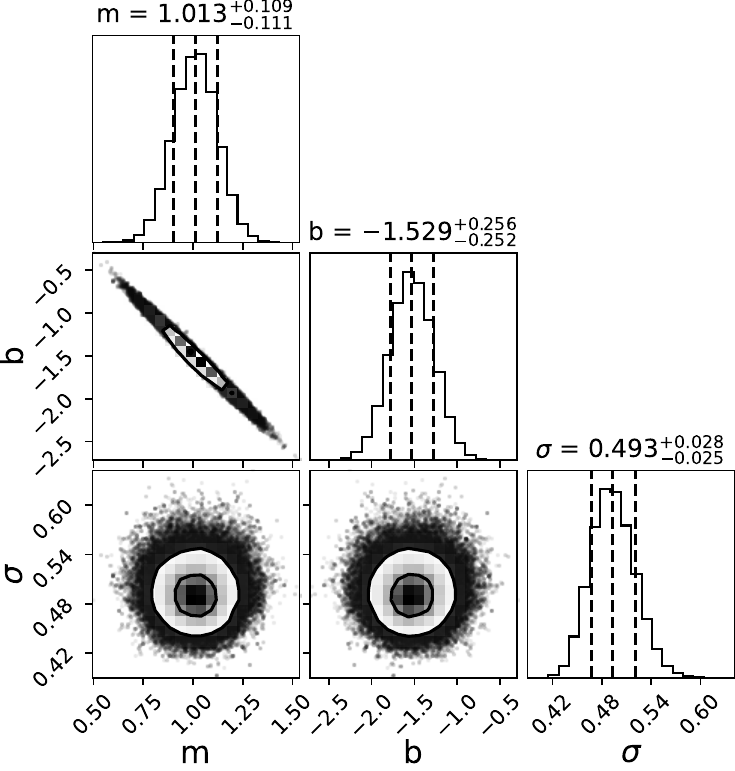}
    \includegraphics[angle=0,width=45mm]{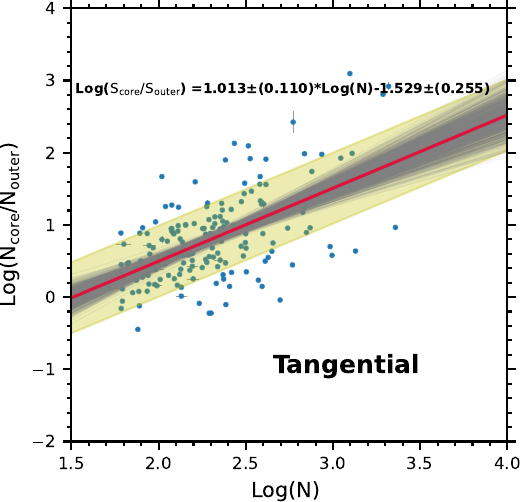}
    \includegraphics[angle=0,width=45mm]{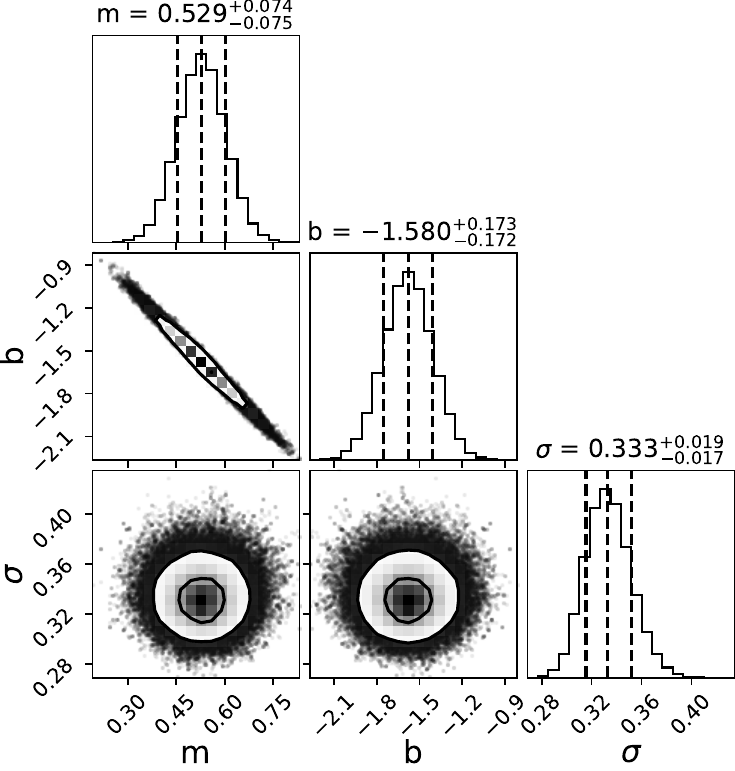}
    \includegraphics[angle=0,width=45mm]{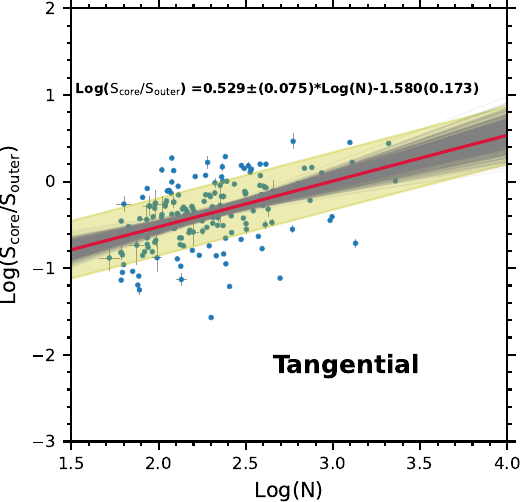}
    
    \caption{Relationships between the morphological stability parameters (N$_{\mathrm{core}}$/$\mathrm{N}_{\mathrm{outer}}$ and $\mathrm{S}_{\mathrm{core}}$/$\mathrm{S}_{\mathrm{outer}}$) and the number of members (N) of the sample OCs in the radial and tangential region, in the same style as Fig.~\ref{fig:all_N_s}.}
    \label{fig:ap_radia_tangential}
\end{figure*}

\begin{figure*}[htbp]
    \centering
    \includegraphics[angle=0,width=45mm]{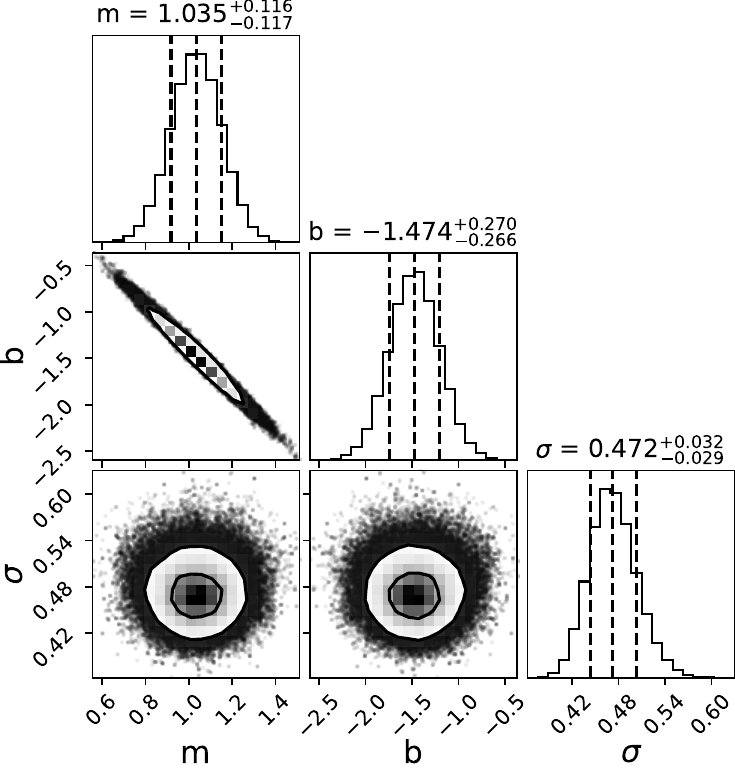}
    \includegraphics[angle=0,width=45mm]{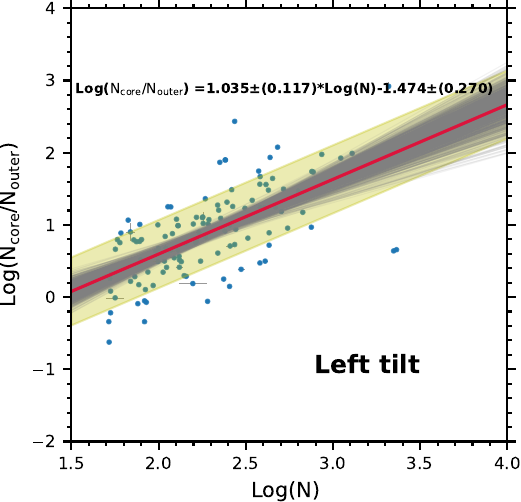}
    \includegraphics[angle=0,width=45mm]{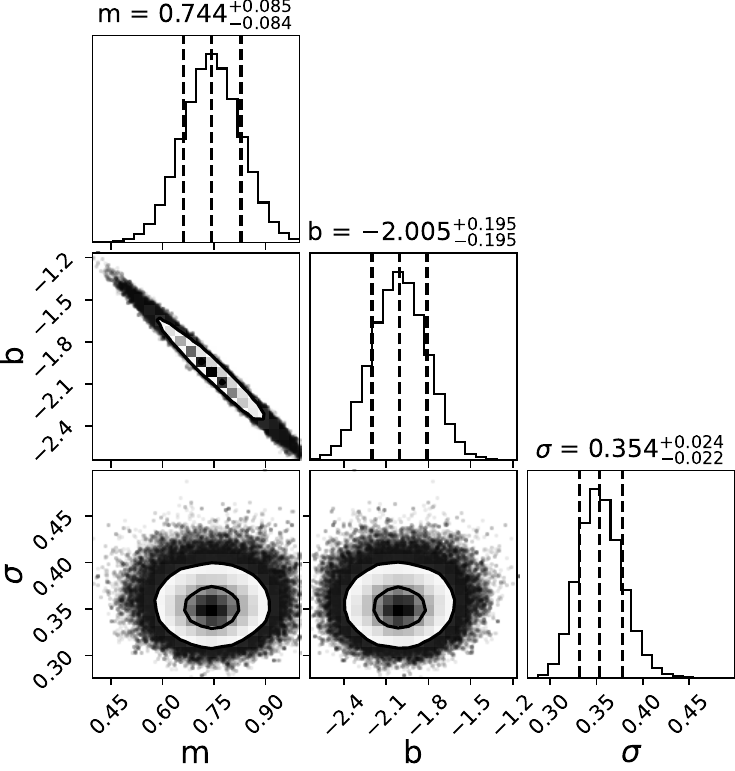}
    \includegraphics[angle=0,width=45mm]{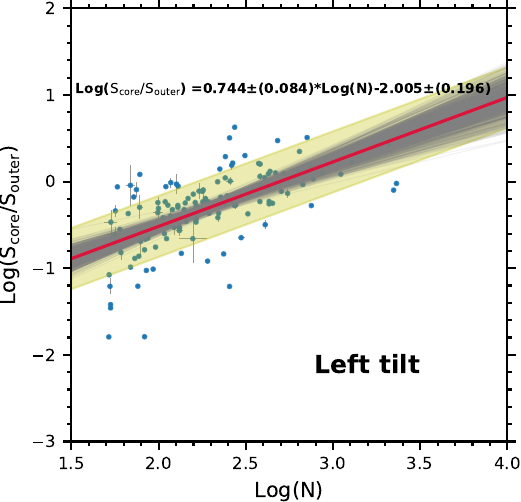}
    \includegraphics[angle=0,width=45mm]{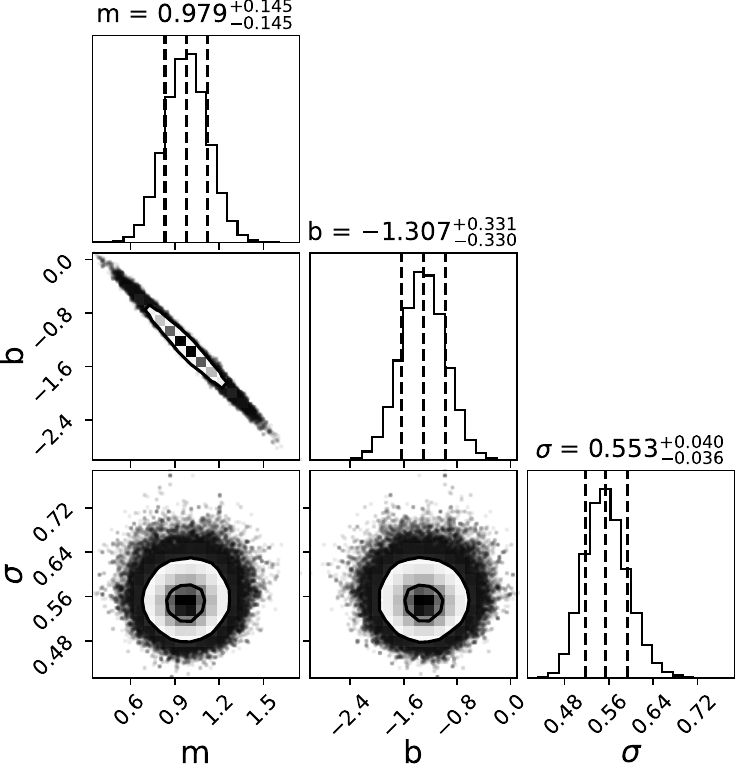}
    \includegraphics[angle=0,width=45mm]{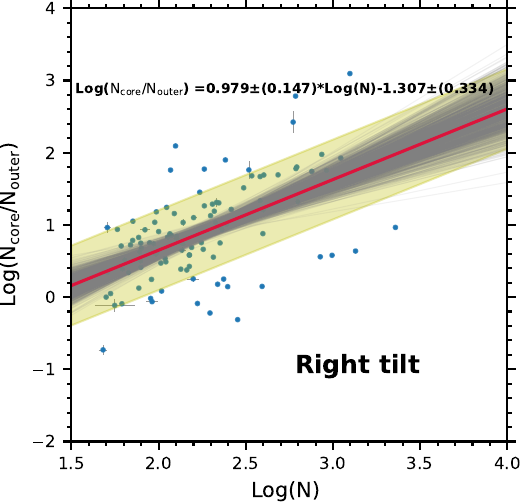}
    \includegraphics[angle=0,width=45mm]{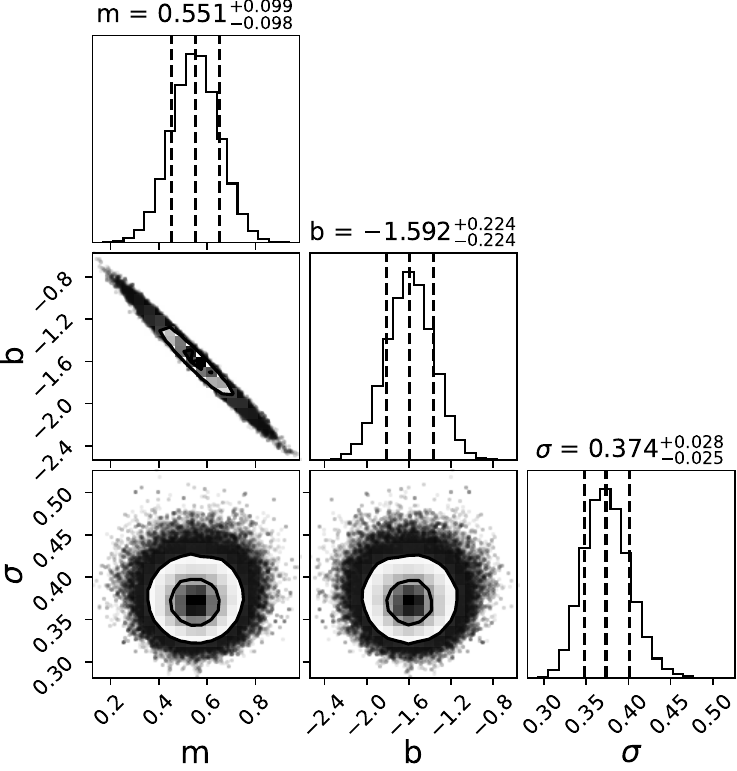}
    \includegraphics[angle=0,width=45mm]{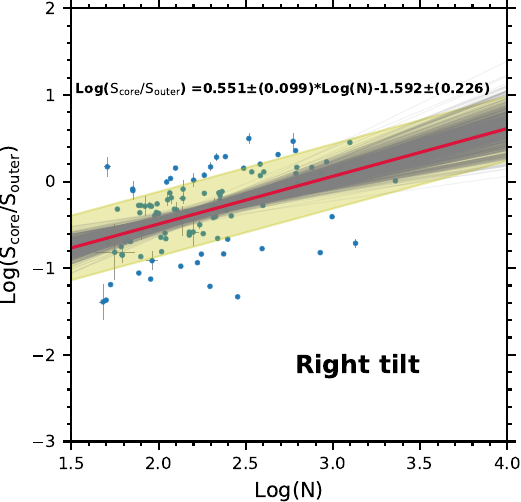}
    \caption{Same as Fig.~\ref{fig:ap_radia_tangential}, but the sample OCs in the ``left\_tilt'' and ``right\_tilt'' regions.}
    \label{fig:13}
\end{figure*}

\begin{figure*}[htbp]
	\centering
    \includegraphics[angle=0,width=160mm]{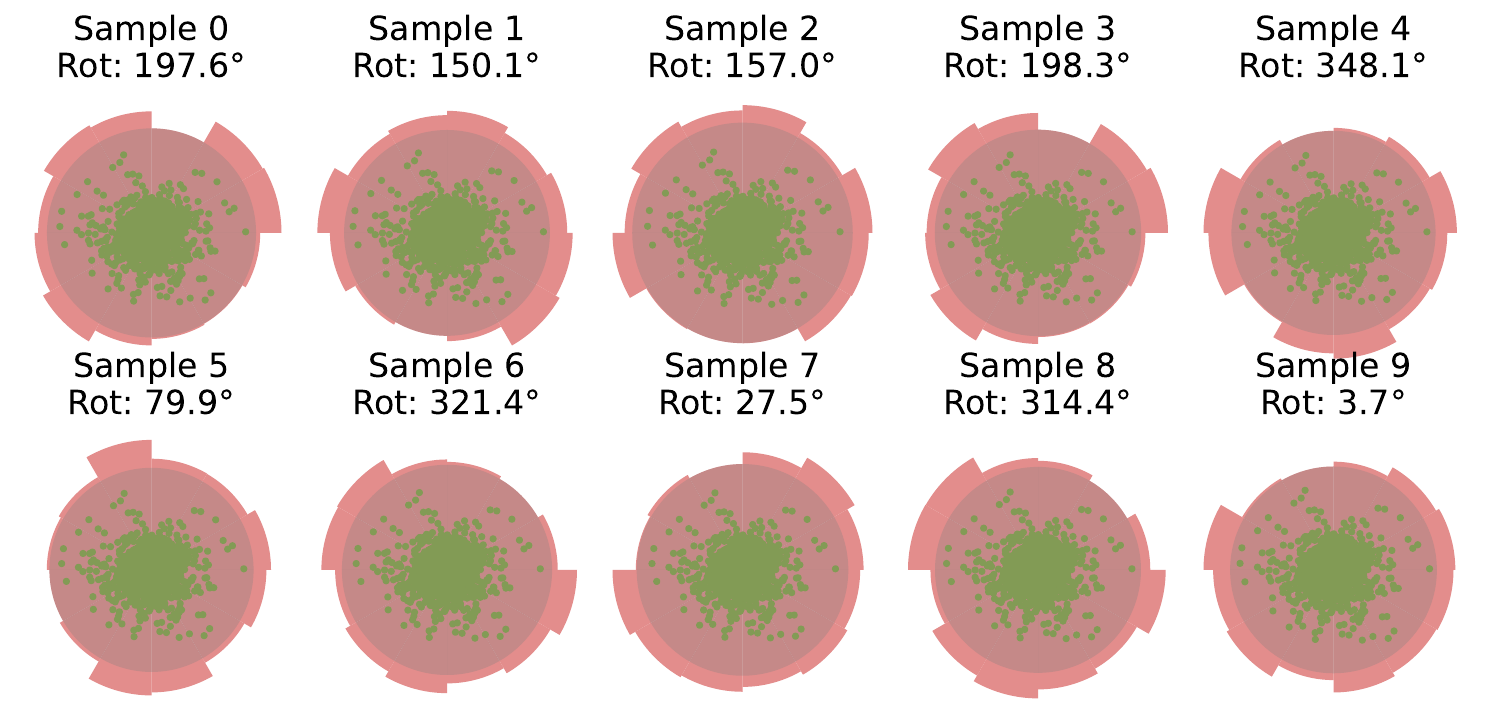}
	\caption{Rose diagrams of \texttt{NGC\_6791}. These were obtained by randomly sampling the starting angle ten times, but with a fixed tidal radius R$_{\text{t}}$.}
	\label{fig: 10_sample_fixed_Rt_rose}
\end{figure*}

\begin{figure*}[htbp]
	\centering
    \includegraphics[angle=0,width=160mm]{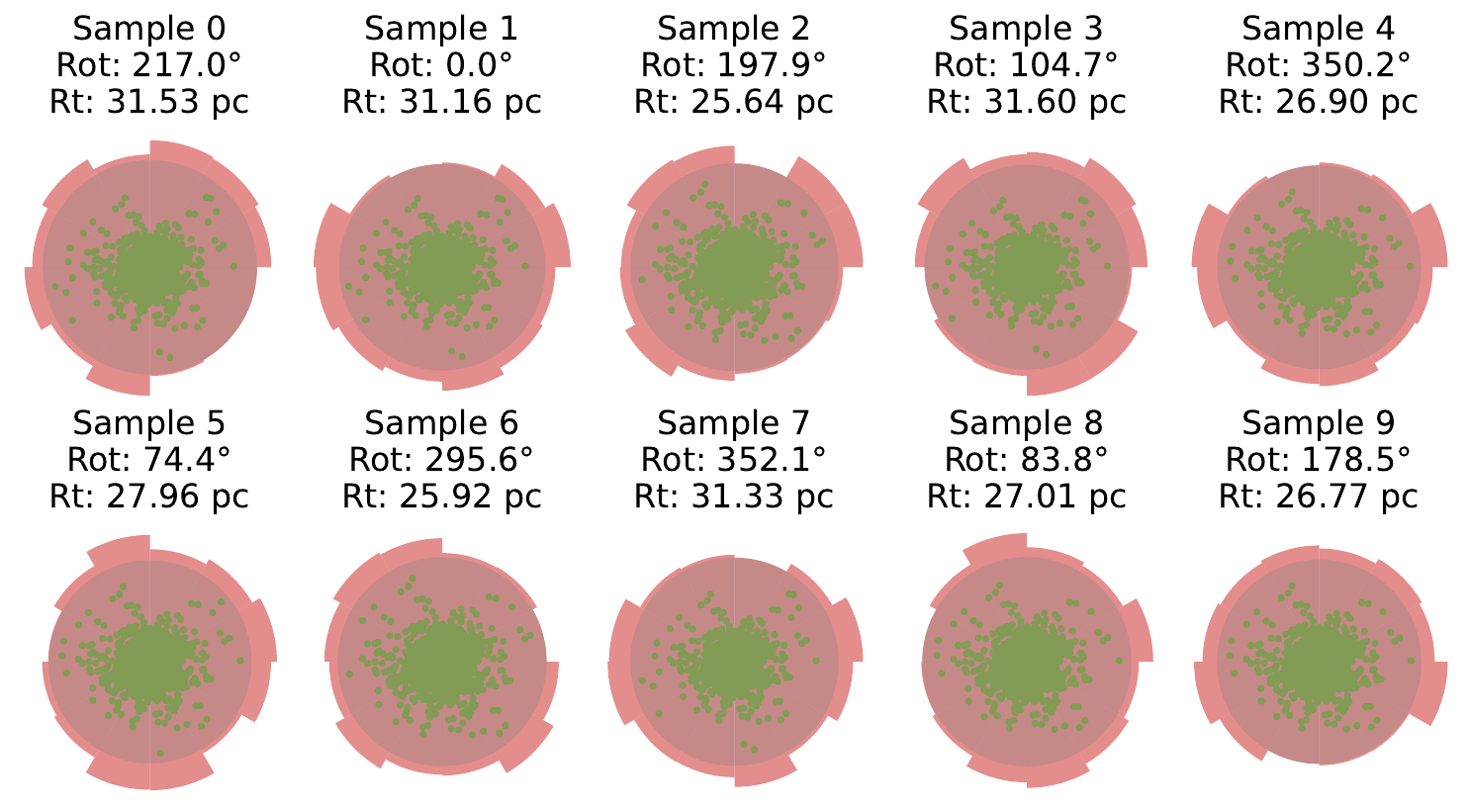}
	\caption{Same as Fig.~\ref{fig: 10_sample_fixed_Rt_rose}, but with ten random sampling iterations conducted simultaneously for the starting angle and tidal radius R$_{\text{t}}$.}
	\label{fig: 10_sample_rose_2}
\end{figure*}

\end{document}